\documentclass[twocolumn]{aastex61}

\accepted{for Publication in The Astrophysical Journal}

\shorttitle{Environment of a $z=6.6$ Quasar}
\shortauthors{Ota et al.}

\begin{document}



\title{Large Scale Environment of a $ z=6.61$ Luminous Quasar Probed by Ly$\alpha$ Emitters and Lyman Break Galaxies\footnote{Based on data collected at Subaru Telescope, which is operated by the National Astronomical Observatory of Japan.}}

\correspondingauthor{Kazuaki Ota}
\email{kota@ast.cam.ac.uk}

\author{Kazuaki Ota}
\altaffiliation{Kavli Institute Fellow}
\affil{Kavli Institute for Cosmology, University of Cambridge, Madingley Road, Cambridge, CB3 0HA, UK}
\affil{Cavendish Laboratory, University of Cambridge, 19 J.J. Thomson Avenue, Cambridge, CB3 0HE, UK}

\author{Bram P. Venemans}
\affiliation{Max-Planck Institut f\"ur Astronomie, K\"onigstuhl 17, D-69117, Heidelberg, Germany}

\author{Yoshiaki Taniguchi}
\affiliation{The Open University of Japan, 2-11, Wakaba, Mihama-ku, Chiba, Chiba 261-8586, Japan}

\author{Nobunari Kashikawa}
\affiliation{National Astronomical Observatory of Japan, 2-21-1 Osawa, Mitaka, Tokyo, 181-8588, Japan}
\affiliation{The Graduate University for Advanced Studies, 2-21-1 Osawa, Mitaka, Tokyo, 181-8588, Japan}

\author{Fumiaki Nakata}
\affiliation{Subaru Telescope, 650 North A'ohoku Place, Hilo, HI 96720, USA}

\author{Yuichi Harikane}
\affiliation{Department of Physics, Graduate School of Science, The University of Tokyo, 7-3-1 Hongo, Bunkyo, Tokyo 113-0033, Japan}
\affiliation{Institute for Cosmic Ray Research, The University of Tokyo, 5-1-5 Kashiwanoha, Kashiwa, Chiba 277-8582, Japan}

\author{Eduardo Ba\~{n}ados}
\altaffiliation{Carnegie-Princeton Fellow}
\affiliation{The Observatories of the Carnegie Institute of Washington, 813 Santa Barbara Street, Pasadena, CA 91101, USA}

\author{Roderik Overzier}
\affiliation{Observat\'{o}rio Nacional, Rua Jos\'{e} Cristino, 77. CEP 20921-400, S\~{a}o Crist\'{o}v\~{a}o, Rio de Janeiro-RJ, Brazil}

\author{Dominik A. Riechers}
\affiliation{Department of Astronomy, Cornell University, 220 Space Sciences Building, Ithaca, NY 14853, USA}

\author{Fabian Walter}
\affiliation{Max-Planck Institut f\"ur Astronomie, K\"onigstuhl 17, D-69117, Heidelberg, Germany}

\author{Jun Toshikawa}
\affiliation{Institute for Cosmic Ray Research, The University of Tokyo, 5-1-5 Kashiwanoha, Kashiwa, Chiba 277-8582, Japan}

\author{Takatoshi Shibuya}
\affiliation{Institute for Cosmic Ray Research, The University of Tokyo, 5-1-5 Kashiwanoha, Kashiwa, Chiba 277-8582, Japan}

\author{Linhua Jiang}
\affiliation{Kavli Institute for Astronomy and Astrophysics, Peking University, Beijing 100871, China}



\begin{abstract}
Quasars (QSOs) hosting supermassive black holes are believed to reside in massive halos harboring galaxy overdensities. However, many observations revealed average or low galaxy densities around $z\gtrsim6$ QSOs. This could be partly because they measured galaxy densities in only tens of arcmin$^2$ around QSOs and might have overlooked potential larger scale galaxy overdensities. Some previous studies also observed only Lyman break galaxies (LBGs; massive older galaxies) and missed low mass young galaxies like Ly$\alpha$ emitters (LAEs) around QSOs. Here we present observations of LAE and LBG candidates in $\sim700$ arcmin$^2$ around a $z=6.61$ luminous QSO using Subaru Telescope Suprime-Cam with narrow/broadbands. We compare their sky distributions, number densities and angular correlation functions with those of LAEs/LBGs detected in the same manner and comparable data quality in our control blank field. In the QSO field, LAEs and LBGs are clustering in 4--20 comoving Mpc angular scales, but LAEs show mostly underdensity over the field while LBGs are forming $30\times60$ comoving Mpc$^2$ large scale structure containing 3--$7\sigma$ high density clumps. The highest density clump includes a bright (23.78 mag in the narrowband) extended ($\gtrsim 16$ kpc) Ly$\alpha$ blob candidate, indicative of a dense environment. The QSO could be part of the structure but is not located exactly at any of the high density peaks. Near the QSO, LAEs show underdensity while LBGs average to $4\sigma$ excess densities compared to the control field. If these environments reflect halo mass, the QSO may not be in the most massive halo, but still in a moderately massive one.
\end{abstract}






\section{Introduction}

Several cosmological models and simulations predict that high redshift quasars (QSOs) hosting supermassive ($M_{\rm BH} \gtrsim 10^9M_{\odot}$) black holes (SMBHs) reside in the most massive dark matter halos and that their environments harbor galaxy overdensities formed by hierachical merging of many galaxies \citep[e.g.,][]{Romano-Diaz11,Costa14}. In contrast, some other simulations and studies suggest that such QSOs may not necessarily be in the most massive halos \citep[e.g.,][]{Overzier09,Angulo12,Fanidakis13,Orsi16}. To investigate the environment in which QSOs hosting SMBHs reside, it is essential to actually observe galaxies around QSOs and examine if they exhibit overdensities. Meanwhile, galaxy overdensities at early cosmic epochs, often called protoclusters, are thought to be progenitors of massive clusters of galaxies seen in the present-day universe \citep{Overzier16}. Hence, observing such galaxy overdensities, their associated galaxies and QSOs, if any, over cosmic time, we can understand how clusters, galaxies and SMBHs have formed and evolved and how overdense environments have affected galaxy formation and evolution. 

Some observations to date have found significant galaxy overdensities around QSOs at various epochs $z\sim 2$--6 \citep[e.g.,][]{Steidel05,Kim09,Capak11,Swinbank12,Husband13,Morselli14,Balmaverde17,Decarli17}. Conversely, other observations revealed average galaxy densities or even underdensities around $z\sim 2$--7 QSOs \citep[e.g.,][]{Francis04,Kashikawa07,Kim09,Banados13,Simpson14,Mazzucchelli17,Kikuta17,Uchiyama17}. This would imply that QSOs may not always be hosted by the most massive halos and/or the densest environments. 

However, some of these observations typically probed areas of at most tens of arcmin$^2$ around QSOs \citep{Kim09,Banados13,Mazzucchelli17,Simpson14}. In regions close to a luminous QSO, the QSO intense ultraviolet (UV) radiation may be able to suppress galaxy formation by evaporating gas in dark halos before it cools and forms stars (QSO negative feedback), possibly resulting in lack of observed galaxies even if an underlying halo excess may exist \citep[e.g.,][]{Efstathiou92,Thoul96,Benson02,Kashikawa07,Okamoto08}. The QSO radiation also ionizes neutral hydrogen in the surrounding intergalactic medium (IGM) which can shield gases that form galaxies in halos from the QSO radiation. At the reionization epoch, the fraction of residual neutral hydrogen in the IGM may vary significantly from one line of sight to another. Thus, various combination of QSO radiation strength and amount of residual neutral hydrogen that changes from site to site may cause a wide variety of galaxy densities (from overdensities to underdensities) observed in the close vicinities of $z \gtrsim 6$ QSOs \citep{Kim09}. To overcome this issue, we have to observe galaxy sky distributions and number densities over much wider areas around $z \gtrsim 6$ QSOs.  

A few previous studies have observed wide areas (hundreds of arcmin$^2$) around $z \gtrsim 6$ QSOs. \citet{Utsumi10} imaged the $z=6.417$ QSO, CFHQS J2329--0301, hosting a black hole with $M_{\rm BH} \sim 2.5 \times 10^8M_{\odot}$ \citep{Willott10b} using the 8.2m Subaru Telescope with its wide field ($27' \times 34'$) prime focus camera, Suprime-Cam \citep{Miyazaki02}, in the broadband $i'$, $z'$ and $z_{\rm R}$ filters. They claim that there is a possible large scale ($\sim 6 \times 6$ physical Mpc$^2$) overdensity of $z'$-band dropout Lyman break galaxies (LBGs) around the QSOs. However, as already pointed out by \citet{Willott11}, most of them are visible in the $i'$-band images, and thus some of them might be located at $z<6$ (unless they have some detectable fluxes left in the spectral trough bluewards of Ly$\alpha$) because the red edge of the $i'$-band is $\lambda \sim 8500$\AA~corresponding to Ly$\alpha$ at $z\sim6$.

Meanwhile, \citet{Morselli14} and \citet{Balmaverde17} imaged $23' \times 25'$ areas around four $z=5.95$--6.41 QSOs all hosting $M_{\rm BH} = 1.0$--$4.9 \times 10^9M_{\odot}$ SMBHs using the Large Binocular Camera at the Large Binocular Telescope or the Wide-field InfraRed Camera (WIRCam) at the Canada-France-Hawaii Telescope (CFHT) and broadbands. They found that all the QSO fields show LBG ($i$-band dropout) densities higher than those in their comparison blank sky fields on a large scale ($\sim 8 \times 8$ physical Mpc$^2$). Hence, these QSOs might live in massive halos embedded in large scale (tens of physical Mpc$^2$) galaxy overdensities.

Nonetheless, these wide-field studies as well as many other previous smaller-field surveys observed only color-selected LBG candidates around $z \gtrsim 6$ QSOs. Their potential redshifts span a wide range $\Delta z \sim 1$ as they are detected by broadband filter dropout technique. On the other hand, a galaxy overdensity at $z\sim6$ seems to have a size of $\Delta z < 0.1$ \citep[e.g., see][for the case of a spectroscopically confirmed $z\sim6$ protocluster]{Toshikawa12,Toshikawa14}. Thus, even if we find an apparent LBG overdensity around a QSO in a projected sky plane, some of the LBGs may not be associated with the overdensity and the QSO. Also, as LBGs tend to be relatively massive older galaxy population, the previous studies missed young low mass galaxies such as Ly$\alpha$ emitters (LAEs) around QSOs. As LAEs are usually observed by detecting their Ly$\alpha$ emission in a narrowband filter, their redshifts span a very narrow range $\Delta z \sim 0.1$. Hence, if we find overdensities of LAEs around QSOs, they are likely associated with the QSOs, possibly forming protoclusters. However, until recently, no QSO at $z>6$ whose redshift matches a bandpass of a narrowband filter targeting $z>6$ Ly$\alpha$ emission has been found, and studies of LAEs around QSOs using narrowbands have been carried out only up to $z\sim 5.7$ \citep[e.g.,][]{Kashikawa07,Banados13,Kikuta17,Mazzucchelli17}. This has made it difficult to reliably estimate real number densities of galaxies of various ages and masses in a narrow redshift range around $z>6$ QSOs. 

Another interesting aspect that has been missed when observing $z>6$ QSOs is the possible impact of galaxy overdensities on the structure of cosmic reionization. There are competing theories arguing whether ionization of neutral hydrogen occurs more rapidly in denser environments where galaxies are clustering, or not \citep[e.g.,][]{Morales10}. If $z>6$ QSOs are associated with galaxy overdensities, they can be laboratories to examine environmental effects on reionization if we can estimate galaxy densities around the QSOs accurately and have any observational probe of reionization. LAEs can be the probe as their Ly$\alpha$ luminosity function (LF) could decline or their spatial distribution could modulate as neutral hydrogen absorbs or scatters Ly$\alpha$ photons from LAEs \citep{Rhoads01,McQuinn07}. However, no previous study has observed LAEs around $z>6$ QSOs and investigated the reionization state in overdense environments. 

Eventually, if we observe sky areas much larger than QSO radiation fields (which could suppress formation of low mass galaxies within $\sim 1$--3 physical Mpc from the QSOs; e.g., see \citet{Kashikawa07} and Section \ref{Feedback} in this paper) and detect both LBGs and LAEs (galaxies in wide ranges of masses and ages) with a wide field camera and a narrowband ($\Delta z \sim 0.1$) matched to the QSO redshifts, we can reveal if $z > 6$ QSOs are embedded in galaxy overdensities (most massive halos) and how overdense environment affects early galaxy formation and reionization. 

Very recently, \citet{Goto17} made a custom narrowband filter for the Subaru Telescope Suprime-Cam whose bandpass matches the redshift of the $z=6.417$ QSO, CFHQS J2329--0301 \citep{Willott10b}, which was observed by \citet{Utsumi10} (see above). Despite their wide-field ($27' \times 34'$) narrowband imaging, they did not detect any LAEs around the QSO. They mentioned that the QSO UV radiation could suppress formation of LAEs in lower halo masses ($< 10^{10}M_{\odot}$) within $\sim1$ physical Mpc from the QSO. However, they could not explain why they did not detect any LAEs over the most of their survey area probably not affected by the QSO radiation. The QSO may not reside in any galaxy overdensity environment even on a large scale as it hosts a relatively less massive black hole with $M_{\rm BH} \sim 2.5 \times 10^8M_{\odot}$ \citep{Willott10b}. However, even if this is the case, it cannot still explain why \citet{Goto17} did not detect any LAEs in their entire survey area. One possibility is the shallowness of their images, especailly the narrowband image, which may have resulted in a potentially lower line sensitivity than expected. This is because some ($\sim$ 1/4) of their narrowband exposures were taken under poor transparency and because the strong skyline existing within the bandpass wavelengths of their narrowband filter may possibly reduce the sensitivity (Goto et al.~2017, private communication). Another possibility is that the range where the QSO UV radiation is effective on suppressing formation of LAEs is wider than expected. 

To clarify environments in which QSOs hosting SMBHs reside at reionization epoch $z\gtrsim6$, we have to conduct wide-field observations of both LAEs and LBGs around QSOs hosting an SMBH with $M_{\rm BH} \geq 10^9M_{\odot}$ to a sufficiently deep flux limit to which we already know how many LAEs and LBGs we can expect to detect if a QSO does not exist based on previous LAE/LBG studies or by carefully designed equivalent observations of LAEs/LBGs in a comparison sky field where there is no QSO.

\begin{figure}
\epsscale{1.17}
\plotone{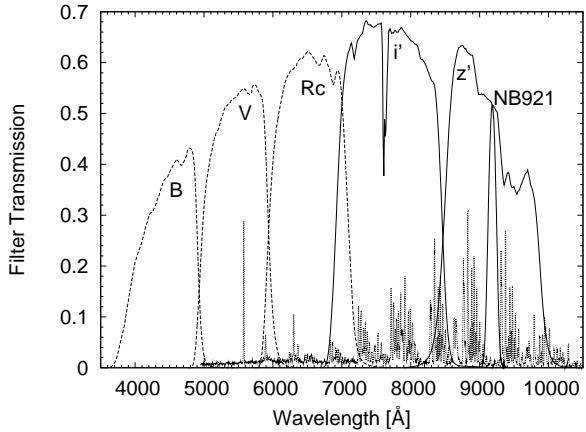}
\caption{Transmission curves of the Suprime-Cam broadband and narrowband filters used for our study ($i'$, $z'$ and NB921; solid curves) as well as the broadbands we did not use but \citet{Taniguchi05} additionally used to select $z\sim6.6$ LAEs in SDF ($B$, $V$ and $R_c$; dashed curves, see Section \ref{LAE-Selections} for details). The transmission curves include the CCD quantum efficiency (MIT-Lincoln Laboratory CCDs for $B$, $V$ and $R_c$, and Hamamatsu CCDs for $i'$, $z'$ and NB921), the reflection ratio of the telescope primary mirror, correction for the prime focus optics and transmission to the atmosphere (airmass sec $z=1.2$). The OH night sky lines are also overplotted with the dotted curve.\label{FilterTransmission}}
\end{figure}
\begin{figure}
\epsscale{1.5}
\hspace*{-2cm}
\plotone{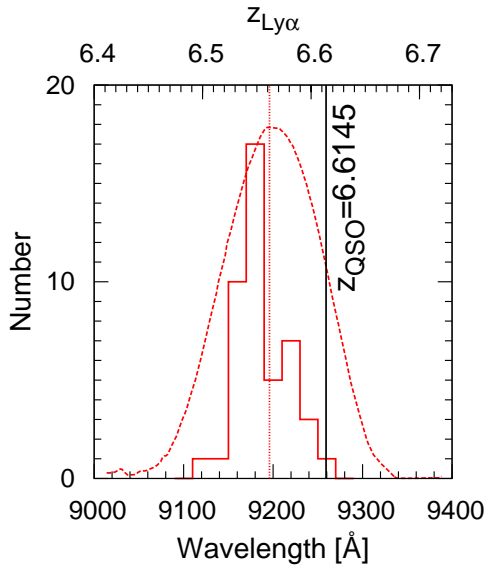}
\caption{The transmission curve of the Subaru Suprime-Cam narrowband NB921 filter (dashed curve) and observed wavelength distribution of the Ly$\alpha$ line peaks of the $z \sim 6.6$ LAEs in the SDF (Control Field) previously detected in the NB921 imaging by \citet{Taniguchi05} and spectroscopically confirmed by \citet{Kashikawa06,Kashikawa11} (solid line). The top axis indicates the redshift of Ly$\alpha$ emission ($z_{{\rm Ly}\alpha}$) corresponding to the wavelength at the bottom axis. The vertical dotted and solid lines are the central wavelength (9196\AA) of the NB921 filter and the redshift of the QSO J0305--3150, respectively.\label{NB921_LyaPeakDist}}
\end{figure}

Recently, \citet{Venemans13} discovered a $z = 6.6145\pm0.0001$ QSO, VIKING J030516.92--315056.0 (hereafter J0305--3150), hosting a $M_{\rm BH} \sim 1.0 \times 10^9M_{\odot}$ SMBH. This redshift has been reliably measured from the [CII] line detected by the Atacama Large Millimeter/submillimeter Array \citep{Venemans16} and fortunately fits in the bandpass of the narrowband filter NB921 ($\lambda_{\rm c} = 9196$\AA~and $\Delta \lambda_{\rm FWHM} = 132$\AA~corresponding to $z\sim6.51$--6.62 Ly$\alpha$ emission; see Figures \ref{FilterTransmission} and \ref{NB921_LyaPeakDist}) of the Subaru Telescope Suprime-Cam. This gives us the first opportunity for a wide-field (hundreds of arcmin$^2$) narrowband and broadband search for both LAEs and LBGs around a QSO hosting an SMBH in the reionization epoch at $z>6$. At this moment, J0305--3150 is the highest redshift QSO for which such observations are possible. Moreover, the red-sensitive CCDs of the Suprime-Cam allow us to detect faint LAEs and LBGs at $z\sim6.6$ to fairly deep limits with modest amounts of observing time. 

%
\vspace*{0.5cm}

In addition, the NB921 filter can detect $z\sim6.6$ LAE candidates in a very narrow redshift range ($\Delta z \sim0.1$) with a fairly low contamination rate. \citet{Kashikawa06,Kashikawa11} spectroscopically identified 42 out of 58 $z\sim6.6$ LAE candidates that \citet{Taniguchi05} photometrically detected in the Subaru Deep Field \citep[SDF,][]{Kashikawa04} using NB921 and found that only $\sim 2$--19\% of them are contaminants. Hence, by observing the region around the $z=6.61$ QSO J0305--3150 in the NB921 filter and using the same LAE selection criteria, we can photometrically detect LAE candidates that are mostly real LAEs at the redshifts very close to that of the QSO. Also, these previous studies have constructed the robust $z\sim6.6$ LAE sample by using the Suprime-Cam broadband and NB921 imaging of the SDF \citep{Taniguchi05,Kashikawa06,Kashikawa11}. The SDF is a general blank field, and there is no $z\sim6.6$ QSO, no over/underdensity of $z\sim6.6$ LAEs and LBGs and no clustering of them (we show this in the subsequent sections). Hence, we can use the SDF and the LAE sample (and the LBG sample we construct in this paper) in this field as the control field and the control sample, the rigorous baseline that can be compared with the LAEs and the LBGs we detect around the $z=6.61$ QSO J0305--3150 to reveal the potential LAE/LBG overdensities, if any. 

Meanwhile, it should be also noted that although the redshift of the $z=6.61$ QSO J0305--3150 is in the bandpass of the NB921 filter, it is in the red side of the bandpass where the sensitivity to LAEs is lower. Figure \ref{NB921_LyaPeakDist} shows the transmission curve of the NB921 filter and observed wavelength distribution of the Ly$\alpha$ line peaks of the $z \sim 6.6$ LAEs in the SDF (control field) previously detected in the NB921 imaging by \citet{Taniguchi05} and spectroscopically confirmed by \citet{Kashikawa06,Kashikawa11}. As seen in the figure, the NB921 filter has a better sensitivity to LAEs at the blue side of its bandpass as it can detect both Ly$\alpha$ emission and more UV continuum fluxes while the red side of the bandpass detects Ly$\alpha$ emission and less UV continuum fluxes. Hence, we have to keep in mind that we might miss detecting some fraction of LAEs around the $z=6.61$ QSO, especially those located at the further side of the QSO. 

In this paper, we present the result of our Subaru Suprime-Cam wide field broadband and narrowband NB921 search for overdensities of both LAEs and LBGs around the $z=6.61$ QSO J0305--3150. This paper is organized as follows. In Section 2, we describe our observations of the QSO field, the data reduction and the control field (SDF) data. Then, in Section 3, we select LAE and LBG candidates in the QSO and the control fields. We compare sky distributions, number densities and clustering of the LAE and LBG candidates in the QSO field with those in the control field to investiagte the possibility of existence of any galaxy overdensities around the QSO in Section 4. We summarize and conclude our study in Section 5. In Appendix A, we check for the nonexistence of $z\sim6.6$ QSOs in the control field. Finally, in Appendix B, we examine the contamination rate of our photometric LAE samples due to not imposing non-detections in the broadbands bluewards of $z\sim6.6$ Ly$\alpha$ as a part of LAE selection criteria. Throughout, we adopt AB magnitudes \citep{Oke74} and a concordance cosmology with $(\Omega_m, \Omega_{\Lambda}, h)=(0.3, 0.7, 0.7)$, unless otherwise specified.  
%


\vspace*{0.5cm}


\begin{deluxetable*}{ccccccccc}
\tabletypesize{\scriptsize}
\tablecaption{Summary of the Imaging Data of the $z\sim 6.6$ QSO Field and the Control Field SDF\label{ImagingData}}
\tablewidth{510pt}
\tablehead{
\colhead{Field} & \colhead{Band} & \colhead{$t_{\rm exp}$$^{\rm c}$} & \colhead{PSF Size$^{\rm d}$} & \colhead{Area$^{\rm e}$} & \colhead{$m_{\rm lim}$$^{\rm f}$} &$N_{\rm LAE}$$^{\rm g}$ & $N_{\rm LBG}$$^{\rm g}$ & \colhead{Observation Date}\\
\colhead{} & \colhead{} & \colhead{(min)} & \colhead{(arcsec)} & \colhead{(arcmin$^2$)} & \colhead{(mag)} & & & \colhead{} 
}
\startdata
QSO$^{\rm a}$ & $i'$  & 128 & 0.91 (0.91) & 697 & 27.0 & 14 & 53 & 2014 Aug 25/27\\
            & $z'$  & 220 & 0.91 (0.83) & 697 & 26.5 &    &    & 2014 Aug 26/27\\
            & NB921 & 380 & 0.91 (0.77) & 697 & 26.5 &    &    & 2014 Aug 22/24/25\\
\hline
SDF$^{\rm b}$ & $i'$  & 801 & 0.98 & 876 & 27.4 & 63 & 32 & 2002 Apr 11/14, May 6, 2003 Mar 31, Apr 2/24/25/29/30\\
        & $z'$  & 504 & 0.98 & 876 & 26.6 &    &    & 2002 Apr 9/14, 2003 Mar 7, Apr 1/28\\
        & NB921 & 899 & 0.98 & 876 & 26.5 &    &    & 2002 Apr 9/11/14, May 6, 2003 Mar 7/8, Apr 24\\
\enddata
\tablenotetext{a}{The images of the QSO field were taken by Suprime-Cam with Hamamatsu red-sensitive fully depleted CCDs \citep{Kamata08}.}
\tablenotetext{b}{The SDF public version 1.0 images \citep{Kashikawa04} taken by Suprime-Cam with MIT-Lincoln Laboratory (MIT-LL) CCDs \citep{Miyazaki02}.}
\tablenotetext{c}{Total exposure times. The differences in exposure times between the images of the QSO field and the Control Field SDF to reach the similar depths originate from the different CCDs of the Suprime-Cam used for the observations of each field.}
\tablenotetext{d}{The FWHM of PSFs. The original images were convolved to have the common PSFs for the aperture photometry purpose. The ones in the parentheses are the original PSF FWHMs of the QSO field images before the convolution.}
\tablenotetext{e}{The image area finally used for our science analysis.}
\tablenotetext{f}{The $3\sigma$ limiting magnitude measured in a $2''$ diameter aperture.}
\tablenotetext{g}{The numbers of LAE and LBG candidates detected in each field.}
\end{deluxetable*}

\section{Observation and Data}
\subsection{The $z=6.61$ QSO Field\label{QSOQSOField}}
We imaged the field centered at the $z=6.61$ QSO J0305--3150 with Subaru Telescope Suprime-Cam in the broadbands $i'$ and $z'$ as well as the narrowband NB921. We aimed to reach depths in these bands that are as similar as possible to those of the control field (SDF) images. The observations were carried out during dark nights on 2013 November 27, and 2014 August 22 and 24--27. The sky conditions were partly clear/cloudy with a seeing of $\sim 0.''8$--$1.''3$ in 2013 and photometric with a seeing of $\sim 0.''5$--$0.''9$ in 2014. We took 240, 120--240 and 1200 second individual exposure frames with the $i'$, $z'$ and NB921 bands, respectively, using eight-point dithering patterns. 

We reduced the exposure frames using the software SDFRED2 \citep{Ouchi04,Yagi02} in the same standard manner as in \citet{Kashikawa04} and \citet{Ota08}, including bias subtraction, flat-fielding, distortion correction, matching of point spread functions (PSFs) between the CCD chips, sky subtraction and masking of the shadow of the auto guider probe. Then, the dithered exposure frames were matched and stacked. We did not eventually use the exposures taken in 2013 as they were obtained under poor transparency conditions. The integration times of these stacked $i'$, $z'$ and NB921 images amount to 2.1, 3.7 and 6.3 hours, respectively. The $i'$ and $z'$ images were then registered to the NB921 image by using the positions of common stellar objects detected in these images. Finally, we corrected the astrometry of the $i'$, $z'$ and NB921 images by matching pixel positions of the stars in the images to the coordinates of them in the USNO-B1.0 catalog \citep{Monet03} with the WCSTools version 3.8.1 \citep{Mink99}. 

Meanwhile, images of the spectrophotometric standard star GD71 \citep{Oke90} taken in all the bands during the observations in 2014 were used to calibrate the photometric zero points. We checked the zero points by comparing the colors of stellar objects detected in the $i'$, $z'$ and NB921 images of the QSO field and 175 Galactic stars calculated from spectra given in \citet{GunnStryker83}\footnote[1]{Taken from ftp://ftp.stsci.edu/cdbs/grid/gunnstryker/} in the $z'-{\rm NB921}$ versus ${\rm NB921}-i'$ diagram. We selected the stellar objects in the QSO field by running SExtractor version 2.8.6 \citep{BA96} on the images and using the criteria with the SExtractor parameters {\tt CLASS\_STAR} (stellarity) $> 0.98$ and {\tt FLAGS} $=0$ (no blending with neighboring objects). We found that the sequence of the stellar objects in the QSO field was offset from that of the \citet{GunnStryker83}'s Galactic stars by $\sim +0.15$--0.20 mag in $z'-{\rm NB921}$ and $\sim -0.15$--0.20 mag in ${\rm NB921}-i'$. Thus, we corrected the zero point of only the NB921 image by $+0.20$ mag and did not correct those of the $i'$ and $z'$ band images, by which colors of the two stellar sequences became consistet within $\sim 0.05$ mag. Finally, the zero points of the QSO field images are ($i'$, $z'$, NB921) $=$ (33.52, 32.30, 32.19) mag ADU$^{-1}$. The summary of our imaging data is given in Table \ref{ImagingData}.


\subsection{The Control Field -- Subaru Deep Field\label{ControlField}}
The objective of this study is to investigate how different the galaxy environment around the $z=6.61$ QSO is from a general field where there is no $z \sim 6.6$ QSO, no over/underdensity of LAEs/LBGs and no clustering of these galaxies. We choose the SDF as our comparison general field (hereafter ``Control Field'') because the previous studies have established a robust sample of $z\sim6.6$ LAEs with low contmination in this field \citep{Kodaira03,Taniguchi05,Kashikawa06,Kashikawa11}. Furthermore, it has been shown that the sky distribution of these LAEs is quite homogeneous and does not show any clustering based on two-point angular correlation function (ACF), two-dimensional Kolmogorov-Smirnov test and the void probability function analyses \citep{Kashikawa06}. The SDF also corresponds to one pointing of the Suprime-Cam and has comparable sky area and survey volume to those of the QSO field (see Table \ref{ImagingData}). 

In addition, we should also note and address the following four points when we adopt the SDF as the Control Field. (i) We confirm that there is not any $z \sim 6.6$ QSOs in the SDF as shown in Appendix A. (ii) Our SDF $z\sim6.6$ LAE sample is slightly different from that in the previous studies of \citet{Taniguchi05} and \citet{Kashikawa06,Kashikawa11} in that we use LAE selection criteria without $B$, $V$ and $R_c$ bands (see Figure \ref{FilterTransmission} and the criteria (\ref{Criteria-1}) and (\ref{Criteria-2}) in Section \ref{LAE-Selections}). However, we confirm that the contamination rate is still low in Appendix B and that the LAE candidates still exhibit a sky distribution without any over/underdensity nor clustering as shown in Sections \ref{NdensityContours} and \ref{ACF}. (iii) We newly detect LBG candidates in the SDF and confirm that their sky distribution shows no over/underdensity nor clustering as shown in Sections \ref{NdensityContours} and \ref{ACF}. (iv) The photometric zero points of the $i'$, $z'$ and NB921 images of the SDF were calibrated by \citet{Kashikawa04} in the similar way as we did for the QSO field images in Section \ref{QSOQSOField} by using colors of the \citet{GunnStryker83}'s Galactic stars. Using these photometric zero points, we confirm that colors of stellar objects in the SDF are consistent with those of the \citet{GunnStryker83}'s Galactic stars and stellar objects in the $z=6.6$ QSO field within $\sim 0.05$ mag in the $z'-{\rm NB921}$ versus ${\rm NB921}-i'$ diagram. Thus, the calibrations of the photometric zero points are consistent between the QSO and Control fields.

%
%
%
%
%
%
%
%

\begin{figure*}
\epsscale{1.17}
\plottwo{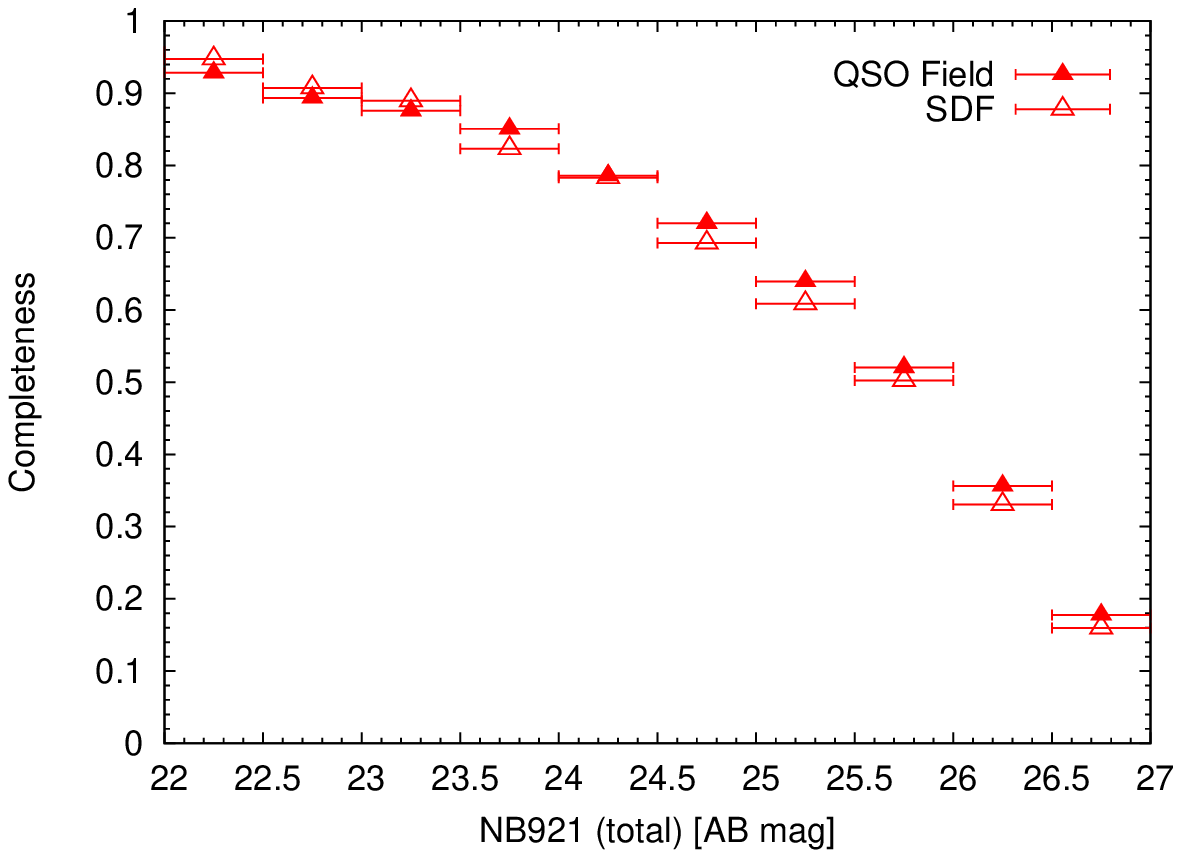}{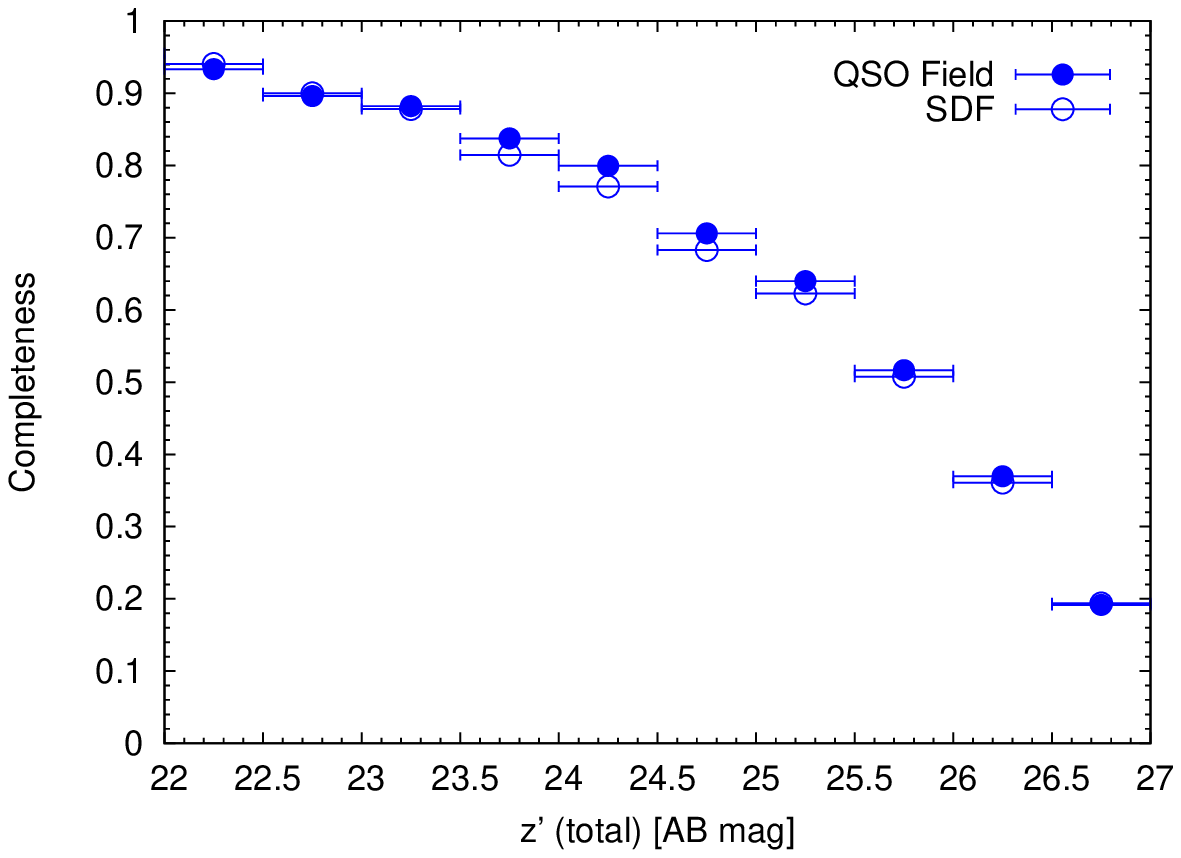}
\caption{Detection completeness of the NB921 (left) and $z'$-band (right) images of the QSO field and the Control Field (SDF) per 0.5 mag bin. We can see that the completenesses in the NB921 or $z'$ band are almost the same between the two fields.\label{Completeness}}
\end{figure*}

\section{Galaxy Candidate Selection}
\subsection{Photometry, Object Catalogs and Image Depths\label{Photometry}}
In order to detect LAEs and LBGs in the QSO field, we performed photometry to make the NB921-detected and $z'$-detected object catalogs. The original PSFs of the $i'$, $z'$ and NB921 images were $0.''91$, $0.''83$ and $0.''77$, respectively. We convolved the PSFs of the $z'$ and NB921 images to match that of the $i'$ band image (the worst of the three) because we have to calculate the $i' - z'$ and $z' -$ NB921 colors of objects by measuring the $i'$, $z'$ and NB921 magnitudes using the same aperture in order to select LAE and LBG candidates (see Section \ref{LAE-Selections}--\ref{LBG-Selections}). As shown in Table \ref{ImagingData}, the final PSFs ($0.''91$) of the QSO field images are slightly better but comparable to those of the Control Field images ($0.''98$).

We used the SExtractor version 2.8.6 \citep{BA96} for source detection and photometry. The Suprime-Cam CCDs have a pixel size of $0.''202$ pixel$^{-1}$. We considered an area larger than five contiguous pixels with a flux (mag arcsec$^{-2}$) greater than $2\sigma$ (i.e. two times the background rms) to be an object. Object detection was first made in the NB921 ($z'$) image, and then photometry was performed in the $i'$, $z'$ and NB921 images to detect LAE (LBG) candidates, using the double-image mode. We measured $2''$ diameter aperture magnitudes of detected objects with {\tt MAG\_APER} parameter and total magnitudes with {\tt MAG\_AUTO}. We used a $2''$ aperture, about twice the PSF of the $i'$, $z'$ and NB921 images of the both QSO and Control fields, to measure colors of objects, especially faint ones, with a good signal-to-noise (S/N) ratio. The NB921-detected and $z'$-detected object catalogs were constructed by combining the photometry in all the $i'$, $z'$ and NB921 bands. 

Meanwhile, we also measured the limiting magnitudes of the images by placing $2''$ apertures in random blank positions excluding the low S/N regions near the edges of the images (see Section \ref{LAE-Selections} and Section \ref{LBG-Selections} for the details of removing such edge regions). They are ($i'$, $z'$, NB921) $=$ (27.0, 26.5, 26.5) at $3\sigma$. As shown in Table \ref{ImagingData}, these depths are only slightly shallower than or comparable to those of the $i'$, $z'$ and NB921 images of the Control Field.

On the other hand, for the Control Field, we use the SDF public version 1.0 images and NB921-detected and $z'$-detected object catalogs\footnote[2]{Available from http://soaps.nao.ac.jp/SDF/v1/index.html} \citep{Kashikawa04} to detect LAE and LBG candidates as the control samples. We do not use the SDF $i'$ and $z'$ band images deeper than these public images \citep{Poznanski07,Graur11,Toshikawa12}. This is because the well-established $z\sim6.6$ LAE sample constructed by the previous studies by the SDF project \citep{Kodaira03,Taniguchi05,Kashikawa06,Kashikawa11} are based on these public images and catalogs and also because the depths of the public $i'$, $z'$ and NB921 images are comparable to those of the $z=6.61$ QSO field images as shown in Table \ref{ImagingData}. 

Based on \cite{Schlegel98}, we estimate the Galactic extinction to be $E(B-V)=0.0123$ ($A_{\lambda}=0.026$, 0.018 and 0.018 mag for $i'$, $z'$ and NB921 bands) in the QSO field and  $E(B-V)=0.0173$ ($A_{\lambda}=0.036$, 0.026 and 0.026 mag for $i'$, $z'$ and NB921 bands) in the Controal Field, respectively. As the amount of Galactic extinction in each band in each field and the difference between the two fields are negligibly small, we do not correct the magnitudes of detected objects for Galactic extinction.    

\subsection{Detection Completeness\label{CompletenessSec}}
What fraction of real objects in an image we can reliably detect by photometry depends on the magnitudes and blending of objects. To examine what fraction of objects in the NB921 and $z'$ images SExtractor can detect or fails to detect to fainter magnitude, we measured the detection completeness of our photometry as it is important to correct for it when we derive the number counts of LAEs and LBGs and the Ly$\alpha$ LFs of LAEs later (see Sections \ref{NC} and \ref{LyaLFSec}). 

Using the IRAF task {\tt starlist}, we first created $\sim 10,000$ artificial objects with the same PSFs as the real objects and random but uniform spatial and magnitude distributions, ranging from 20 to 27 mag. We spread them over the NB921 and $z'$ images of the QSO field by using the IRAF task {\tt mkobject} allowing them to blend with themselves and real objects. Then, SExtractor was run for source detections in exactly the same way as our actual photometry. Finally, we calculated the ratio of the number of detected artificial objects to that of created ones to obtain the detection completeness. We repeated this procedure ten times and averaged the obtained completeness. The result is shown in Figure \ref{Completeness}. The completeness of the QSO field images are $\sim 52$\% at our LAE detection limit of NB921 $= 26.0$ and $\sim 36$\% at our LBG detection limit of $z'= 26.1$ (see Sections \ref{LAE-Selections} and \ref{LBG-Selections} for the LAE and LBG detection limits). 

We also estimated the detection completeness of the NB921 and $z'$ band images of the Control Field in the same manner and also show them in Figure \ref{Completeness}. As seen in the figure, detection completeness of the NB921 and $z'$ images of the QSO field are fairly comparable to those of the Control Field, as expected from the same/similar limiting magnitudes of the QSO and Control field NB921 and $z'$ images (see Table \ref{ImagingData}). This also means that differences in the impact of object blending on the completeness between the QSO and Control field images are negligibly small. Eventually, we can fairly compare LAEs and LBGs selected in the QSO field and the Control Field with the comparable detection completeness. The detection completeness is corrected when the number counts of LAEs and LBGs and the Ly$\alpha$ LFs of LAEs are derived in Sections \ref{NC} and \ref{LyaLFSec}.

\subsection{Selection of $z \simeq 6.6$ LAE Candidates in the QSO Field\label{LAE-Selections}} 
We use the photometric criteria similar to the ones adopted for the previous NB921 $z\sim6.6$ LAE survey in the Control Field SDF \citep{Taniguchi05,Kashikawa06,Kashikawa11} to newly select $z\sim6.6$ LAE candidates in both $z=6.61$ QSO field and Control Field. We choose to use these criteria because they have been already proven to be reliable by yielding the robust $z\sim6.6$ LAE sample in the Control Field with a low contamination rate ($\sim 2$--19\%) confirmed by spectroscopy as mentioned earlier in Section 1 \citep[see also][]{Kashikawa11}. Another reason is that to investigate the sky distribution, number density and clustering of LAEs around the QSO, we compare the LAE sample in the QSO field with that in the SDF. Hence, we should use exactly the same selection criteria to detect LAEs in both fields for fair and rigorous comparison. 

We use the following criteria (all the magnitudes are those measured in a $2''$ aperture) to newly select $z\sim6.6$ LAE candidates in the both QSO and Control fields.
\begin{eqnarray}
i' - z' > 1.3 \nonumber\\
z' - {\rm NB921} > 1.0 \nonumber\\ 
z' - {\rm NB921} > 3\sigma_{\rm SDF} \nonumber\\ 
{\rm NB921} \leq 26.0
\label{Criteria-1}
\end{eqnarray}
or else
\begin{eqnarray}
i' > i'_{2\sigma, {\rm SDF}} \nonumber\\  
z' > i'_{2\sigma, {\rm SDF}} - 1.3 \nonumber\\  
z' - {\rm NB921} > 1.0 \nonumber\\  
z' - {\rm NB921} > 3\sigma_{\rm SDF} \nonumber\\  
{\rm NB921} \leq 26.0
\label{Criteria-2}
\end{eqnarray}

As the $z\sim6.6$ Ly$\alpha$ emission appears in the middle of the $z'$ band (see the relative locations of the NB921 and $z'$ bands in wavelength in Figure \ref{FilterTransmission}) and fluxes of LAEs bluewards of Ly$\alpha$ are absorbed by the IGM \citep{Madau95}, they should have red $i' - z'$ colors. Also, the LAEs should show NB921 flux excess against the continuum band ($z'$ band). The criterion $z' - $ NB921 $> 3\sigma_{\rm SDF}$ means the $3\sigma$ NB921 flux excess against the $z'$ band flux in the SDF $z'$ and NB921 images: $z' - $ NB921 $> -2.5 \log [(f_{\rm NB} - 3 \sqrt{\sigma_{z', {\rm SDF}}^2 + \sigma_{\rm NB, {\rm SDF}}^2})/f_{\rm NB}]$. Here, $f_{\rm NB}$ is the NB921 flux. $\sigma_{z', {\rm SDF}}$ and $\sigma_{\rm NB, SDF}$ are the fluxes corresponding to the $1 \sigma$ limiting magnitudes of the SDF $z'$ and NB921 images, respectively. Also, $i'_{2\sigma, {\rm SDF}}=27.87$ is the $2\sigma$ limiting magnitude of the SDF $i'$ band image. We limit our LAE sample to NB921 $\leq 26.0$ (5$\sigma$ in both QSO and Control fields). 

We adopted $\sigma_{z', {\rm SDF}}$, $\sigma_{\rm NB, SDF}$ and $i'_{2\sigma, {\rm SDF}}$, not the $1\sigma$ fluxes and $2\sigma$ magnitude of our QSO field images, although the depths of the QSO field $i'$ and $z'$ band images are slightly shallower than those of the SDF images (see Table \ref{ImagingData}). However, as shown in Figure \ref{Completeness}, as for the $z'$ and NB921 images, the differences in detection completeness (including the effect of object blending) between the QSO field and the SDF at $< 27$ mag are negligible. Hence, the difference in depths does not cause any significant bias on the selection of LAEs in the QSO field against that in the SDF. 

Meanwhile, using Equations (6) and (7) in their paper, \citet{Taniguchi05} estimated that the narrowband excess criterion $z'-{\rm NB921} > 1.0$ corresponds to the rest-frame Ly$\alpha$ equivalent width (EW) threshold of EW$_0({\rm Ly}\alpha) > 7$\AA~for an LAE with its Ly$\alpha$ emission located at the center of the NB921 bandpass (i.e., $\lambda_{{\rm Ly}\alpha} = 9196$\AA~or $z_{{\rm Ly}\alpha} = 6.5625$). If we assume that redshift of an LAE is the same as that of the $z=6.61$ QSO (i.e., $z_{{\rm Ly}\alpha} = 6.61$ and replacing $\Delta\lambda_{\rm NB921}/2$ in Equation (7) in \citet{Taniguchi05} by $9196{\rm \AA}+\Delta\lambda_{\rm NB921}/2 - 1216{\rm \AA}(1+z_{{\rm Ly}\alpha})$), the EW threshold is EW$_0({\rm Ly}\alpha) > 15$\AA. Hence, the NB921 image of the QSO field has a better sensitivity to LAEs at the front side of the $z=6.61$ QSO as discussed in Section 1 and Figure \ref{NB921_LyaPeakDist}.

The criteria (\ref{Criteria-1}) and (\ref{Criteria-2}) are exactly same as the ones previously used by \citet{Taniguchi05} to detect $z\sim 6.6$ LAEs in the SDF except that they also additionally used the null detections in the wavebands bluewards of $z\sim6.6$ Ly$\alpha$ to reduce the contaminations from low-$z$ interlopers: $B > B_{3\sigma}$, $V > V_{3\sigma}$ and $R_c > R_{c 3\sigma}$, where $B$, $V$ and $R_c$ are the magnitudes in the $B$, $V$ and $R_c$ band filters for Suprime-Cam, respectively (see Figure \ref{FilterTransmission}), and the $B_{3\sigma}$, $V_{3\sigma}$ and $R_{c 3\sigma}$ are the $3\sigma$ limiting magnitudes of the $B$, $V$ and $R_c$ images of the SDF, respectively. As we did not take $B$, $V$ and $R_c$ band images of the $z=6.61$ QSO field, we use the above LAE selection criteria (\ref{Criteria-1}) and (\ref{Criteria-2}) without $B > B_{3\sigma}$, $V > V_{3\sigma}$ and $R_c > R_{c 3\sigma}$ to select $z\sim 6.6$ LAEs in the QSO field and re-select those in the Control Field SDF for a fair comparison. This would increase contaminations. However, in Appendix B, we have evaluated the increase in the number of contaminants due to the lack of $B$, $V$ and $R_c$ bands and confirmed that it is small.

\begin{figure*}
\epsscale{1.17}
\includegraphics[angle=0,scale=0.69]{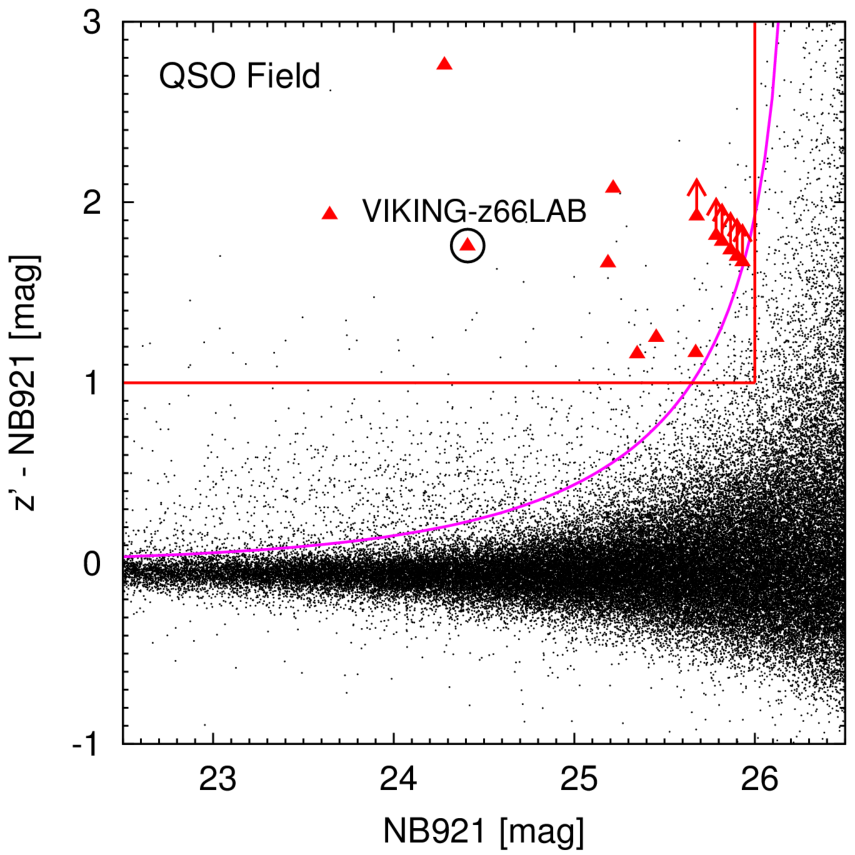}
\includegraphics[angle=0,scale=0.69]{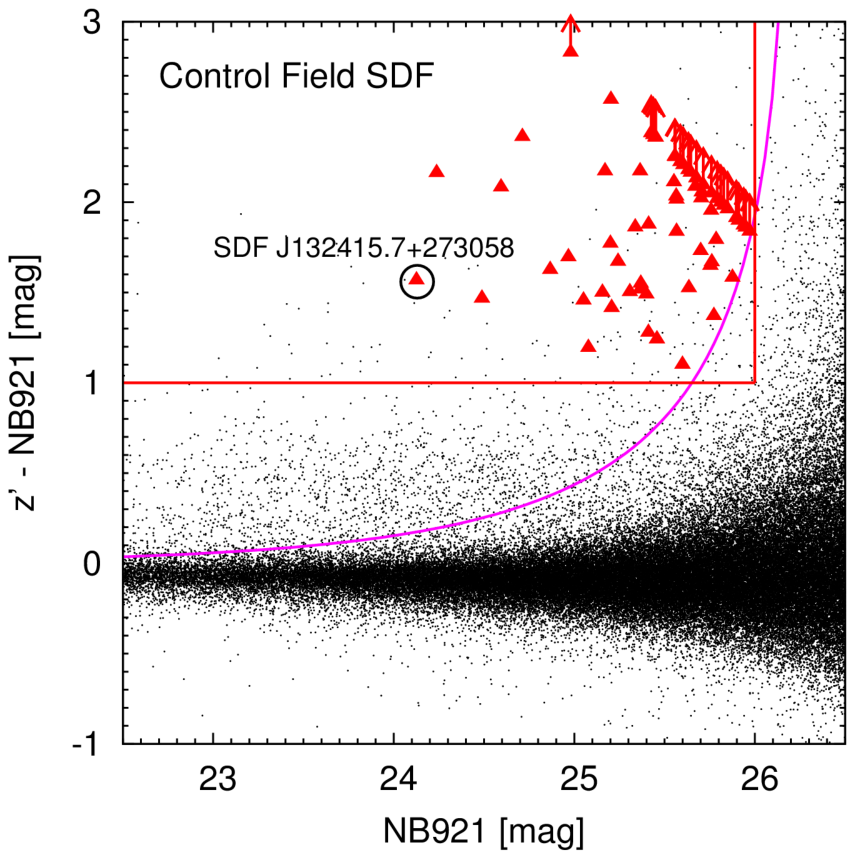}
\includegraphics[angle=0,scale=0.69]{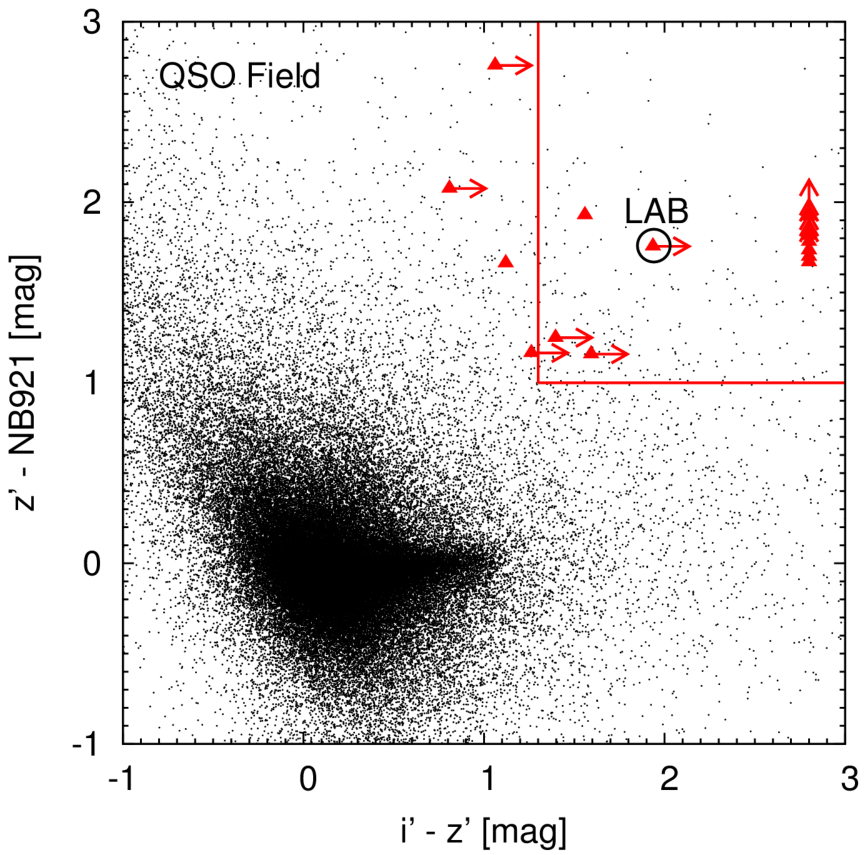}
\hspace{0.4cm}
\includegraphics[angle=0,scale=0.69]{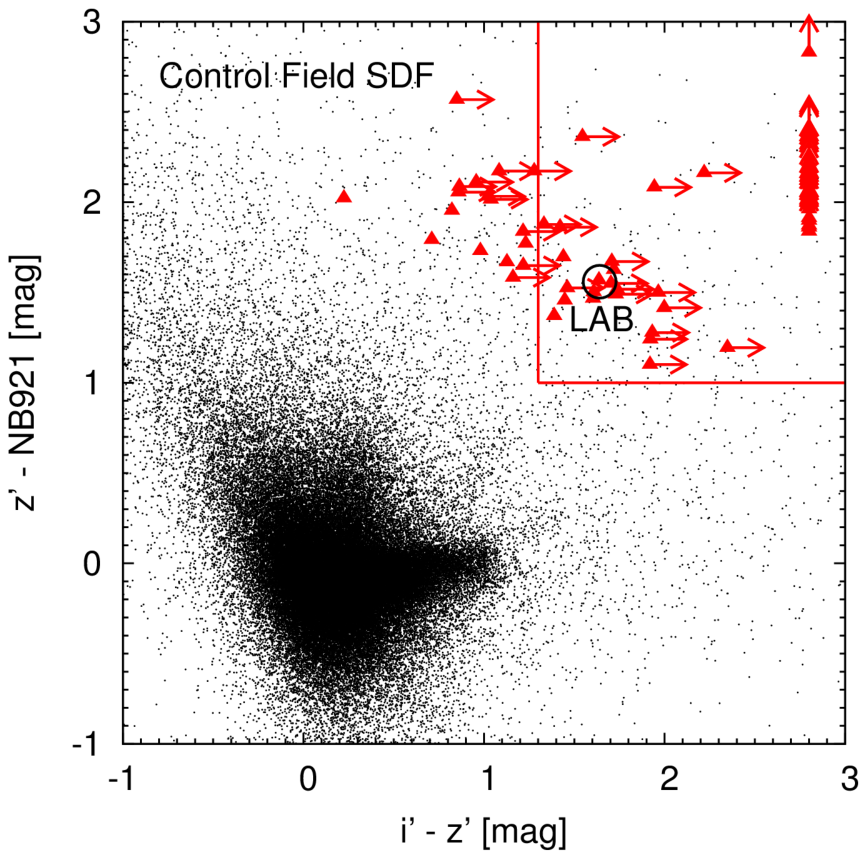}
\caption{Upper panels: $z' - {\rm NB921}$ color as a function of NB921 ($2''$ aperture) magnitude of all the objects detected in our NB921 images of the QSO and Control fields (shown by dots). The solid curves show the $3\sigma$ error track of $z' - {\rm NB921}$ color of the objects in our Control Field, SDF, not that of the objects in the QSO field (see Section \ref{LAE-Selections} for details). The horizontal solid lines are a part of our LAE color selection criteria, $z'-{\rm NB921} > 1.0$. The vertical solid lines indicate the limiting magnitude, ${\rm NB921} = 26.0$ ($5\sigma$ in both QSO and Control fields). The selected $z\sim6.6$ LAE candidates are denoted by the triangles with the arrows showing the $1\sigma$ limits on $z'-{\rm NB921}$ colors. By the open circles, we denote the LAB candidate, VIKING-z66LAB, found near the highest galaxy density peak in the south-west of the QSO field and the spectroscopically confirmed $z=6.541$ LAB, SDF J132415.7+273058 found in the Control Field SDF \citep{Kodaira03,Taniguchi05} (see Figures \ref{NumberDensityContours-QSO-SDF-each-own-mean-and-sigma}, \ref{NumberDensityContours-QSO-SDF-mean-and-sigma_in_SDF} and \ref{IsophotalArea} and Section \ref{LAB}). Lower panels: $z'-{\rm NB921}$ versus $i'-z'$ plots of all the objects detected in our NB921 images of the QSO and Control fields (shown by dots). The upper right rectangles surrounded by the solid line indicate parts of our LAE selection criteria (\ref{Criteria-1}), $i'-z'>1.3$ and $z'-{\rm NB921} > 1.0$ (see Section \ref{LAE-Selections} for details). The selected $z\sim6.6$ LAE candidates are denoted by the triangles with the arrows showing the $1\sigma$ limits on their colors. The LAE candidates with their $z'$ band magnitudes fainter than $1\sigma$ are placed at $i'-z'=2.8$. The LAB candidate, VIKING-z66LAB, and the LAB, SDF J132415.7+273058, are denoted by the open circles. Our $z' - {\rm NB921}$ versus $i'-z'$ diagram of the $z\sim6.6$ LAEs in the Control Field plotted here looks slightly different from the one plotted in Figure 3 in \citet{Taniguchi05} as we adopt the $1\sigma$ limits for $i'-z'$ colors while they did not.\label{CMDand2CD}}
\end{figure*}

To select $z\sim6.6$ LAE candidates in the $z=6.61$ QSO field, we applied the criteria (\ref{Criteria-1}) and (\ref{Criteria-2}) to our NB921-detected object catalog made in Section \ref{Photometry}. However, the criteria yielded a large number of objects, and most of them are located in the noisy regions near the edges of the NB921 image where the S/N is low. This implies that most of them could be noise. Such low S/N edge regions originate from the dithering of the exposure frames taken during the observations. 

To examine if they are noise, we created the negative NB921 image by multiplying each pixel value by $-1$, performed source detection runnig SExtractor and limited the detected objects to NB921 $\leq 26.0$. Each edge of the NB921 image was dominated with negative detections, which are considered noise. These edge regions coincide with the locations where most of the sources selected with the $z\sim6.6$ LAE candidate criteria (\ref{Criteria-1}) and (\ref{Criteria-2}) distribute. Hence, we trimmed these edge regions off the original NB921 image. Then, running SExtractor on this NB921 image as well as $i'$ and $z'$ images with the same edge regions trimmed, we constructed the NB921-detected object catalog again and applied the criteria (\ref{Criteria-1}) and (\ref{Criteria-2}) to it. In this process, if $i'$ and/or $z'$ band magnitudes of a source in the catalog are fainter than $1\sigma$ (i.e., $i'> i'_{1\sigma, {\rm SDF}}$ and/or $z'> z'_{1\sigma, {\rm SDF}}$), we replaced them by $i'_{1\sigma, {\rm SDF}}$ and/or $z'_{1\sigma, {\rm SDF}}$. 

To further remove spurious sources, we visually inspected $i'$, $z'$ and NB921 images of each source that satisfies the selection criteria. We especially removed obviously spurious sources such as columns of bad pixels, pixels saturated with bright stars, noise events of deformed shapes, and scattering pixels having anomalously large fluxes. Finally, we were left with 14 sources that are the final $z\sim 6.6$ LAE candidates in the QSO field. The color-magnitude ($z'-{\rm NB921}$ versus NB921) and two color ($z'-{\rm NB921}$ versus $i'-z'$) diagrams of the LAE candidates and all the NB921-detected objects in the QSO field are plotted in Figure \ref{CMDand2CD}.

To remove the noise in the process of the LAE candidate selection, we trimmed the low S/N edge regions off the NB921 image. After this, the total area of the image became 697 arcmin$^2$ (or $29.'8 \times 23.'4$). The comoving distance along the line of sight corresponding to the redshift range $6.51 \leq z \leq 6.62$ for LAEs covered by the FWHM of the NB921 filter is 41 Mpc. Therefore, we have probed a total of $\sim 1.7 \times 10^5$ Mpc$^3$ volume for our $z\sim6.6$ LAE selection in the QSO field.

The limiting magnitude of $i'$-band image of the QSO field ($i'_{2\sigma, {\rm QSO}}=27.4$ mag at $2\sigma$) is shallower than that of the Control Field ($i'_{2\sigma, {\rm SDF}} = 27.87$ mag). Thus, the criteria $i' > i'_{2\sigma, {\rm SDF}}$ and $z' > i'_{2\sigma, {\rm SDF}} - 1.3$ in the LAE selection criteria (\ref{Criteria-2}) might be very stringent to selecting LAEs in the QSO field and could result in missing detecting some LAEs. To investigate this issue, we repeat the LAE selection with the criteria (\ref{Criteria-1}) and (\ref{Criteria-2}) but using $i' > i'_{2\sigma, {\rm QSO}}$ and $z' > i'_{2\sigma, {\rm QSO}} - 1.3$. We detect 4 additional objects after removing spurious sources by visual inspection. All of them are faintly visible in the $i'$-band image. Usually, a large majority of $z=6.6$ LAEs are not visible in the $i'$-band due to IGM absorption of their fluxes at the wavelengths bluewards of $z=6.6$ Ly$\alpha$ while a small minority of $z=6.6$ LAEs, especially bright ones, are visible in the $i'$-band if detecatable fluxes are left at the wavelengths bluewards of $z=6.6$ Ly$\alpha$ \citep[e.g.,][]{Taniguchi05}. Thus, we cannot tell whether the additionally detected 4 objects are $z=6.6$ LAEs or lower-$z$ interlopers. Nonetheless, even if we assume that all of them are $z=6.6$ LAEs, we stick to only the LAE candidates selected using $i'_{2\sigma, {\rm SDF}}$ based on the following two reasons. (1) Adding the four objects to our LAE sample would not change our conclusion of this study. More specifically, the LAE number density excess contours shown later in Figures \ref{NumberDensityContours-QSO-SDF-each-own-mean-and-sigma} and \ref{NumberDensityContours-QSO-SDF-mean-and-sigma_in_SDF} would not change much because of the locations of the four objects in the QSO field; ($\Delta$DEC [arcmin], $\Delta$RA [arcmin]) $=$ (4.8, 2.6), (17.7, 2.0), (25.0, 0.2) and (25.9, 7.8). (2) We want to use exactly the same LAE selection criteria for both QSO and Control fields for consistency despite the difference in depth in the $i'$-band between the two fields.

\subsection{Selection of LBGs in the QSO Field\label{LBG-Selections}}
\begin{figure}
\epsscale{1.22}
\plotone{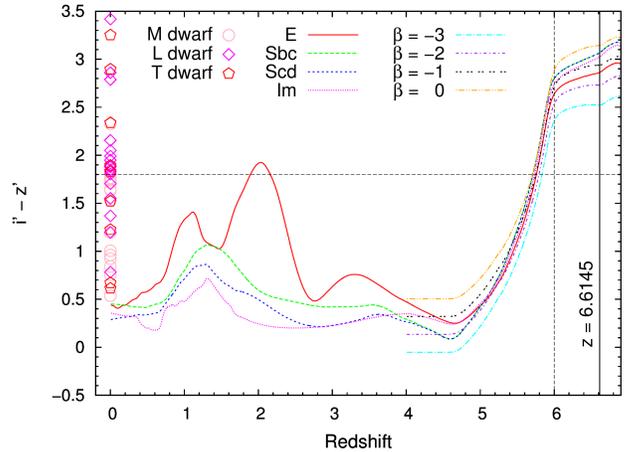}
\caption{$i'-z'$ color (Suprime-Cam $i'$ and $z'$ bands) as a function of redshift of model LBGs, various types of galaxies and M/L/T type dwarf stars. The colors of LBGs are calculated assuming power-law spectra $f_{\lambda} \propto \lambda^{\beta}$ with several different UV continuum slopes $\beta$ and no Ly$\alpha$ emission. We applied IGM absorption to the spectra using the \citet{Madau95} prescription. The colors of E (elliptical), Sbc, Scd and Im (irregular) galaxies are calculated using \citet{Coleman80} template spectra. Also, colors of M/L/T dwarfs (types M3--M9.5, L0--L9.5 and T0--T8) are calculated using their observed spectra provided by \citet{Burgasser04,Burgasser06a,Burgasser06b,Burgasser08,Burgasser10} and \citet{Kirkpatrick10} at the SpeX Prism Spectral Libraries (see footnote 3). The horizontal and vertical dashed lines denote our color cut $i'-z' > 1.8$ for $z>6$ LBG selection and the corresponding lower redshift cut $z=6$, respectively. The vertical solid line is the redshift of the QSO.\label{i-z_vs_redshift}}
\end{figure}

In addition to LAEs, we also investigate LBGs around the $z=6.61$ QSO. To detect them, we first examined the expected colors of LBGs and potential contaminants and determined the LBG selection criteria. Figure \ref{i-z_vs_redshift} shows $i'-z'$ color (convolved with the Suprime-Cam $i'$ and $z'$ filters) as a function of redshift of model LBGs as well as other types of galaxies and M/L/T type dwarf stars that can be contaminants. We modeled LBGs with power-law spectra $f_{\lambda} \propto \lambda^{\beta}$ with UV continuum slopes $\beta=-3$ to 0, IGM absorption applied using the \citet{Madau95} prescription and no Ly$\alpha$ emission. The $z\sim 6.6$ LBGs are expected to have colors of $i'-z' \sim 2.5$--3.1. In Figure \ref{i-z_vs_redshift}, we also plot colors of E (elliptical), Sbc, Scd and Im (irregular) galaxies using \citet{Coleman80} template spectra as well as M/L/T dwarfs (types M3--M9.5, L0--L9.5 and T0--T8) using their real spectra provided by \citet{Burgasser04,Burgasser06a,Burgasser06b,Burgasser08,Burgasser10} and \citet{Kirkpatrick10} at the SpeX Prism Spectral Libraries\footnote[3]{http://pono.ucsd.edu/{\textasciitilde}adam/browndwarfs/spexprism/library.html}. While low-$z$ Sbc, Scd and Im galaxies show bluer colors of $i'-z' \lesssim 1.1$, low-$z$ ellipticals could have red colors up to $i' - z' \sim 2.0$ due to their 4000\AA~Balmer breaks. Moreover, M/L/T dwarfs exhibit a wide range of colors $i'-z' \sim 0.5$--3.5.

Based on these color information, we selected the LBG candidates in the QSO field by applying the following $i'$-dropout criteria (all the magnitudes are those measured in a $2''$ aperture) to the $z'$-detected object catalog constructed in Section \ref{Photometry}.
\begin{eqnarray}
i' > i'_{2\sigma, {\rm SDF}} \nonumber\\
25.0  \leq z' \leq 26.1 \nonumber\\
i' - z' > 1.8 
\label{Criteria-3}
\end{eqnarray}
If an LBG is at $z\sim6.6$ (i.e., if it is associated with the $z=6.61$ QSO), then its Lyman break is located at $\sim9240$\AA~in the middle of the $z'$ band wavelength range (see Figure \ref{FilterTransmission}). As fluxes of the LBG bluewards of the $z\sim6.6$ Lyman break should be absorbed by IGM, we require the null detection ($< 2\sigma$) in $i'$ band (the first criterion where $i'_{2\sigma, {\rm SDF}}=27.87$ mag is the $2\sigma$ limiting magnitude of the $i'$ band image of the Control Field SDF). This limits the expected redshift of the selected LBGs to $z>6$ as the red edge of the Suprime-Cam $i'$ band is at $\sim8500$\AA~corresponding to $z\simeq 6$ Ly$\alpha$. 

Also, we limit our LBG sample to $z'\leq 26.1$ where $z'=26.1$ is the 4$\sigma$ (5$\sigma$) limiting magnitude of the $z'$ band image of the QSO field (Control Field SDF). We adopt $z' \leq 26.1$ ($5\sigma$ in the SDF) as well as $i'_{2\sigma, {\rm SDF}}$, not the $5\sigma$ and $2\sigma$ limiting magnitudes of the QSO field images, because we should fairly compare the LBGs in the QSO field and the Control Field by selecting them with exactly the same criteria, although the depths of the QSO field images are slightly shallower than those of the SDF images (see Table \ref{ImagingData}). As shown in Figure \ref{Completeness}, as for the $z'$ band images, the difference in detection completeness (including the effect of object blending) between the QSO field and the SDF at $< 27$ mag is small and almost negligible. Hence, the difference in depth does not cause any significant bias on the selection of LBGs in the QSO field against that in the Control Field. 

The two criteria, $i' > i'_{2\sigma, {\rm SDF}} = 27.87$ and $z' \leq 26.1$, automatically require that the LBG candidates should have an $i' - z' > 1.77$ color. Figure \ref{i-z_vs_redshift} shows that the $z\sim 6.6$ LBGs are expected to have colors of $i'-z' \sim 2.5$--3.1 due to their Lyman breaks. To cover the possible variety of $i'-z'$ colors of real LBGs, we adopt the inclusive color cut of $i' - z' > 1.8$. This color cut, together with $i' > i'_{2\sigma, {\rm SDF}}$, detects LBGs at $6 < z < 7$. This has been previously independently confirmed by \citet{Toshikawa12}, who also used a similar color cut $i' - z' > 1.5$ to select and study $i'$-dropout galaxies in the SDF. They produced various galaxy spectra using the population synthesis model of \citet{BC03} and determined a color cut of $i' - z' > 1.5$ which detects Lyman breaks at $5.6 \lesssim z \lesssim 6.9$.

Figure \ref{i-z_vs_redshift} shows that low redshift ellipticals could be contaminants having colors of $i' - z' > 1.8$ due to their 4000\AA~Balmer breaks in principle. However, \citet{Ota05} and \citet{Toshikawa12} have already shown that low-$z$ ellipticals actually have $i' - z' < 1.5$ colors and do not contaminate the $i'$-dropout LBG samples by examining $i' - z'$ colors of $z \sim 1$--4 extremely red objects (EROs consisting of old ellipticals and dusty starbursts) detected in the Suprime-Cam $i'$ and $z'$ bands by \citet{Miyazaki03}. Also, \citet{Toshikawa12,Toshikawa14} carried out spectroscopy of 31 $i' - z' > 1.5$ $i'$-dropouts and detected no low redshift elliptical. Therefore, we consider contamination of our $i' - z' > 1.8$ LBG sample by low redshift ellipticals negligible.


\begin{figure*}
\includegraphics[angle=0,scale=0.69]{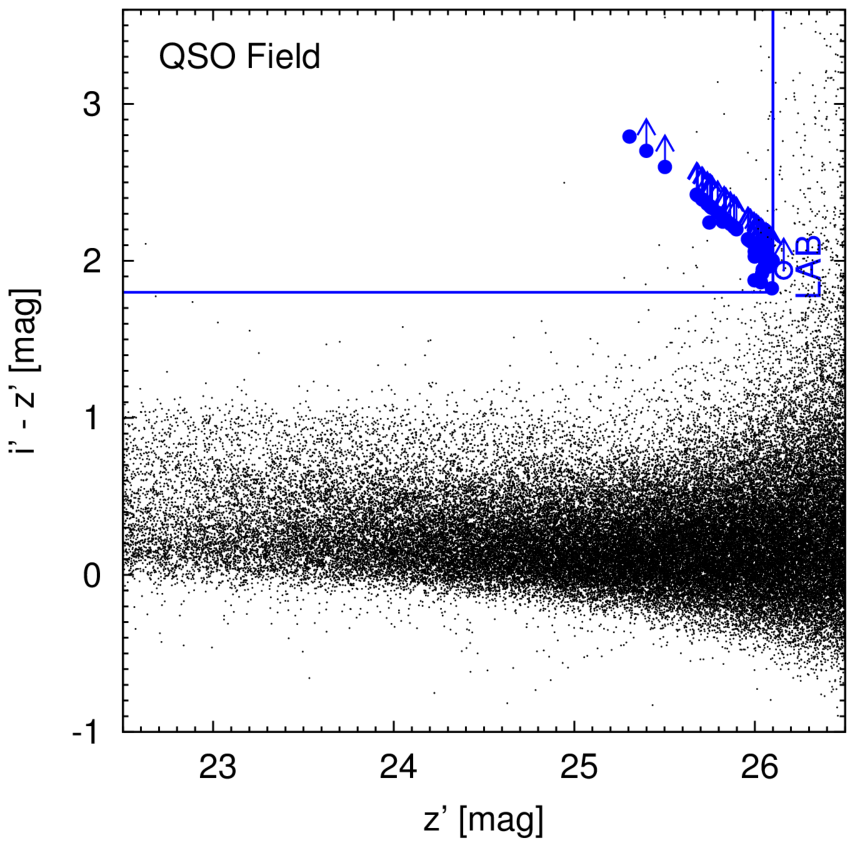}
\includegraphics[angle=0,scale=0.69]{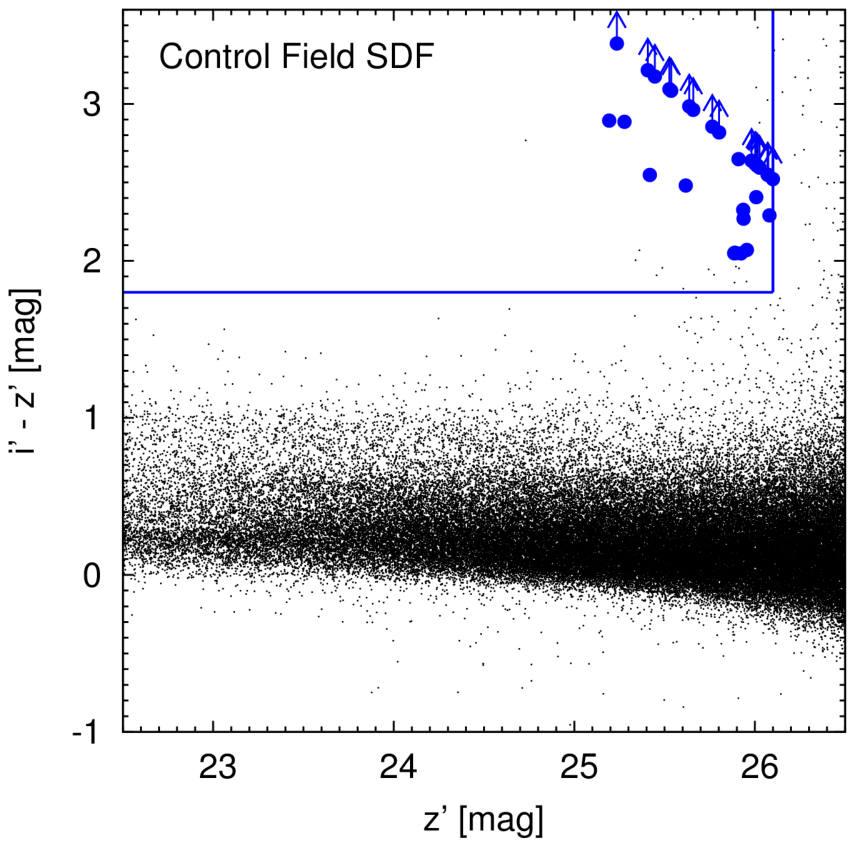}
\caption{Left: $i' - z'$ color as a function of $z'$ ($2''$ aperture) magnitude of all the objects detected in the $z'$ band image of the $z=6.61$ QSO field (shown by dots). The horizontal solid line is a part of our LBG selection criteria, $i' - z' > 1.8$. The vertical solid line indicates the limiting magnitude, $z' = 26.1$. The selected LBG candidates in the QSO field are denoted by the filled circles with the arrows showing the $1\sigma$ limits on $i' - z'$ colors. The LAB candidate, VIKING-z66LAB, found near the highest galaxy density peak in the south-west of the QSO field (see Figures \ref{NumberDensityContours-QSO-SDF-each-own-mean-and-sigma} and \ref{NumberDensityContours-QSO-SDF-mean-and-sigma_in_SDF} and Section \ref{LAB}) is also plotted by the open circle and labeled so. Its $i'-z'$ color is comparable to those of the LBG candidates, but it is not selected as an LBG candidate as its $z'$ band magnitude is slightly fainter than the limiting magnitude. Right: The same diagram as the left panel but for all the objects (dots) and the LBG candidates (circles) detected in the $z'$ band image of the Control Field SDF.\label{CMD_LBGs_both_fields}}
\end{figure*}

Moreover, we also limit our LBG sample to $z' \geq 25.0$ to reduce the contamination by dwarf stars. Figure \ref{i-z_vs_redshift} shows that $i'-z'$ colors of M/L/T dwarfs can have colors of $i' - z' > 1.8$ to contaminate our LBG sample. Again, \citet{Toshikawa12,Toshikawa14}, who studied $i' - z' > 1.5$ $i'$-dropouts in the SDF, estimated that the contamination rate of M/L/T dwarfs in their sample was high at $z' < 25.0$ but as low as only $\sim6$\% at $z' \geq 25.0$. As mentioned earlier, they also carried out spectroscopy of 31 $i' - z' > 1.5$ $i'$-dropouts and detected no dwarf, either. Hence, we adopt the criterion $z' \geq 25.0$ in our LBG selection criteria (\ref{Criteria-3}) to reduce the number of contaminating M/L/T dwarfs in our sample. However, actually, no $i' - z' > 1.8$ object exists at $z' < 25.0$ in the QSO field. 


To avoid spurious LBG detections at the low S/N edge regions (due to dithering) of the $z'$ band image of the QSO field, we performed a negative image test with the $z'$ band image. This is the test similar to the one we did for the NB921 image of the QSO field when selecting the $z\sim 6.6$ LAE candidates (see Section \ref{LAE-Selections}). Based on this test, we identified the borders of the low S/N edge regions and trimmed them off the $z'$ band image. Running SExtractor on this $z'$ band image as well as the $i'$ and NB921 band images with the same edge regions trimmed, we then created the $z'$-detected object catalog again and applied the LBG selection criteria (\ref{Criteria-3}) to it. 

To further remove the spurious sources, we visually inspected $i'$, $z'$ and NB921 images of each source that satisfies the selection criteria in the same way as we did for the selection of $z\sim6.6$ LAE candidates. Finally, we were left with 53 $z>6$ LBG candidates in the QSO field. We show their color-magnitude diagram ($i'-z'$ versus $z'$) in the left panel of Figure \ref{CMD_LBGs_both_fields}.  

The limiting magnitude of $i'$-band image of the QSO field ($i'_{2\sigma, {\rm QSO}}=27.4$ mag at $2\sigma$) is shallower than that of the Control Field ($i'_{2\sigma, {\rm SDF}} = 27.87$ mag). Thus, the criterion $i' > i'_{2\sigma, {\rm SDF}}$ in the LBG selection criteria (\ref{Criteria-3}) might be very stringent to selecting LBGs in the QSO field and could result in missing detecting some LBGs. To investigate this issue, we repeat the LBG selection with the criteria (\ref{Criteria-3}) but relaxing the first criterion to $i' > i'_{2\sigma, {\rm QSO}}=27.4$ mag. We detect 12 more objects after removing spurious sources by visual inspection. Out of the 12, 8 are faintly visible in the $i'$-band image, so they are not $z >$ 6 LBGs but either $z < 6$ LBGs or interlopers.
The remaining four are not seen in the $i'$-band image and thus could be $z > 6$ LBGs. Thus, at most only four LBG candidates are missed. Nonetheless, we stick to only the LBG candidates selected using $i' > i'_{2\sigma, {\rm SDF}}$ based on the following four reasons. (1) There still remains a possibility that the four objects might become visible in the $i'$-band image of the QSO field and not be $z > 6$ LBG candidates if the image is as deep as the $i'$-band image of the Control Field. (2) Adding the four objects to our LBG sample would not change our conclusion of this study. More specifically, the LBG number density excess contours shown later in Figures \ref{NumberDensityContours-QSO-SDF-each-own-mean-and-sigma} and \ref{NumberDensityContours-QSO-SDF-mean-and-sigma_in_SDF} would not change much because of the locations of the four objects in the QSO field; ($\Delta$DEC [arcmin], $\Delta$RA [arcmin]) $=$ (21.0, 17.5), (9.6, 4.3), (6.9, 9.0) and (12.8, 16.4). (3) We want to use exactly the same LBG selection criteria for both QSO and Control fields for consistency despite the difference in depth in the $i'$-band between the two fields. (4) The majority (8/12) of the extra objects detected using the relaxed criterion $i' > i'_{2\sigma, {\rm QSO}}=27.4$ mag are visible in the $i'$-band image and not the $z>6$ LBGs we want to include in our LBG sample. Thus, this criterion is not stringent enough. 

\subsection{The LAE and LBG Samples in the Control Field\label{SDFsample}}
In Appendix B, we have examined the validity of our $z \sim 6.6$ LAE selection criteria (\ref{Criteria-1}) and (\ref{Criteria-2}) using only the $i'$, $z'$ and NB921 images of the SDF and not using the SDF $B$, $V$ and $R_c$ images. Using these criteria, we have successfully re-selected the same 58 photometric $z\sim 6.6$ LAE candidates as the ones the previous work by the SDF project \citep{Taniguchi05} had selected and also 5 additional objects (cases 1--5 in Figure \ref{woBVR_Objects}). As we show in Figures \ref{NumberDensityContours-QSO-SDF-each-own-mean-and-sigma} and \ref{ACF_LAEs_LBGs_SDF_vs_QSOField} and Sections \ref{NdensityContours} and \ref{ACF}, these 63 $z\sim6.6$ LAE candidates in the SDF neither show any significant over/underdensity nor clustering. Hence, we adopt the 63 SDF LAE candidates as our final Control Field LAE sample. We show their color-magnitude ($z'-{\rm NB921}$ versus NB921) and two-color ($z'-{\rm NB921}$ versus $i'-z'$) diagrams in Figure \ref{CMDand2CD}.  

On the other hand, in exactly the same way as we did to construct the QSO field LBG sample in Section \ref{LBG-Selections}, we selected 32 LBG candidates in the Control Field by applying the selection criteria (\ref{Criteria-3}) to the SDF public version 1.0 $z'$-detected object catalog. As we show in Figures \ref{NumberDensityContours-QSO-SDF-each-own-mean-and-sigma} and \ref{ACF_LAEs_LBGs_SDF_vs_QSOField} and Sections \ref{NdensityContours} and \ref{ACF}, the LBG candidates in the SDF neither show any significant over/underdensity nor clustering. Hence, we adopt these SDF LBG candidates as our final Control Field LBG sample. We show their color-magnitude diagram ($i'-z'$ versus $z'$) in the right panel of Figure \ref{CMD_LBGs_both_fields}. The numbers of the LAE and LBG candidates detected in the QSO and Control fields are listed in Table \ref{ImagingData}.

\section{Result and Discussion\label{Result}}
As we have constructed LAE and LBG samples in the QSO and Control fields in a consistent manner to comparable depths, completeness and effective survey areas/volumes, in the subsequent sections we derive and compare their sky distributions, number density contours, number counts and clustering properties in both fields to elucidate if the $z=6.61$ QSO resides in a galaxy overdensity indicative of a massive halo.   

\subsection{Sky Distributions and Number Density Excess Contours of LAEs and LBGs\label{NdensityContours}}
If there are any galaxy overdensities in the vicinity of the QSO, we can identify them by comparing sky distributions and surface number density excess contours of the LAEs and the LBGs over the entire QSO and Control fields as the observed sky areas covering these fields are quite large (one Suprime-Cam FoV each, $\sim 34' \times 27'$ or $\sim 11 \times 9$ physical Mpc$^2$ at $z=6.6$). We can draw surface number densitiy excess contours by measuring the number density of galaxies at many positions in an image, deriving its mean and dispersion $\sigma$ and plotting the mean $\pm$ $1 \sigma$, mean $\pm$ $2 \sigma$ and so on. Previous studies that found no significant galaxy overdensities around QSOs derived the contours based on the mean and $\sigma$ measured in the field in which the contours are drawn \citep[e.g.,][]{Kashikawa07,Kikuta17}. In our study, we have the Control Field that includes no QSO at $z \sim 6.6$ but has comparable depth, completeness and area to those of the $z=6.61$ QSO field. Hence, in the subsequent sections, we first show the number density excess contours of the LAEs and the LBGs in each of the QSO and Control fields based on the mean and the $\sigma$ measured in each field just like previous studies. Then, we show the same contours of the LAEs and the LBGs in the QSO field but based on the mean and the $\sigma$ measured in the Control Field. In this way, we try to elucidate how much more significant the number density excess of the LAEs and the LBGs in the QSO field are than those in a general blank field.


\begin{figure*}
\includegraphics[angle=0,scale=0.68]{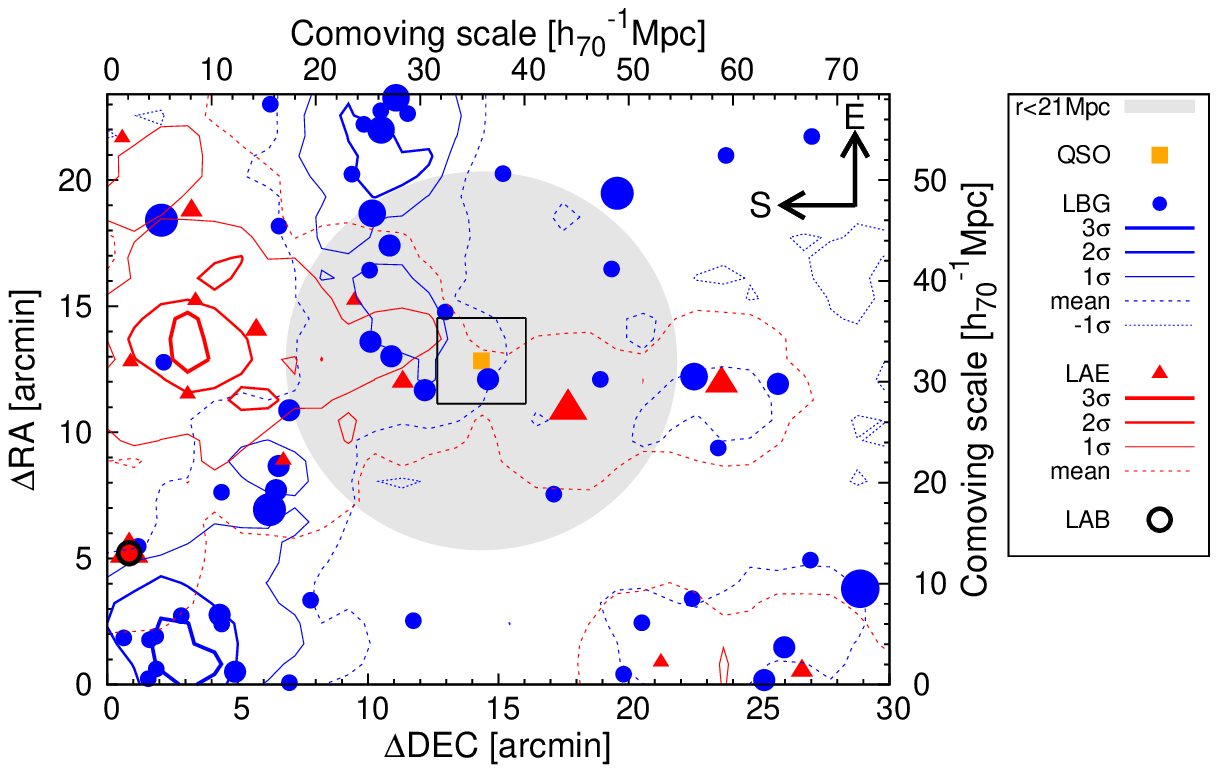}
\includegraphics[angle=0,scale=0.75]{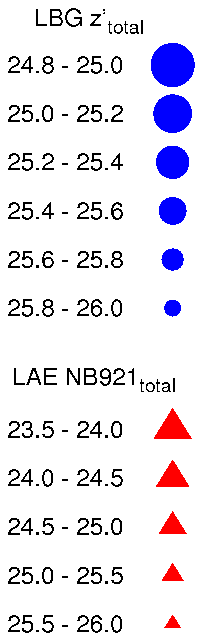}
\hspace*{-1.2cm}
\includegraphics[angle=0,scale=0.73]{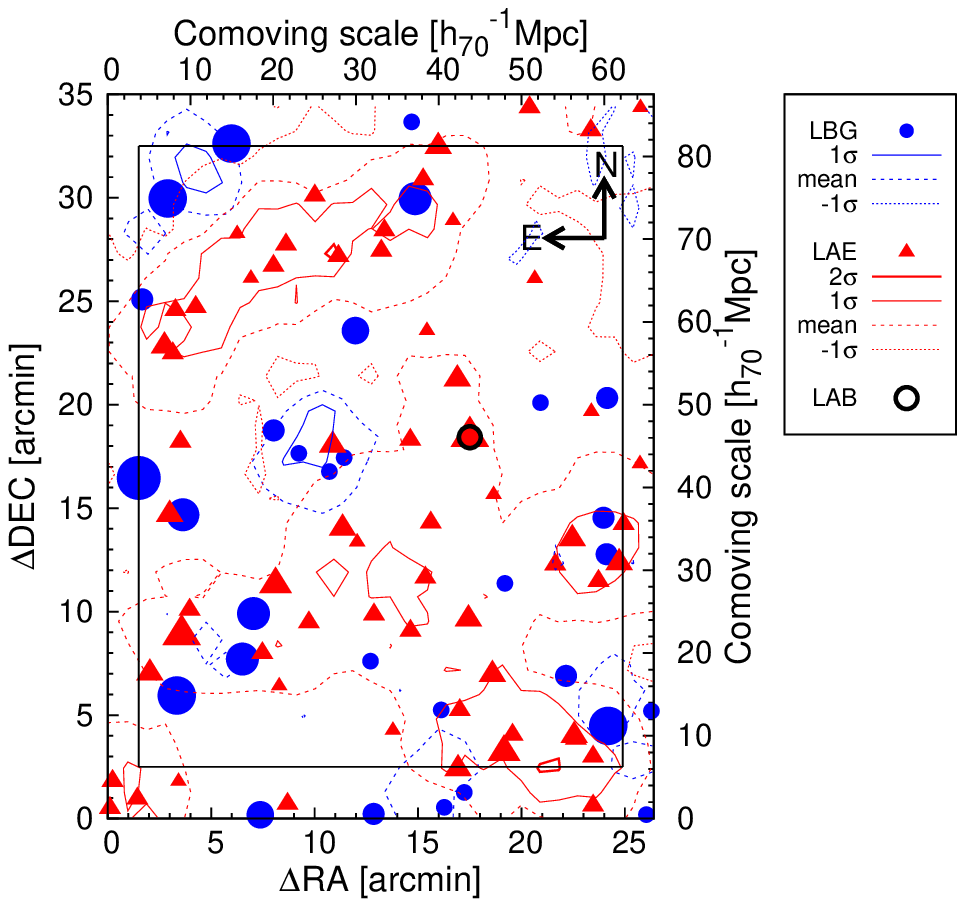}
\caption{Sky distributions and surface number density excess contours of the LAEs (red triangles and contours) and the LBGs (blue circles and contours) in the $z=6.61$ QSO field (left panel) and the Control Field SDF (right panel). The symbols for the LAEs (LBGs) are scaled with their NB921 ($z'$) total magnitudes. East (North) is up, and South (East) to the left in the QSO (Control) field. The black solid line rectangle in the Control Field corresponds to the size of the QSO field. In each field, the surface number densities of LAEs or LBGs were measured by randomly distributing 100,000 comoving 8 Mpc radius circles and counting the numbers of LAEs or LBGs in them. Then, we derived their mean and dispersion $\sigma$ for LAEs or LBGs in each field. In the both panels, the red and blue dashed contours indicate the mean numbers of LAEs and LBGs in a circle, respectively. The dotted contours denote mean$-1\sigma$ number deficits while the solid thin to thick contours show mean$+1\sigma$, mean$+2\sigma$ and mean$+3\sigma$ number excess. We do not plot the mean$-1\sigma$ contour of the LAEs in the QSO field as it is below zero. The orange square in the left panel is the location of the $z=6.61$ QSO. The grey shade shows the ``proximity'' region within 21 comoving Mpc ($\sim3$ physical Mpc) distances in projection from the QSO where the 21 comoving Mpc is half of the comoving distance along the line of sight corresponding to the redshift range $\Delta z \sim 0.1$ for LAEs covered by the FWHM of the NB921 filter. If an LAE/LBG in this region is also located within $\Delta z \sim 0.05$ from the QSO, it is likely associated with the QSO. The large black open square around the QSO in the left panel is the $\sim 12$ arcmin$^2$ field of view (FoV) of the {\it Hubble Space Telescope} (HST) Advanced Camera for Surveys (ACS) often used by previous studies to search for galaxy overdensities around $z \gtrsim 6$ QSOs, which resulted in finding a variety of galaxy densities \citep[e.g.,][]{Kim09,Simpson14}. The black open circles in the both panels show locations of a bright extended LAB candidate, VIKING-z66LAB (QSO field), and an LAB, SDF J132415.7+273058 (Control Field) (see Sections \ref{NC} and \ref{LAB}).\label{NumberDensityContours-QSO-SDF-each-own-mean-and-sigma}} 
\end{figure*}

\subsubsection{Density Contours in the QSO and Control Fields Based on the Mean and the Dispersion in Each Field\label{NdensityContours_ownMeanSigma}}
Figure \ref{NumberDensityContours-QSO-SDF-each-own-mean-and-sigma} shows sky distributions of the LAEs and the LBGs and their surface number density excess contours in the QSO and the Control fields. In each of the QSO and the Control fields, we spread circles of an 8 comoving Mpc radius to 100,000 homogeneously distributed random positions, measured the number of LAEs or LBGs in each circle and derived the mean and the dispersion $\sigma$ for the LAEs or the LBGs. Then, we drew the surface number density contours of mean$- 1\sigma$, mean, mean$+ 1\sigma$, mean$+ 2\sigma$, mean$+ 3\sigma$, ... for the LAEs or the LBGs in each field in Figure \ref{NumberDensityContours-QSO-SDF-each-own-mean-and-sigma}.

We chose 8 comoving Mpc for the radius of the circle as it is not too small to encompass LAEs or LBGs in sparse regions in each field (especially those in the right half of the QSO field; see Figure \ref{NumberDensityContours-QSO-SDF-each-own-mean-and-sigma}) and not too large compared to the size of each field. The number of circles (100,000) are large enough to cover the entirety of each field. The numbers of LAEs or LBGs counted in the circles at the positions near the edges of the images (regions within 8 comoving Mpc from the edges) may affect the derived values of means and $\sigma$'s to some extent because parts of the circles are outside the image. To examine this effect, we calculated means and $\sigma$'s by spreading circles over the image avoiding the regions within 8 comoving Mpc from the edges. For both LAEs and LBGs, means and $\sigma$'s changed only very slightly from those derived including the edge regions. The amount of changes is very small and negligible. Accordingly, the derived number density excess contours of LAEs and LBGs look very similar to those derived including the edge regions. Hence, we did not make any corrections to the numbers of LAEs or LBGs counted in the circles at the positions near the edges of the images. 

As seen in the right panel of Figure \ref{NumberDensityContours-QSO-SDF-each-own-mean-and-sigma}, both the LAEs and the LBGs in the Control Field exhibit no significant overdensity but show mostly mean $\pm$ $1 \sigma$ densities with a few small peaks of $2\sigma$ number density excess. This means that a general blank sky field like this Control Field is almost flat with only $1 \sigma$ fluctuation in surface number density distributions of LAEs and LBGs on $\sim 9 \times 11$ physical Mpc$^2$ ($\sim 67 \times 86$ comoving Mpc$^2$) scale. 

On the other hand, the left panel of Figure \ref{NumberDensityContours-QSO-SDF-each-own-mean-and-sigma} shows the sky distributions and number density contours of the LAEs and LBGs in the QSO field. The $z=6.61$ QSO is located at the center of the field. We define and show the ``proximity'' region around the QSO by the gray shade in the similar way to the one used in the previous study of environments around $z\sim5$ QSOs/radio galaxy conducted by \cite{Kikuta17}. \cite{Kikuta17} observed two QSOs and a radio galaxy at $z\sim5$ by the Subaru Suprime-Cam and narrow/broadband filters, detected LAE and LBG candidates in the QSO/radio galaxy fields and analyzed their sky distributions and number density contours in the similar way to ours. They defined the proximity regions around the QSOs/radio galaxy whose sizes (3 or 5 physical Mpc) were determined by the FWHMs of the two narrowband filters they used ($\sim72$\AA~or 120\AA) and are sufficiently small to detect galaxies associated with the QSOs/radio galaxy. 

In the left panel of Figure \ref{NumberDensityContours-QSO-SDF-each-own-mean-and-sigma}, we also show the proximity region within 21 comoving Mpc ($\sim 3$ physical Mpc) distances in projection from the $z=6.61$ QSO with the gray shade. Here, the 21 comoving Mpc is half of the comoving distance along the line of sight corresponding to the redshift range $\Delta z \sim 0.1$ for LAEs covered by the FWHM of the NB921 filter. Hence, if an LAE or an LBG in this region is also located within $\Delta z \sim 0.05$ from the QSO (though we need to take the spectrum of it to know this), it is likely associated with the QSO. We find 3 LAE candidates and 11 LBG candidates in the proximity region. In this region, the number densities of both LAE and LBG candidates show only mean to mean$+ 1\sigma$ values. Hence, we see apparently no significant overdensities of LAEs and LBGs in the proximity of the QSO.

However, the sky distributions and number density contours of the LAEs and the LBGs in the entire QSO field look quite different from those in the Control Field on a larger scale. In the right (north) half of the QSO field, LAEs and LBGs are very sparse mostly exhibiting the densities between mean and mean$- 1\sigma$ (underdensity at $1\sigma$ level). Meanwhile, in the left (south) half of the QSO field, the number densities of both LAEs and LBGs are between mean and mean$+ 3\sigma$ and higher than those in the Control Field on a large scale. Both of these LAE and LBG overdensities in the QSO field also show filamentary structures side by side extending from east to west (top to bottom in the figure). The LAE structure also extends from south to north (left to right in the figure). The LAE structure includes a $3\sigma$ density peak between east and west while the LBG structure contains a $3\sigma$ density peak in the west. These large scale structures of LAEs and LBGs with weak overdensities partly include the proximity region of the QSO and the QSO itself though the densities of LAEs and LBGs in the proximity region are mean to mean$+ 1\sigma$. Hence, the QSO might possibly be associated with the filamentary large scale structures of LAEs and LBGs, and such structures seem to be highly biased compared to the relatively flat large scale structures of LAEs and LBGs with low density fluctuations in a general blank field like our Control Field.

Interestingly, we found that there is a bright (${\rm NB921}_{\rm total} \sim 23.78$ mag) and extended (diameter $\gtrsim 3''$ or 16 physical kpc) Ly$\alpha$ Blob (LAB) candidate near the $3\sigma$ LBG overdensity peak in the west (hereafter VIKING-z66LAB, see Section \ref{LAB} for further details). This LAB candidate is one of the LAE candidates we detected by the LAE selection criteria (\ref{Criteria-1}) and (\ref{Criteria-2}) and the most extended one. LABs often show evidence of galaxy interaction or merging within their extended Ly$\alpha$ clouds \citep[e.g.,][]{Ouchi09} and tend to be found in/around dense environments like protoclusters or galaxy overdensity regions \citep[e.g.,][]{Steidel00,Chapman04,Colbert06,Matsuda11,Bridge12,Yang12,Yang14,Badescu17}. VIKING-z66LAB appears to show a sign of interaction or merging of two sources in the $z'$-band image (see Figure \ref{LABFig}). Thus, the existence of this LAB candidate could support the validity of the overdensity of LBGs.

For comparison, in the left panel of Figure \ref{NumberDensityContours-QSO-SDF-each-own-mean-and-sigma}, we also depict the $\sim 12$ arcmin$^2$ field of view (FoV) of the {\it Hubble Space Telescope} (HST) Advanced Camera for Surveys (ACS) often used by previous studies to search for galaxy overdensities around $z \gtrsim 6$ QSOs, which resulted in finding high, average and even low galaxy densities \citep[e.g.,][]{Kim09,Simpson14}. The ACS FoV includes only one LBG candidate in our QSO field and is much smaller than the size of the QSO proximity region and the large scale structures of LAEs and LBGs. This suggests that it is difficult to see a positional relation between a $z>6$ QSO and large scale spatial and number density distributions of galaxies within a small FoV, and that large area imaging by a wide-field camera is essential to exploring the large scale galaxy environment of a QSO. This was also pointed out by \citet{Morselli14} who observed $z\sim6$ $i$-dropout LBGs around four $z\gtrsim6$ QSOs with the wide-field ($\sim 23' \times 25'$) Large Binocular Camera (LBC) on the Large Binocular Telescope. Three out of the four QSO fields had been also previously observed by \citet{Kim09} using the HST ACS (one pointing each) and known to show overdensity, average density and underdensity of $i$-dropout LBGs, respectively, within the ACS FoV. However, \citet{Morselli14} found that the LBG number densities in all the four QSO fields are higher than that in a blank field when seen in the LBC FoV and emphasized that wide-field imaging can capture possible large-scale galaxy overdensities around QSOs. 


\begin{figure*}
\epsscale{1.18}
\plotone{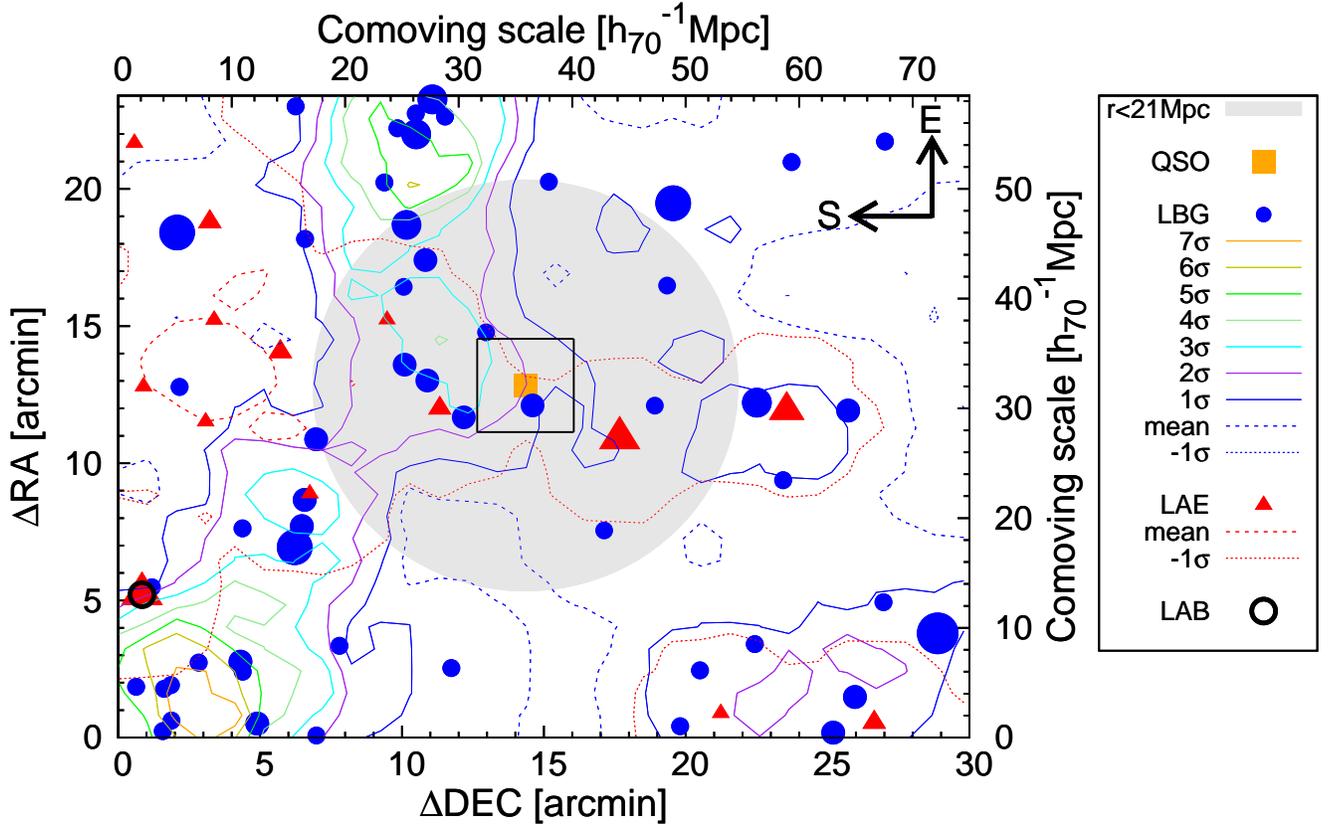}
\caption{Sky distributions and surface number density excess contours of the LAEs (red triangles and contours) and the LBGs (blue circles and blue or color coded contours) in the $z=6.61$ QSO field. The symbols for the LAEs (LBGs) are scaled with their NB921 ($z'$) total magnitudes as in Figure \ref{NumberDensityContours-QSO-SDF-each-own-mean-and-sigma}. The surface number densities of LAEs or LBGs were measured by randomly distributing 100,000 comoving 8 Mpc radius circles in the QSO field and counting the numbers of LAEs or LBGs in them. Then, we draw the contours of mean$_{\rm SDF}-1\sigma_{\rm SDF}$ (dotted contours), mean$_{\rm SDF}$ (dashed contour), mean$_{\rm SDF}+1\sigma_{\rm SDF}$, mean$_{\rm SDF}+2\sigma_{\rm SDF}$, mean$_{\rm SDF}+3\sigma_{\rm SDF}$ ... number excess (solid contours) for the LAEs or the LBGs. Note that the mean$_{\rm SDF}$ and the $\sigma_{\rm SDF}$ are the mean number of LAEs or LBGs in a circle and the dispersion both measured in the Control Field SDF in the right panel of Figure \ref{NumberDensityContours-QSO-SDF-each-own-mean-and-sigma}, not the mean and the dispersion measured in the QSO field. In this way, we can elucidate how much more significant the number density excess of the LAEs or the LBGs in the QSO field are than those in the Control Field. The orange square and the grey shade are the $z=6.61$ QSO and its proximity region. The large black open square around the QSO is the FoV of the HST ACS often used by previous studies to search for galaxy overdensities around $z \gtrsim 6$ QSOs, resulting in finding high, average and even low galaxy densities. The black open circle is a bright extended LAB candidate VIKING-z66LAB (see Sections \ref{NC} and \ref{LAB}).\label{NumberDensityContours-QSO-SDF-mean-and-sigma_in_SDF}}
\end{figure*}
                                                      
\subsubsection{Density Contours in the QSO Field Based on the Mean and the Dispersion in the Control Field\label{NdensityContours_SDFMeanSigma}}
The number density excess contours of the LAEs and the LBGs in the $z=6.61$ QSO field shown in the left panel of Figure \ref{NumberDensityContours-QSO-SDF-each-own-mean-and-sigma} are based on the means and the $\sigma$'s of the LAEs and the LBGs measured in the QSO field itself. However, as we discussed above, the spatial and number density distributions of the LAEs and the LBGs in the QSO field are quite different from and apparently highly biased compared to those in the Control Field. Hence, it might not be ideal to use the means and the $\sigma$'s of LAEs and LBGs measured in the QSO field to draw the number density excess contours, which previous studies did. 

On the other hand, the Control Field represents a blank field without any extremely biased galaxy spatial and number density distributions as seen in Figure \ref{NumberDensityContours-QSO-SDF-each-own-mean-and-sigma} and described in Section \ref{NdensityContours_ownMeanSigma}. Also, as we will show in Section \ref{NC}, the cosmic variance of $z\sim6.6$ LAEs in the Control Field is only $\sigma_v \sim 0.19$. Moreover, using the Subaru Suprime-Cam and the NB921 filter, \citet{Ouchi10} detected 207 $z\sim 6.6$ LAE candidates in the $\sim 1.0$ deg$^2$ Subaru/{\it XMM-Newton} Deep Survey (SXDS) field (5 Suprime-Cam pointings). They compared their surface number density (number/0.5 mag/arcmin$^2$ versus NB921 magnitude) with that of 58 $z\sim 6.6$ LAE candidates \citet{Taniguchi05} detected in SDF (Control Field) and showed that they are almost consistent \citep[see Figure 5 in][]{Ouchi10}. Hence, the area of the Control Field is large enough not to be affected much by cosmic variance, and it is more appropriate to use its means and $\sigma$'s to plot the number density excess contours. Therefore, we also draw the contours of the LAEs and the LBGs in the QSO field based on the means and the $\sigma$'s of LAEs and LBGs measured in the Control Field; i.e., mean$_{\rm SDF}-1\sigma_{\rm SDF}$, mean$_{\rm SDF}$, mean$_{\rm SDF}+1\sigma_{\rm SDF}$, mean$_{\rm SDF}+2\sigma_{\rm SDF}$, ... and so on in Figure \ref{NumberDensityContours-QSO-SDF-mean-and-sigma_in_SDF}.

In this figure, we see the shapes of LAE and LBG number density contours similar to those we have found in the left panel of Figure \ref{NumberDensityContours-QSO-SDF-each-own-mean-and-sigma}, but the LAE contours show much lower densities while the LBG contours exhibit higher significances of overdensities compared to the Control Field. The large scale structure of LAEs in the QSO field traces only mean$_{\rm SDF}-1\sigma_{\rm SDF}$ to mean$_{\rm SDF}$, equivalent to the mean to even underdensity of the Control Field. Conversely, the number density of LBGs in their filamentary structure varies from mean$_{\rm SDF}+ 1\sigma_{\rm SDF}$ to the highest peak at mean$_{\rm SDF}+7\sigma_{\rm SDF}$ level in the west (lower left in the figure). The LAB candidate VIKING-z66LAB is located in the place of $\sim1$--$3\sigma_{\rm SDF}$ excess of LBGs. There is also $\sim5$--$6\sigma_{\rm SDF}$ excess of LBGs at the eastern part of the LBG large scale structure. Even in the right (northern) half of the QSO field, where LAEs and LBGs are relatively sparse, LBGs are mostly exhibiting the densities typical of the Control Field (mean$_{\rm SDF}$ to mean$_{\rm SDF}+1\sigma_{\rm SDF}$). In this region, the density of the LAEs is entirely below the mean of the Control Field ($\leq$ mean$_{\rm SDF}-1\sigma_{\rm SDF}$). 

The large scale structures of the LAEs and the LBGs again partly include the QSO itself and its proximity region. The density of the LBGs in the proximity region varies from mean$_{\rm SDF}$ to mean$_{\rm SDF}+ 4\sigma_{\rm SDF}$ and mean$_{\rm SDF}+ 1\sigma_{\rm SDF}$ to mean$_{\rm SDF}+ 2\sigma_{\rm SDF}$ right at the position of the QSO. Hence, the number density of LBGs is moderately high in the vicinity of the QSO compared to that in a general field (Control Field). However, the density of the LAEs in the QSO proximity region is entirely below mean$_{\rm SDF}$ including the position of the QSO. Thus, the number density of LAEs in the vicinity of the QSO is below the mean of a general field.

Eventually, the sky distributions and number density excess contours of LAE and LBG candidates lead to three important implications. (1) The number density of the LAE candidates in the proximity of the $z=6.6$ QSO is below the mean density of the LAE candidates in a general blank field at $\sim 1\sigma$ level. (2) The number density of the LBG candidates in the proximity of the $z=6.6$ QSO varies from the mean value to the $4\sigma$ excess of the LBG candidate density in a general blank field. (3) The $z=6.6$ QSO is included in the filamentary large scale structure of LBG candidates and might be associated with the structure but is not located exactly at the highest density peaks of LBG candidates in the structure. Therefore, there still remains the possibility that galaxy (LBG) overdensities exist in the proximity of the QSO. However, we cannot tell if this is real unless we spectroscopically confirm the redshifts of a significant fraction of the LBG candidates in the both QSO and Control fields and draw the contours of mean$_{\rm SDF} - 1\sigma$, mean$_{\rm SDF}$, mean$_{\rm SDF} + 1\sigma$, mean$_{\rm SDF} + 2\sigma$ ... of the LBGs at the confirmed redshifts close to that of the QSO.     


\begin{figure*}
\includegraphics[angle=0,scale=0.69]{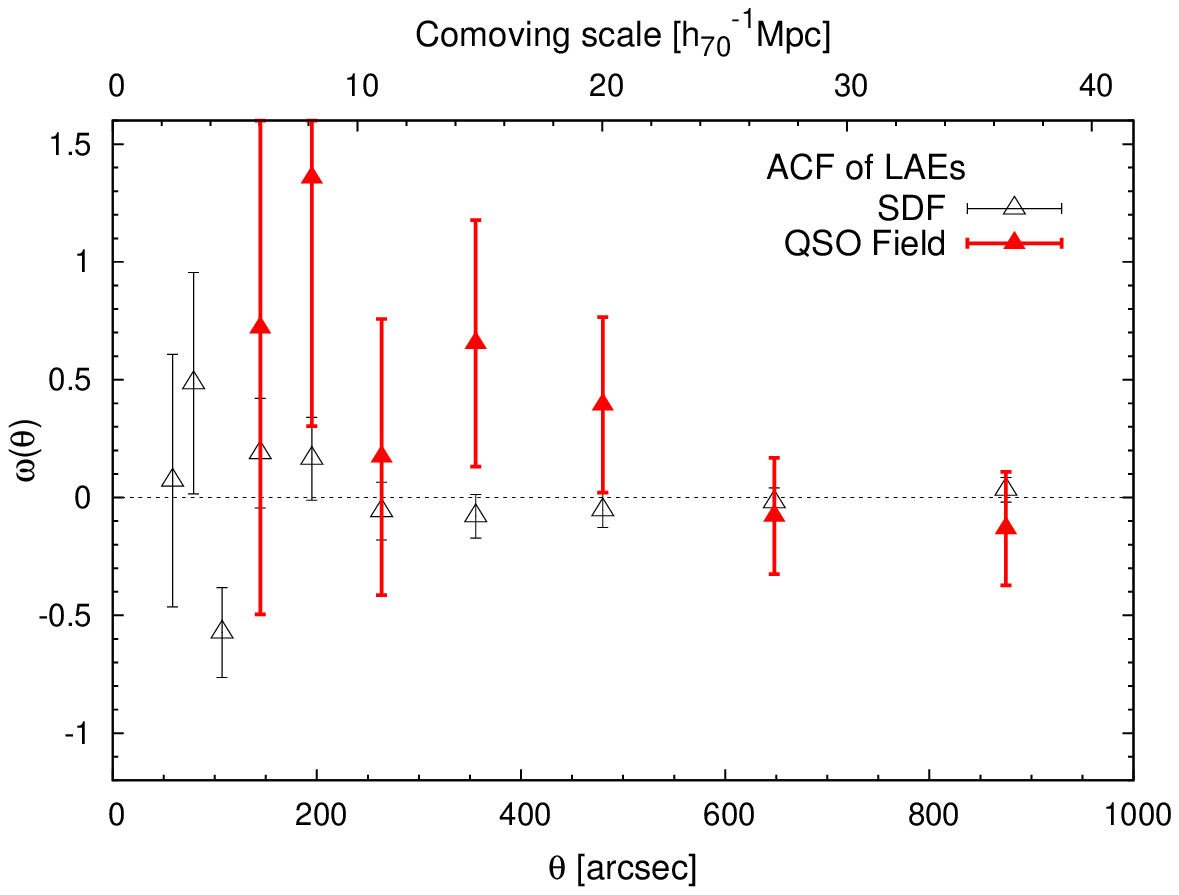}
\includegraphics[angle=0,scale=0.69]{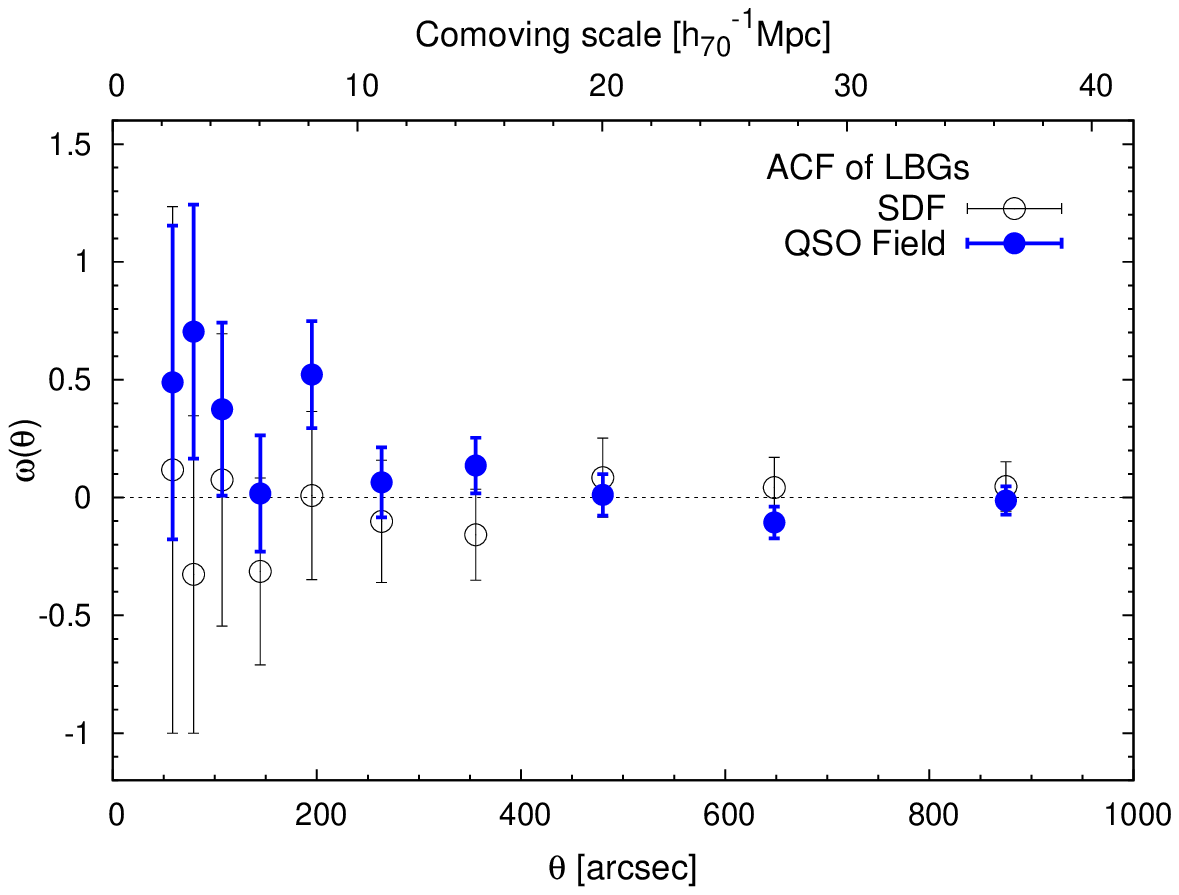}
\caption{Left: ACFs of the $z\sim6.6$ LAE candidates in the Control Field SDF and the QSO field shown by open and filled triangles, respectively. Error bars show the $1\sigma$ Poisson errors. There are no LAE-LAE pairs in the QSO field with separation angles corresponding to the first three $\theta$ bins. Right: ACFs of the LBG candidates in the Control Field SDF and the QSO field shown by open and filled circles, respectively. Error bars show the $1\sigma$ Poisson errors.\label{ACF_LAEs_LBGs_SDF_vs_QSOField}}
\end{figure*}

\subsection{Clustering of LAEs and LBGs\label{ACF}}
We have seen that sky and number density distributions of the LAE and LBG candidates in the QSO field are quite extreme showing their large scale overdensity structures mostly in half side of the entire field. This is in stark contrast to a general blank field (Control Field) where LAE and LBG candidates distribute much more uniformly with small fluctuations. Namely, the LAE and LBG candidates in the QSO field look much more clustered than those in the Control Field. To further quantitatively investigate this trend, we derive and compare two-point angular correlation functions (ACFs) of the LAE and LBG candidates in the QSO and Control fields in Figure \ref{ACF_LAEs_LBGs_SDF_vs_QSOField}. ACF is the estimator of clustering strength defined by \citet{LandySzalay93} as follows.
\begin{equation}
\omega (\theta) = \frac{DD(\theta) - 2DR(\theta) + RR(\theta)}{RR(\theta)}
\label{ACF-equation}
\end{equation}
where the $DD(\theta)$, $DR(\theta)$, and $RR(\theta)$ are the number of galaxy-galaxy, galaxy-random and random-random pairs having angular separations between $\theta$ and $\theta+\delta\theta$. We generated 100,000 random points in each of the QSO and the Control fields to reduce the Poisson noise in random pair counts and normalized $DD(\theta)$, $DR(\theta)$, and $RR(\theta)$ by the total number of pairs in each pair count. We created the random points having exactly the same boundary conditions as the LAE or LBG samples in the QSO and Control fields by avoiding the regions of their NB921 or $z'$ band images masked, removed or trimmed during the data reduction and the LAE/LBG selection. 

We estimated the Poisson errors for the ACFs \citep{LandySzalay93} as
\begin{equation}
\sigma_{\omega} (\theta) = \frac{1+\omega (\theta)}{\sqrt{DD(\theta)}}
\label{ACF-equation}
\end{equation}
For a large number of sources, the Poisson errors tend to underestimate true errors for an ACF compared to the bootstrap or the Jackknife technique \citep{Ling86,Harikane16,Harikane17}. However, in the case of our galaxy samples, the numbers of LAE and LBG candidates are small, and the Poisson errors would not underestimate the errors of the ACFs much \citep[see e.g.,][]{Khostovan17}.
  
Figure \ref{ACF_LAEs_LBGs_SDF_vs_QSOField} shows ACFs of the LAE and LBG candidates in the QSO and the Control fields. Both LAE and LBG candidates in the QSO field exhibit clustering signals while those in the Control Field do not. In the QSO field, LAEs are clustering especially in a wide range of angular distances, 8--20 comoving Mpc while LBGs in small angular distance, 4-8 comoving Mpc. This result, together with the implications obtained in the previous sections, suggests that LAEs and LBGs are clustering in different angular scales and forming large scale structures separately that contain the QSO and its proximity region at their near-edge locations as well as a few high density clumps/peaks. However, it should be noted that the different clustering angular scale between the LAEs and the LBGs in the QSO field could be simply due to the effect of the different volumes or redshift ranges probed by the LAE and LBG selections. In any case, the clustering properties of the LAE and LBG candidates in the QSO field are clearly different from those in a general blank field (Control Field). 

\subsection{Number Counts of LAEs and LBGs\label{NC}}
So far, we have found that distributions of LAE and LBG candidates are spatially quite different between the QSO and the Control fields. Next, we examine if there are any differences in number {\bf counts} of LAEs or LBGs per brightness (surface number density per total NB921 or $z'$ band magnitude) between these fields.  

When we performed photometry in Section \ref{Photometry}, we used the {\tt MAG\_AUTO} parameter of SExtractor to measure total NB921 and $z'$ magnitudes of the NB921-detected and $z'$-detected objects. However, the {\tt MAG\_AUTO} does not always measure a total magnitude of an object accurately especially when it has close neighbors and/or blends with them or noise. Hence, we visually inspected all the LAE and LBG candidates in the QSO and Control fields, also checked their SExtractor paremeter {\tt FLAGS} and split them into isolated LAEs/LBGs and blended LAEs/LBGs. We checked the SExtractor {\tt FLAGS} value of each LAE/LBG candidate because {\tt FLAGS} $=$ 1 means that an object has neighbors or bad pixels affecting its {\tt MAG\_AUTO} photometry while {\tt FLAGS} $=$ 2 means that an object was originally blended with another one but deblended by SExtractor when performing photometry \citep{BA96}. We defined the isolated LAEs/LBGs as those (1) having {\tt FLAGS} $=$ 0 and (2) visually appearing not to blend with any objects or noise. We considered the LAE/LBG candidates not satisfying either or both of the conditions (1) and (2) to be blended. 

For the isolated LAEs/LBGs, we adopted {\tt MAG\_AUTO} measurements as their total NB921/$z'$ magnitudes. For the blended LAEs/LBGs, we applied aperture corrections to their $2''$ aperture NB921/$z'$ magnitudes (those measured by the SExtractor {\tt MAG\_APER} parameter) to estimate their total NB921/$z'$ magnitudes. We estimated the aperture corrections by taking the medians of the differences between the total and $2''$ aperture NB921/$z'$ magnitudes of the isolated LAEs/LBGs in each field. We found 7 (31) LAE and 29 (14) LBG candidates isolated in the QSO (Control) field. The aperture corrections were $-0.06$ ($-0.24$) mag for NB921 magnitudes of the LAEs and $-0.14$ ($-0.14$) mag for $z'$ magnitudes of the LBGs in the QSO (Control) field. The origin of the difference between the NB921 aperture corrections for the LAEs in the QSO and the Control fields is not clear but possibly comes from combination of differences in PSF sizes of the NB921 images ($0.''91$ and $0.''98$) and intrinsic LAE sizes (mostly in Ly$\alpha$ emission) between the two fields.     

On the other hand, as mentioned earlier and described in Section \ref{LAB}, we found that one of the LAE candidates in the QSO field is a bright extended LAB candidate, VIKING-z66LAB, whose angular diameter is $\gtrsim 3''$ (see Figure \ref{LABFig} for its images and see Figures \ref{NumberDensityContours-QSO-SDF-each-own-mean-and-sigma} and \ref{NumberDensityContours-QSO-SDF-mean-and-sigma_in_SDF} for its location in the QSO field). Also, the LAEs in the Control Field include one bright extended LAB whose angular diameter is $\gtrsim 3''$. This is the $z=6.541$ LAE, SDF J132415.7+273058, previously spectroscopically confirmed by \citet{Kodaira03} (see Figure \ref{NumberDensityContours-QSO-SDF-each-own-mean-and-sigma} for its location in the Control Field). They have {\tt FLAGS} $> 0$ and are not isolated. In addition, as they are exceptionally much more extended than other LAE candidates whose NB921 magnitudes were used to estimate the aperture correction values, the aperture correction method would underestimate the total NB921 magnitudes of VIKING-z66LAB and SDF J132415.7+273058. Hence, we estimated their total NB921 magnitudes in different ways as follows. 

VIKING-z66LAB has 4 neighbors and more or less blends with all of them, resulting in {\tt FLAGS} $=$ 3 ($=1+2$) in the NB921 image convolved to have a PSF FWHM of $=0.''91$ for the $2''$ aperture photometry. Hence, we ran SExtractor on the ${\rm PSF}=0.''77$ NB921 image of the QSO field that is the original image before the convolution for the aperture photometry. In this image, VIKING-z66LAB slightly blends with one of the four neighbors and has only {\tt FLAGS} = 2. This means that VIKING-z66LAB originally blended with the neighbor but SExtractor automatically corrected parts of the {\tt MAG\_AUTO} elliptical aperture which are contaminated by the neighbor by mirroring the opposite, cleaner side of the measurement ellipse \citep{BA96}. If the {\tt MAG\_AUTO} measurement had been also affected by more than 10\% of the integrated area due to the blending with the neighbor, SExtractor would have also returned {\tt FLAGS} = 1 (resulting in {\tt FLAGS} = 1 + 2 = 3). Thus, we adopt the NB921 magnitude of VIKING-z66LAB measured by {\tt MAG\_AUTO} on the ${\rm PSF}=0.''77$ NB921 image as its total NB921 magnitude (NB921$_{\rm total}=23.78$), as it would be more robust and reliable than the one measured by the aperture correction method. Meanwhile, SDF J132415.7+273058 has three neighbors and slightly blends with two of them, resulting in {\tt FLAGS} $=$ 1 in the NB921 image of the Control Field SDF (PSF FWHM $=0.''98$). We used a $4''$ aperture that mostly encompasses the entirety of SDF J132415.7+273058 but minimizes the contaminations by the fluxes of the neighbors to measure its total NB921 magnitude, which is NB921$_{\rm total}=23.69$.


\begin{figure*}
%
\includegraphics[angle=0,scale=0.69]{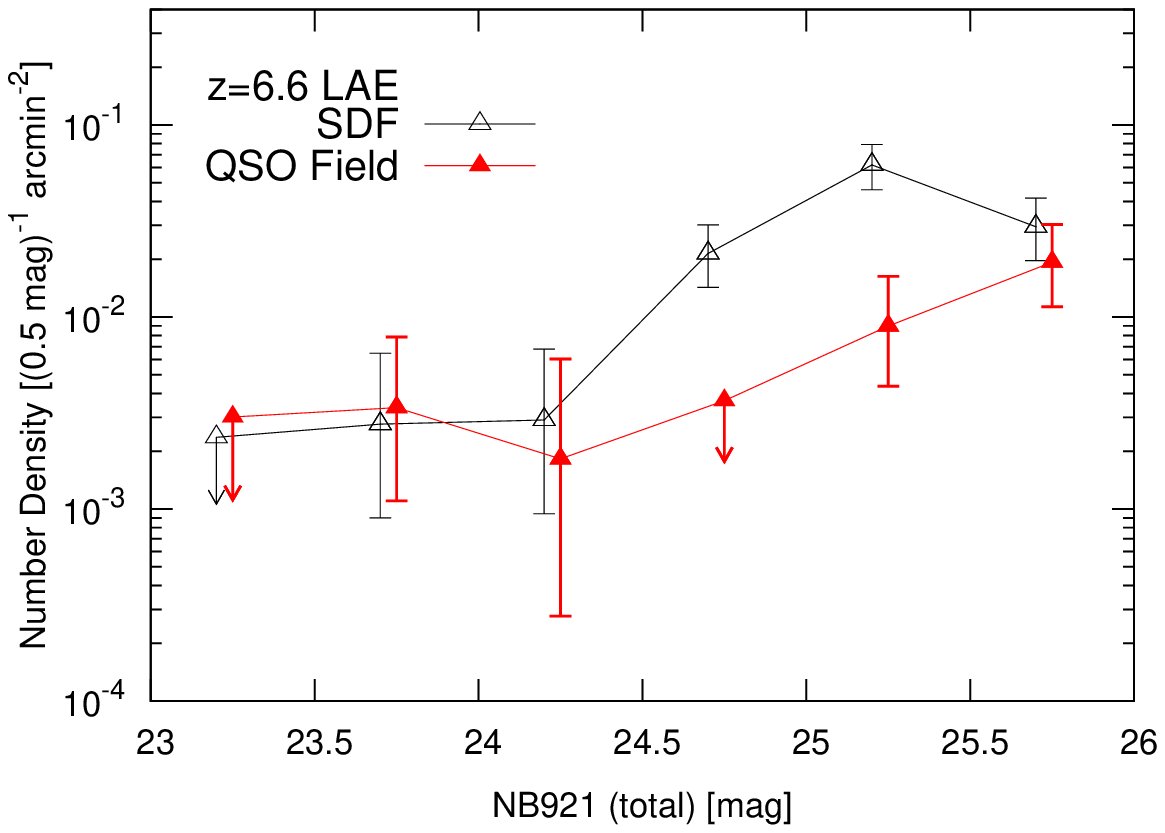}
\includegraphics[angle=0,scale=0.69]{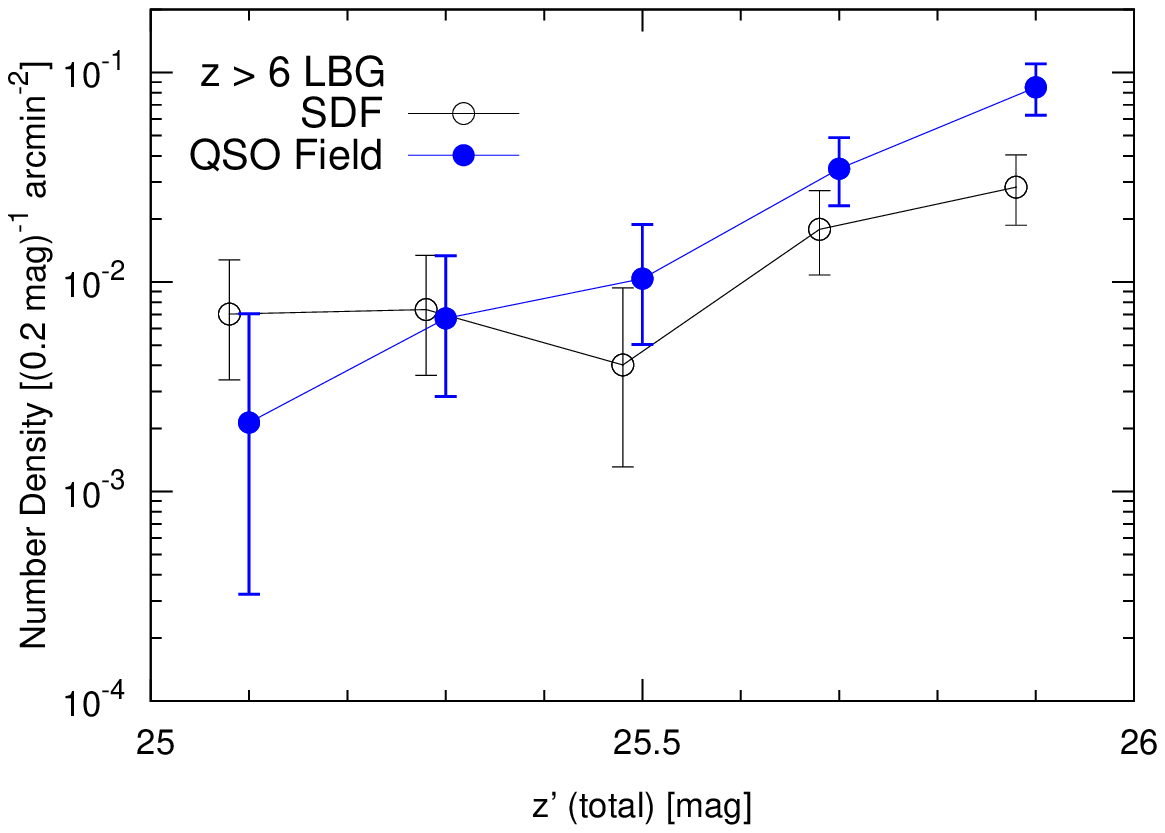}
\caption{Left: Surface number densities of the $z \sim6.6$ LAE candidates in the Control Field SDF (black open triangle) and the QSO field (red filled triangle) as a function of NB921 total magnitude. The data points for the two samples are slightly horizontally shifted for clarity. The error bars include Poisson errors \citep{Gehrels86} and cosmic variance. Detection completeness is corrected for each 0.5 mag bin by using the data in the left panel of Figure \ref{Completeness}. For the bins where no LAE candidate is detected, the upper limits are shown by the arrows. Right: Surface number densities of the $z > 6$ LBG candidates in the Control Field SDF (black open circle) and the QSO field (blue filled circle) as a function of $z'$-band total magnitude. The data points for the two samples are slightly horizontally shifted for clarity. Error bars include the Poisson error and cosmic variance. Detection completeness is corrected for each 0.2 mag bin by using the data in the right panel of Figure \ref{Completeness}.\label{NumberCounts}}
\end{figure*}

We derived the surface number densities of the LAE (LBG) candidates in the QSO and the Control fields by counting their numbers in each 0.5 NB921 (0.2 $z'$) total magnitude bin, correcting them for the NB921 ($z'$) detection completeness estimated in Section \ref{CompletenessSec} and shown in Figure \ref{Completeness} and dividing them by the effective survey area of each of the QSO and the Control fields. For the error of the number density in each bin, we include Poisson errors for small number statistics and cosmic variance estimated in the same way as in \citet{Ota08,Ota10}. We use the Poisson upper and lower limits listed in the second columns of Tables 1 and 2 in \citet{Gehrels86}. For the cosmic variance $\sigma_v$ estimate, \citet{Ota08,Ota10} used the relation, $\sigma_v = b \sigma_{\rm DM}$, adopting a bias parameter of $b=3.4\pm1.8$ derived from the sample of 515 $z\sim5.7$ LAEs detected by \citet{Ouchi05} in the $\sim 1.0$ deg$^2$ SXDS field and the dark matter variance $\sigma_{\rm DM}=0.053$ at $z=6.6$ obtained by using an analytic cold dark matter model \citep{Sheth99,MoWhite02} and their survey volumes. In this study, we use a bias parameter of $b=3.6\pm0.7$ derived from the sample of 207 $z\sim6.6$ LAEs detected by \citet{Ouchi10} in the SXDS field. As our QSO field and Control Field (SDF) survey volumes are similar to those of \citet{Ota08,Ota10} (all of them are the volumes based on one Suprime-Cam pointing), we adopt the same $\sigma_{\rm DM}=0.053$ value as they used. This gives a cosmic variance estimate of $\sigma_v \sim 0.19$ for each of the QSO and the Control fields. We also corrected the errors for the detection completeness estimated in Section \ref{CompletenessSec} and shown in Figure \ref{Completeness}. 

In the left panel of Figure \ref{NumberCounts}, we compare surface number densities of LAE candidates per total NB921 magnitude (0.5 mag bin) in the QSO and the Control fields. The LAE number densities are consistent between the two fields at the brighter (23.0--24.5 mag) and the faintest (25.5--26.0 mag) NB921 magnitudes. On the other hand, the LAE number density in the QSO field is significantly lower than that in the Control Field at the intermediate NB921 magnitudes (24.5--25.5 mag) beyond statistical errors and cosmic variance. This deficit of LAEs in the QSO field is also clearly visible when we compare sky distributions of LAEs in the QSO and Control fields in Figure \ref{NumberDensityContours-QSO-SDF-each-own-mean-and-sigma} (see the sizes and corresponding NB921 total magnitudes of the red triangle symbols in the figure). 
                                                                                                          
We also confirm that the brightest LAE candidate (one in NB921 = 23.5--24.0 mag bin) is not located in any LBG overdense regions in the QSO field and thus seem to be irrelevant to overdense environment (see Figures \ref{NumberDensityContours-QSO-SDF-each-own-mean-and-sigma} and \ref{NumberDensityContours-QSO-SDF-mean-and-sigma_in_SDF}). However, note that the bright extended LAB candidate VIKING-z66LAB (the second brightest LAE candidate in NB921 = 23.5--24.0 mag bin) is located close to the LBG overdensity region containing the highest density peak in the lower left (south-west) of the QSO field and might be possibly associated with the overdense environment (see Figures \ref{NumberDensityContours-QSO-SDF-each-own-mean-and-sigma} and \ref{NumberDensityContours-QSO-SDF-mean-and-sigma_in_SDF} and Section \ref{LAB}). This is consistent with the observational trend that LABs have been often found in/around dense environments such as protoclusters to date \citep[e.g.,][]{Steidel00,Chapman04,Colbert06,Matsuda11,Bridge12,Yang12,Yang14,Badescu17}. In contrast, as mentioned earlier, there is the $z=6.541$ LAB, SDF J132415.7+273058, in the average LAE and LBG density region in the Control Field. Thus, LABs can be also found in a normal environment in a general field, and the positional relation between VIKING-z66LAB and the overdensity region in the QSO field could be alternatively the product of chance.

On the other hand, the right panel of Figure \ref{NumberCounts} compares surface number densities of LBG candidates per total $z'$ magnitude (0.2 mag bin) in the QSO and the Control fields. In contrast to the case of LAEs, there is a clear excess in the faintest ($z'$ = 25.8--26.0 mag) LBGs in the QSO field compared to the Control field. Though consistent within errors, there is also a trend of excess in LBGs at $z'$ = 25.4--25.8 mag in the QSO field against the Control Field. Otherwise, the LBG number densities in both fields are consistent at $z'$ = 25.2--25.4 mag or the QSO field exhibits deficit of LBGs at the brightest magnitudes $z'$ = 25.0--25.2 mag (though consistent within the errors). This agrees with the fact that fainter LBG candidates at $z'$ = 25.4--26.0 mag are forming the overdensities in the QSO field as seen in Figures \ref{NumberDensityContours-QSO-SDF-each-own-mean-and-sigma} and \ref{NumberDensityContours-QSO-SDF-mean-and-sigma_in_SDF} (see the sizes and corresponding $z'$-band total magnitudes of the blue circle symbols in the figures).

\subsection{Ly$\alpha$ Luminosity Functions of LAEs\label{LyaLFSec}}
Many of the LBG candidates are expected to be bright in the rest frame UV continuum and very faint in Ly$\alpha$ emission or having no Ly$\alpha$ emission because of their dropout selection method. Some fraction of LBGs may exhibit strong Ly$\alpha$ emisson. However, at $z>6$, the fraction of Ly$\alpha$ emitting LBGs is observed to be low, possibly due to the attenuation of Ly$\alpha$ emission by neutral hydrogen \citep{Stark10,Stark11,Pentericci11,Pentericci14,Ono12,Schenker12,Schenker14,Tilvi14,Caruana12,Caruana14,Treu12,Treu13,Furusawa16}. Thus, the $z'$ band mostly detects UV continua of the LBG candidates except for low fraction of LBGs exhibiting strong Ly$\alpha$ emission. Hence, their surface number density as a function of $z'$ band magnitude shown in the right panel of Figure \ref{NumberCounts} mostly reflects the surface number density as a function of only UV luminosity with low contamination by Ly$\alpha$ emitting LBGs. We cannot remove contamination by such Ly$\alpha$ fluxes.
 
\begin{figure*}
\epsscale{1.5}
\hspace*{-2cm}
\includegraphics[angle=0,scale=0.9]{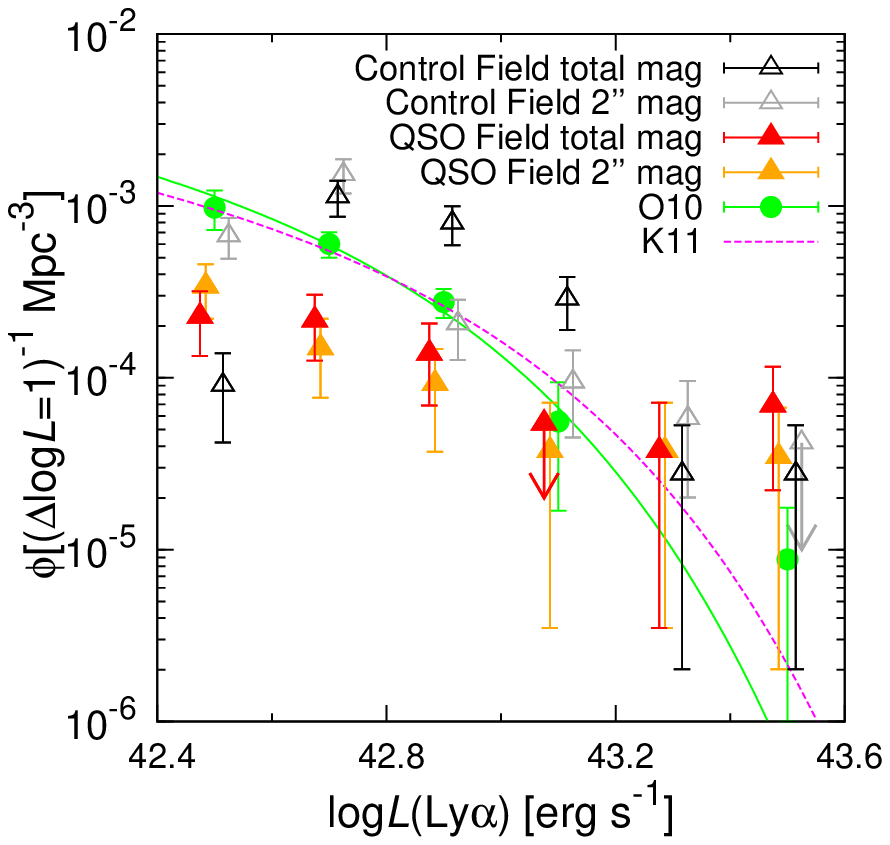}
\hspace*{-2cm}
\includegraphics[angle=0,scale=0.9]{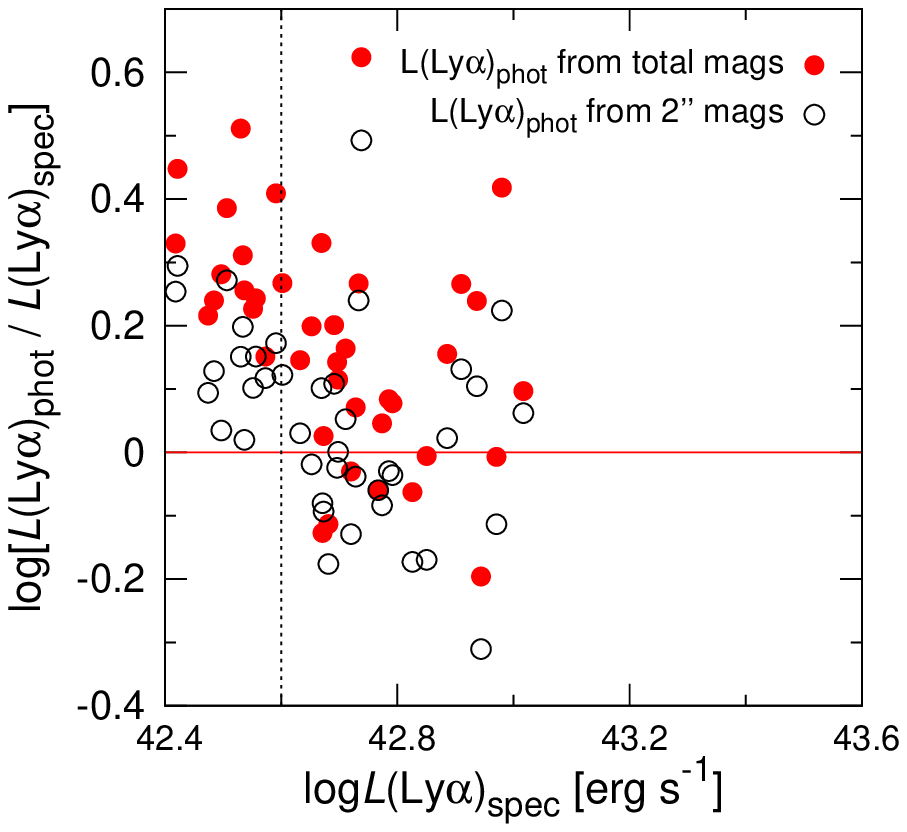}
\caption{(Left) The differential Ly$\alpha$ LFs of the $z\sim6.6$ LAE candidates in the Control Field SDF and the QSO field with their Ly$\alpha$ luminosities $L({\rm Ly}\alpha)_{\rm phot}$'s photometrically estimated from their NB921 and $z'$-band total magnitudes (black open and red filled triangles for the Control and QSO fields, respectively) or $L({\rm Ly}\alpha)_{\rm phot}$'s photometrically estimated from their NB921 and $z'$-band $2''$ aperture magnitudes (grey open and orange filled triangles for the Control and QSO fields, respectively). The data points of the LFs are slightly horizontally shifted from each other for clarity. The errors include both Poission error and cosmic variance. For the Ly$\alpha$ luminosiy bins where no LAE candidate is detected, upper limits are shown by the arrows. We also plot previous measurements of the $z\sim6.6$ LAE Ly$\alpha$ LFs (their best-fit Schechter functions assuming the faint end slope of $\alpha=-1.5$) by \citet[][K11, the magenta dashed curve]{Kashikawa11} and \citet[][O10, the green solid curve, their data points are also shown by the green filled circles]{Ouchi10}. (Right) Ratios of a photometrically estimated Ly$\alpha$ luminosity $L({\rm Ly}\alpha)_{\rm phot}$ to a spectroscopically measured Ly$\alpha$ luminosity $L({\rm Ly}\alpha)_{\rm spec}$ as a function of $L({\rm Ly}\alpha)_{\rm spec}$ of the $z\sim 6.6$ LAEs spectroscopically confirmed in the Control Field SDF. $L({\rm Ly}\alpha)_{\rm phot}$'s were estimated by us from either NB921 and $z'$-band total or $2''$ aperture magnitudes of the LAEs using Equation (\ref{Eqn_LyaUVLum}) while $L({\rm Ly}\alpha)_{\rm spec}$'s were measured from spectra of the LAEs by \citet{Taniguchi05}, \citet{Kashikawa06} and \citet{Kashikawa11}. The horizontal solid line corresponds to $L({\rm Ly}\alpha)_{\rm phot}/L({\rm Ly}\alpha)_{\rm spec} = 1.0$. The vertical dashed line shows the Ly$\alpha$ luminosity $\log L({\rm Ly}\alpha)_{\rm spec}$/[erg s$^{-1}$] $= 42.6$ below which the photometric measurements based on NB921 and $z'$-band total magnitudes largely overestimate real Ly$\alpha$ luminosities measured from spectra such that $L({\rm Ly}\alpha)_{\rm phot}/L({\rm Ly}\alpha)_{\rm spec} \sim 1.5$--3.3 or $\log [L({\rm Ly}\alpha)_{\rm phot}/L({\rm Ly}\alpha)_{\rm spec}] \sim 0.2$--0.5. This overestimation causes many LAEs to move from the faintest Ly$\alpha$ luminosity bin to the next three brighter bins in the Ly$\alpha$ LF of LAEs in the Control Field, making the photometrically estimated LF (black open triangles) higher in the three bins and lower in the faintest bin than the mostly spectroscopically estimated LF (magenta dashed curve) in the left panel.\label{LyaLF}}
\end{figure*}

On the other hand, most of the LAE candidates are expected to be bright in Ly$\alpha$ emission and very faint in the UV continuum because of their narrowband NB921 excess selection method. Hence, the NB921 band detects both their Ly$\alpha$ emission and UV continua (except for LAEs with an undetectably faint UV continuum). Thus, their surface number density as a function of NB921 magnitude shown in the left panel of Figure \ref{NumberCounts} reflects the number density as a function of a mixture of Ly$\alpha$ and UV luminosities. However, we can estimate Ly$\alpha$ and UV luminosities of the LAE candidates separately from their NB921 and $z'$ band total magnitudes and derive the number density as a function of only Ly$\alpha$ luminosity. This is because these bands both cover $z\sim6.6$ Ly$\alpha$ emission and UV continuum redwards of it. To see the trend of the number density of LAEs as a function of only Ly$\alpha$ luminosity, we derive and compare Ly$\alpha$ luminosity functions (LFs) of the LAE candidates in the QSO and Control fields in Figure \ref{LyaLF}. 

We followed the same method as the one used by \citet{Kashikawa11} to estimate Ly$\alpha$ luminosities of our $z\sim6.6$ LAE candidates from their NB921 and $z'$ magnitudes. \citet{Kashikawa11} used the following formula to estimate the Ly$\alpha$ line flux ($f_{\rm line}$ in erg s$^{-1}$ cm$^{-2}$) and the rest frame UV continuum flux density ($f_c$ in erg s$^{-1}$ cm$^{-2}$ Hz$^{-1}$ in observer's frame) of the $z \sim 6.6$ LAEs they detected in the SDF (Control Field) from their narrowband NB921 (NB) and broadband $z'$ (BB) magnitudes, $m_{\rm NB}$ and $m_{\rm BB}$:
\begin{equation}
m_{\rm NB,BB} + 48.6 = -2.5\log\frac{\int^{\nu_{{\rm Ly}\alpha}}_0 (f_c + f_{\rm line})T_{\rm NB,BB}d\nu/\nu}{\int T_{\rm NB,BB}d\nu/\nu} 
\label{Eqn_LyaUVLum}
\end{equation}
where $\nu_{{\rm Ly}\alpha}$ is the observed frequency of Ly$\alpha$, and $T_{\rm NB}$ and $T_{\rm BB}$ are the transmission bandpasses of the NB921 (NB) and $z'$ (BB) filters as a function of observed frequency, respectively (see Figure \ref{FilterTransmission}). \citet{Kashikawa11} used $2''$ aperture NB921 and $z'$ magnitudes of each LAE for $m_{\rm NB}$ and $m_{\rm BB}$. They also used the central frequency of the NB921 filter for $\nu_{{\rm Ly}\alpha}$ if an LAE is not spectroscopically identified. Moreover, they assumed that an SED of an LAE has a constant $f_c$ (i.e., flat continuum), $\delta$-function Ly$\alpha$ emission profile (i.e., flux value of $f_{\rm line}$ at $\nu_{{\rm Ly}\alpha}$ and 0 otherwise) and zero flux at the wavelength bluewards of Ly$\alpha$ due to the IGM absorption. Also, if an LAE was not detected in $z'$-band, $z'$-band $1\sigma$ limiting magnitude was used for $m_{\rm BB}$. 

\citet{Kashikawa11} compared the Ly$\alpha$ fluxes of 45 spectroscopically identified $z\sim6.6$ LAEs in the Control Field estimated photometrically from their $2''$ aperture NB921 and $z'$ magnitudes using the equation (\ref{Eqn_LyaUVLum}) with those measured from their spectra, and confirmed that they are in fairly good agreement within a factor of two (see Figure 5 in their paper and the right panel of Figure \ref{LyaLF} in the present paper). Because they used a large spectroscopic $z\sim6.5$ LAE sample including LAEs with bright to faint Ly$\alpha$ luminosities, the validity of the method was statistically proven to be highly reliable. 

We used the central frequency of the NB921 filter for $\nu_{{\rm Ly}\alpha}$, $2''$ aperture NB921 and $z'$ magnitudes of our $z\sim6.6$ LAE candidates in the QSO and Control fields and equation (\ref{Eqn_LyaUVLum}) to estimate their Ly$\alpha$ fluxes ($f_{\rm line}$). Then, we converted the fluxes to the Ly$\alpha$ luminosities. Also, we estimated the number density of LAE candidates by dividing their observed differential numbers in each Ly$\alpha$ luminosity bin by the effective survey volume of the QSO field or the Control Field \citep[see Section \ref{LAE-Selections} and][for the detials of the survey volumes]{Taniguchi05}. Moreover, we estimated the errors on the LAE number densities including the Poisson errors and cosmic variance in the same way as we did in Section \ref{NC}. Finally, we corrected the number densities and the errors for the detection completeness estimated in Section \ref{CompletenessSec} and shown in Figure \ref{Completeness} by number weighting according to the NB921 magnitude. 


In the left panel of Figure \ref{LyaLF}, we compare the differential Ly$\alpha$ LFs of the LAE candidates in the QSO and the Control fields (orange filled and grey open triangles, respectively). We also plot the previous measurement of the $z\sim6.6$ LAE Ly$\alpha$ LF (its best-fit Schechter function assuming the faint end slope of $\alpha=-1.5$) in the Control Field SDF derived by \citet{Kashikawa11}. They used spectroscopically measured Ly$\alpha$ fluxes for the spectroscopically identified LAEs ($\sim 80$\% of the LAE candidates) and Ly$\alpha$ fluxes photometrically inferred from Equation (\ref{Eqn_LyaUVLum}) for the remaining unidentified LAE candidates ($\sim 20$\% of the LAE candidates). We see that their LF (magenta dashed curve in the figure) and our LF in the Control Field are almost consistent. This suggests that we have been able to successfully photometrically reproduced the LF in the Control Field that was accurately measured by spectroscopy. However, strictly speaking, the number density of LAEs in the second faintest bin of our LF is larger than that of the \citet{Kashikawa11}'s spectroscopic LF due to the following two reasons. (1) The second faintest bin of our LF includes four out of the five additional LAE candidates detected without imposing non-detections in wavebands ($B$, $V$ and $R_c$) bluewards of $z\sim6.6$ Ly$\alpha$ (the case 1--5 objects in Figure \ref{woBVR_Objects}, see Sections \ref{SDFsample} and \ref{lackBVR}) while the \citet{Kashikawa11} LF does not contain them. (2) For the faintest LAEs with spectroscopically measured Ly$\alpha$ luminosities $\log L({\rm Ly}\alpha)_{\rm spec}$/[erg s$^{-1}$] $\sim$ 42.4--42.6 in the faintest LF bin, Equation (\ref{Eqn_LyaUVLum}) and $2''$ aperture NB921 and $z'$ magnitudes photometrically tend to overestimate their Ly$\alpha$ luminosities $L({\rm Ly}\alpha)_{\rm phot}$ by a factor of $\log L({\rm Ly}\alpha)_{\rm phot}/L({\rm Ly}\alpha)_{\rm spec} \sim 0.1$--0.3 (see the right panel of Figure \ref{LyaLF}). This causes some LAEs in the faintest bin to move to the second faintest bin, decreasing the LAE number density in the faintest bin and increasing the LAE number density in the second faintest bin when comparing the photometric LF with the LF mostly spectroscopically derived by \citet{Kashikawa11}. Nonetheless, despite the larger LAE number in the second faintest bin, our photometric LF is entirely well consistent with the \citet{Kashikawa11}'s spectroscopic LF. Thus, we presume that our photometric LF of LAE candidates in the QSO field estimated by using Equation (\ref{Eqn_LyaUVLum}) also well reproduces the realistic LF in the QSO field that can be ideally derived by spectroscopic measurements of Ly$\alpha$ luminosities of LAEs. 

In the left panel of Figure \ref{LyaLF}, we see that the LFs in the QSO and Control fields (orange filled and grey open triangles) are consistent at the four brightest Ly$\alpha$ luminosity bins $\log L({\rm Ly}\alpha)$/[erg s$^{-1}] = 42.8$--43.6 within statistical error and cosmic variance. Although consistent within errors, there is a trend that the number density of LAEs in the QSO field is lower than that in the Control Field at the intermediate Ly$\alpha$ luminosities $\log L({\rm Ly}\alpha)$/[erg s$^{-1}] = 42.8$--43.2. Moreover, the number density of the LAEs in the QSO field is lower than that in the Control Field beyond statistical error and cosmic variance at the fainter Ly$\alpha$ luminosities $\log L({\rm Ly}\alpha)$/[erg s$^{-1}] = 42.4$--42.8. Therefore, the number density of the LAEs at the intermediate to faint Ly$\alpha$ luminosities is lower in the QSO environment than a general blank field. This is consistent with the same trend we can see when we compare sky distributions and number density contours of LAEs in the QSO and Control fields in Figure \ref{NumberDensityContours-QSO-SDF-each-own-mean-and-sigma}.

Moreover, in the left panel of Figure \ref{LyaLF}, we also plot the $z\sim6.6$ LAE Ly$\alpha$ LF (its data points and best-fit Schechter function assuming the faint end slope of $\alpha=-1.5$) derived by \citet{Ouchi10} although their LAE selection criteria used to select $\sim$ 78\% of their LAEs are slightly different from ours (they did not impose $z'-{\rm NB921}>3\sigma$ to select LAEs in the SXDS field). Their LF is based on the $z\sim6.6$ LAEs detected in the SDF and SXDS fields (6 Suprime-Cam pointing) whose area is 6 times larger than that of our Control Field SDF. Thus, their LF represents more typical trend reducing the field-to-field variance effect. \citet{Kashikawa11}'s LF in the Control Field is consistent with \citet{Ouchi10}'s LF. As our photomerically derived Control Field LF is mostly consistent with \citet{Kashikawa11}'s LF (see text above), it is also compatible with \citet{Ouchi10}'s LF. This means that the Control Field represents a typical field for $z\sim6.6$ LAEs, which supports validity of our choice of the SDF as a control field for this study. This further ensures that the LAE number density is significantly lower in the QSO field than a typical blank field.

Finally, for comparison, we also photometrically derive the Ly$\alpha$ LFs of LAE candidates in the QSO and Control fields by using their ``total'' NB921 and $z'$ magnitudes (rather than $2''$ aperture magnitudes) for $m_{\rm NB}$ and $m_{\rm BB}$ in Equation (\ref{Eqn_LyaUVLum}) to estimate their Ly$\alpha$ fluxes $f_{\rm line}$ and then their Ly$\alpha$ luminosities $L({\rm Ly}\alpha)_{\rm phot}$. We do this because $2''$ aperture magnitudes may underestimate Ly$\alpha$ $+$ UV continuum fluxes of each LAE possibly missing detecting some fluxes lost outside of the $2''$ aperture. We plot the derived Ly$\alpha$ LFs in the QSO and Control fields (red filled and black open triangles, respectively) in the left panel of Figure \ref{LyaLF}. As for the QSO field, the LFs derived from total and $2''$ aperture NB921 and $z'$ magnitudes (red and orange filled triangles) are well consistent with each other. Although consistent within errors, there is a trend that the number density of LAEs in the total magnitude LF is lower in the faintest Ly$\alpha$ luminosity bin and higher in the next two bins than that in the $2''$ aperture magnitude LF. This is because total magnitudes and Equation (\ref{Eqn_LyaUVLum}) give Ly$\alpha$ luminosities higher than those estimated using $2''$ aperture magnitudes, making some LAEs move from the faintest bin to the next two bins when comparing the total magnitude LF with the $2''$ aperture magnitude LF.

On the other hand, as for the Control Field, the LFs derived from total and $2''$ aperture NB921 and $z'$ magnitudes (black and grey open triangles) are consistent only at the two brightest Ly$\alpha$ luminosity bins and the second faintest bin. The number density of LAEs in the total magnitude LF is lower in the faintest Ly$\alpha$ luminosity bin and higher in the third and fourth faintest bins than that in the $2''$ aperture magnitude LF. Also, the number densities of LAEs in the second to fourth faintest Ly$\alpha$ luminosity bins in the total magnitude LF are higher than those of the LF mostly spectroscopically derived by \citet{Kashikawa11}. This is mainly because of the following reason. For the faintest LAEs with spectroscopically measured Ly$\alpha$ luminosities $\log L({\rm Ly}\alpha)_{\rm spec}$/[erg s$^{-1}$] $\sim$ 42.4--42.6 in the faintest LF bin, Equation (\ref{Eqn_LyaUVLum}) and total NB921 and $z'$ magnitudes photometrically tend to overestimate their Ly$\alpha$ luminosities $L({\rm Ly}\alpha)_{\rm phot}$ by a factor of $\log L({\rm Ly}\alpha)_{\rm phot}/L({\rm Ly}\alpha)_{\rm spec} \sim 0.2$--0.5 (see the right panel of Figure \ref{LyaLF}). This causes many LAEs in the faintest bin to move to the second to fourth faintest bins, decreasing the LAE number density in the faintest bin and increasing the LAE number density in the second to fourth faintest bins when comparing the total magnitude LF with the LF spectroscopically derived by \citet{Kashikawa11}. Despite the difference between the LFs derved from total and $2''$ aperture NB921 and $z'$ magnitudes, both LFs in the QSO and Control fields show almost the same trend that the number density of LAEs with intermediate to faint Ly$\alpha$ luminosities is lower in the QSO field than the Control Field.

\subsection{Can QSO Feedback Suppress the Formation of LAEs and LBGs?\label{Feedback}}
In Section \ref{NdensityContours} and Figures \ref{NumberDensityContours-QSO-SDF-each-own-mean-and-sigma} and \ref{NumberDensityContours-QSO-SDF-mean-and-sigma_in_SDF}, we found that LAE candidates are sparse (3 LAEs) while there are more LBG candidates (12 LBGs) in the QSO proximity region ($< 3$ physical Mpc from the QSO). Also, the proximity region is located at the lower density edge of the large scale structures of the LAE and LBG candidates. The UV radiation from the QSO might have suppressed formation of LAEs (lower mass galaxies) while it may have not affected formation of LBGs (higher mass galaxies) to finally form these biased sky and density distributions of LAEs and LBGs. To examine this scenario, we quantitatively estimate the strength of the QSO radiation around it and see if it has any effects on the formation of LAEs and LBGs in the proximity region. 

We follow the same method taken by \cite{Kashikawa07}. We assume that the QSO spectrum can be approximated by a power low, $F^Q_{\nu} \propto \nu^{\beta}$, where $\beta$ is the UV continuum slope of the QSO spectrum. Then, the local flux density at the Lyman limit frequency, $\nu_{\rm L}$, at a radius $r$ from the QSO is given by
\begin{equation}
F^Q_{\nu}(\nu_{\rm L}, r) = \frac{L_{\nu}(\nu_{\rm L})}{4 \pi r^2} 
\label{QSO-flux}
\end{equation}
where $L_{\nu}(\nu_{\rm L}) = 4 \pi D_{\rm L}^2 F^Q_{\nu}(\nu_{\rm L}, D_{\rm L}) (1+z)$ is the QSO luminosity at $\nu_{\rm L}$ and $D_{\rm L}$ is the luminosity distance. We estimate the continuum slope $\beta$ of the $z=6.61$ QSO from its magnitudes at the rest frame UV wavelengths by using 
\begin{equation}
\beta = \frac{m_1 - m_2}{2.5 \log (\lambda_{c,1}/\lambda_{c,2})}
\label{UV-slope}
\end{equation} 
where $m_1$, $m_2$, $\lambda_{c,1}$ and $\lambda_{c,2}$ are apparent magnitudes and central wavelengths of broadband filters 1 and 2. The QSO was observed in the VISTA $Y$, $J$, $H$ and $K_{\rm s}$ bands in the VIKING survey. All of them cover the rest-frame UV wavelengths redwards of Ly$\alpha$ and do not include the Ly$\alpha$ emission and the continuum trough bluewards of Ly$\alpha$. The central wavelengths are 1.020, 1.252, 1.645 and 2.147 $\mu$m for $Y$, $J$, $H$ and $K_{\rm s}$ bands, respectively\footnote[4]{http://casu.ast.cam.ac.uk/surveys-projects/vista/technical/ filter-set}. \citet{Venemans13} measured the magnitudes of the QSO in these bands to be $Y=20.89$, $J=20.68$, $H=20.72$ and $K_{\rm s}=20.27$ AB mag. We calculate $\beta$'s from the combinations of ($Y$, $J$), ($J$, $H$) and ($H$, $K_{\rm s}$) and the Equation (\ref{UV-slope}) and adopt the average of the three, $\beta = -0.79$, as the UV continuum slope of the QSO. 

From this $\beta$ and the absolute magnitude of the QSO at a rest-frame wavelength of 1450\AA, $M_{1450} = -25.96$ AB mag, measured by \citet{Venemans13}, we obtain  $L_{\nu}(\nu_{\rm L}) \sim 8.5 \times 10^{31}$ erg s$^{-1}$ Hz$^{-1}$. For the radius of the proximity region, $r = 3$ physical Mpc, we obtain $F^Q_{\nu}(\nu_{\rm L}, 3 {\rm pMpc}) \sim 7.9 \times 10^{20}$ erg s$^{-1}$ cm$^{-2}$ Hz$^{-1}$ from the Equation (\ref{QSO-flux}).  
 
We assume the UV intensity in the form of a power-law spectrum,
\begin{equation}
J(\nu) = J_{21} \times (\nu/\nu_{\rm L})^{\alpha} \times 10^{-21} {\rm erg} {\rm s}^{-1} {\rm cm}^{-2} {\rm Hz}^{-1} {\rm sr}^{-1}
\label{UV-intensity}
\end{equation}
where $J_{21}$ is an isotropic UV intensity at the Lyman limit and $\alpha$ is the continuum slope. As $J(\nu_{\rm L}) = F^Q_{\nu}(\nu_{\rm L}, r)/4 \pi$, we obtain $J_{21} \sim 6.3$ for the UV intensity at the edge of the proximity region ($r = 3$ physical Mpc). 

On the other hand, \citet{Calverley11} measured the UV background (UVB) at $z=4.6$--6.4 using QSO proximity effect. They derived the correlation between the HI photoionization rate by the UVB and redshift, $\log \Gamma_{\rm bkg} \sim -0.87 z - 7.7$ (see Figure 10 in their paper). This gives $\log \Gamma_{\rm bkg} \sim -13.45$ at $z=6.61$ that corresponds to the UVB intensity of $J_{21} \sim 0.013$ (using Equation (8) in their paper and Equation (\ref{UV-intensity}) in this paper). Hence, the QSO radiation is 485 times stronger than the UVB at $r = 3$ physical Mpc from the $z=6.61$ QSO. Meanwhile, the LAE candidate nearest to the QSO is located at the projected distance $r \sim 1$ physical Mpc from the QSO. This is the minimum possible distance between the observed LAE and the QSO. At $r \sim 1$ physical Mpc, we estimate the QSO radiation intensity to be $J_{21} \sim 56.4$. 
 
Does this affect the formation of LAEs and LBGs? We examine this by using Figure 8 in \citet{Kashikawa07} that predicts the delay time $t_{\rm delay}$ of star formation as a function of radiation intensity $J_{21}$ for a given virial mass of a halo $M_{\rm vir}$. \citet{Kashikawa07} derived this relation between $t_{\rm delay}$, $J_{21}$ and $M_{\rm vir}$ by performing radiation-hydrodynamic simulations to examine the effect of radiation of the $z=4.87$ QSO they observed on star formation. In the simulations, gas was set to collapse (in the absence of thermal pressure) at $z=4.87$. We assume that the same relation also holds in the case of gas that collapses at $z=6.61$. When $J_{21} \sim 6.3$ ($r = 3$ physical Mpc from the $z=6.61$ QSO), star formation is suppressed in a halo with $M_{\rm vir} < 10^{10} M_{\odot}$. If $J_{21} \sim 56.4$ ($r = 1$ physical Mpc from the QSO), star formation is suppressed in a halo with $M_{\rm vir} < 3 \times 10^{10} M_{\odot}$. 


\begin{figure*}
\epsscale{1.17}
\plottwo{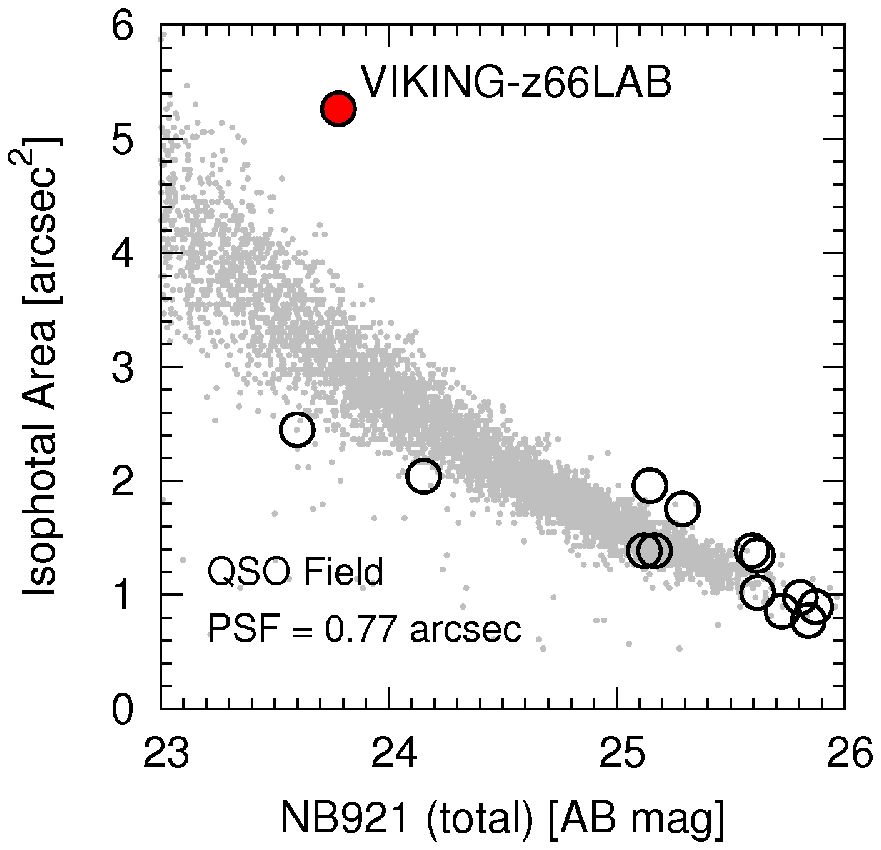}{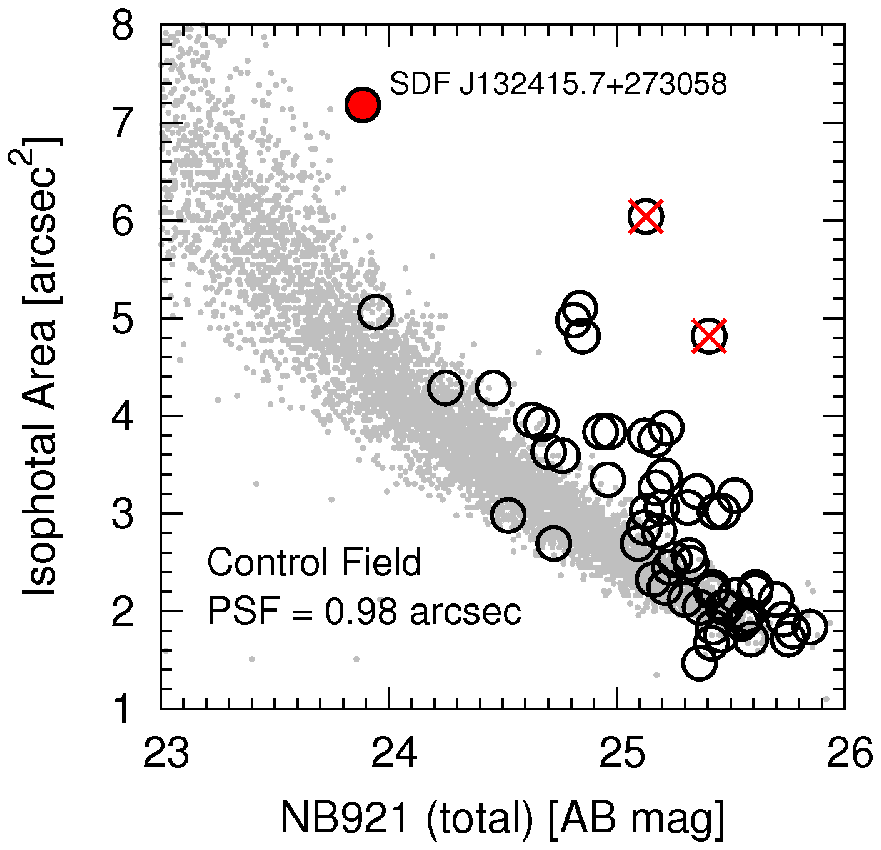}
\caption{Isophotal area as a function of NB921 total magnitude of $z\sim6.6$ LAE candidates (open circles) and point sources (gray dots) measured with the PSF FWHM $= 0.''77$ NB921 image of the QSO field (left panel) and the PSF FWHM $= 0.''98$ NB921 image of the Control Field SDF (right panel) and measured by the SExtractor parameter {\tt ISOAREA\_IMAGE}. The isophotal area is defined as an area corresponding to the pixels with values above $2\sigma$ sky fluctuation in each image. The point sources were selected with the SExtractor parameters {\tt CLASS\_STAR} (stellarity) $> 0.9$ and {\tt FLAGS} $=0$. The $z\sim6.6$ LAB candidate, VIKING-z66LAB, and the spectroscopically confirmed $z=6.541$ LAB, SDF J132415.7+273058 \citep{Kodaira03,Taniguchi05}, are denoted by the red filled circles. The $z\sim6.6$ LAE candidates with the red crosses are located in very noisy regions (blended with noises) in the NB921 image and their isophotal area measurements are unreliable.\label{IsophotalArea}}
\end{figure*}


\begin{figure}
\epsscale{1.18}
\plotone{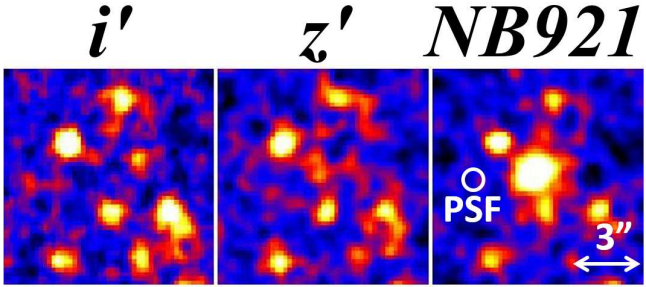}
\caption{The $10'' \times 10''$ $i'$, $z'$ and NB921 images of the LAB candidate, VIKING-z66LAB, found near the most overdense region in the QSO field. It is extended in NB921 (Ly$\alpha$ + UV continuum) with its angular diameter $\gtrsim 3''$ or $\gtrsim 16$ physical kpc. It appears that at least two objects are interacting at the position of VIKING-z66LAB in the $z'$-band image. All the images have been convolved to have the same PSF FWHM of $0.''91$. This PSF size is displayed in the NB921 image to clearly show how extended VIKING-z66LAB is.\label{LABFig}}
\end{figure}


\begin{deluxetable*}{ccccccccccc}
\tabletypesize{\scriptsize}
\tablecaption{Photometric Properties of the $z\sim 6.6$ LAB candidate VIKING-z66LAB in the QSO Field\label{LABPhotometry}}
\tablewidth{510pt}
\tablehead{
\colhead{R.A.(J2000)} & \colhead{Decl.(J2000)} & \colhead{$i'_{2''}$} & \colhead{$z'_{2''}$} & \colhead{$z'_{\rm total}$$^{\rm b}$} & \colhead{NB921$_{2''}$} & \colhead{NB921$_{\rm total}$$^{\rm c}$} & \colhead{stellarity$^{\rm d}$} & \colhead{$L({\rm Ly}\alpha)$$^{\rm e}$} & \colhead{$M_{\rm UV}$$^{\rm e}$} & \colhead{EW$_0$$^{\rm f}$}\\
\colhead{} & \colhead{} & \colhead{(mag)} & \colhead{(mag)} & \colhead{(mag)} & \colhead{(mag)} & \colhead{(mag)} & \colhead{} & \colhead{(erg s$^{-1}$)} & \colhead{(mag)} & \colhead{(\AA)}
}
\startdata
03:04:41.091 & $-$32:04:27.30 & $>$28.1$^{\rm a}$ & 26.16 & 25.96 & 24.41 & 23.78 & 0.01 & $2.6 \times 10^{43}$ & $-20.63$ & 164\\
\enddata
\tablecomments{Units of coordinate are hours: minutes: seconds (right ascension) and degrees: arcminutes: arcseconds (declination) using J2000.0 equinox. The $i'_{2''}$, $z'_{2''}$ and NB921$_{2''}$ are the aperture magnitudes (we first detected VIKING-z66LAB in the NB921 image and then measured the $2''$ aperture magnitudes in the $i'$, $z'$ and NB921 images using the SExtractor double image mode). VIKING-z66LAB has colors of $i'-z' > 1.94$ ($1\sigma$ limit) and $z'-{\rm NB921} =1.75$ calculated from those $2''$ aperture magnitudes and/or its $1\sigma$ limit.}
\tablenotetext{a}{$1\sigma$ limit.}
\tablenotetext{b}{The total $z'$-band magnitude. We detected VIKING-z66LAB in the $z'$-band image, then measured the $2''$ aperture $z'$-band magnitude (in this case, 26.10 mag) using the SExtractor single image mode and finally applied the aperture correction of $-0.14$ mag to obtain $z'_{\rm total}$. We used the aperture correction estimated for LBGs as VIKING-z66LAB is more likely a Ly$\alpha$ emitting LBG (see text). We did not use the $z'_{2''}=26.16$ measured by the SExtractor double image mode to estimate $z'_{\rm total}$ because SExtractor adopts the detection position (X,Y) in the NB921 image as the position of the $2''$ aperture placed in the $z'$-band image but in this case VIKING-z66LAB is not located exactly at the center of the $2''$ aperture with some fraction of its flux is lost out of the aperture. When we use the SExtractor single image mode, the $2''$ aperture is placed at the detection position in the $z'$-band image and VIKING-z66LAB is located at the center of the $2''$ aperture minimizing the flux loss.}
\tablenotetext{c}{The total NB921 magnitude measured by SExtractor {\tt MAG\_AUTO} in the ${\rm PSF}=0.''77$ NB921 image of the QSO field (the original image before the convolution for the $2''$ aperture photometry). See Section \ref{NC} for the details.}
\tablenotetext{d}{The star/galaxy classifier index measured in the ${\rm PSF}=0.''77$ NB921 image and given as {\tt CLASS\_STAR} parameter by SExtractor. It is 0 for a galaxy, 1 for a star, or any intermediate value for more ambiguous objects \citep{BA96}.}
\tablenotetext{e}{The Ly$\alpha$ luminosity $L({\rm Ly}\alpha)$ and the absolute UV continuum magnitude $M_{\rm UV}$ estimated from the NB921$_{\rm total}$ and $z'_{\rm total}$ magnitudes by using the equation (\ref{Eqn_LyaUVLum}).}
\tablenotetext{f}{The rest-frame Ly$\alpha$ equivalent width calculated from $L({\rm Ly}\alpha)$ and $M_{\rm UV}$.}
\end{deluxetable*}

Using a semi-analytic model of galaxy formation, \citet{Garel15} predicted that halo masses of typical ($L({\rm Ly}\alpha) = 10^{42}$--$10^{43}$ erg s$^{-1}$) and bright ($L({\rm Ly}\alpha) = 10^{43}$--$10^{44}$ erg s$^{-1}$) LAEs at $z=6.6$ are $\log (M_{\rm h}/M_{\odot}) = 10.8_{-0.2}^{+0.4}$ and $\log (M_{\rm h}/M_{\odot}) = 11.5_{-0.1}^{+0.1}$, respectively. Meanwhile, \citet{Ouchi10} estimated a halo mass of a $z\sim 6.6$ LAE from the clustering of 207 $z\sim 6.6$ LAE candidates detected over 1 deg$^2$ sky of SXDS field in the Suprime-Cam NB921 band to the same depth as our observations of the $z=6.61$ QSO (NB921 $<$ 26.0 or $L({\rm Ly}\alpha) > 2.5 \times 10^{42}$ erg s$^{-1}$). They estimated the minimum, average and maximum halo masses to be $\log (M_{\rm h}^{\rm min}/M_{\odot}) = 9.9_{-0.6}^{+0.4}$, $\log (M_{\rm h}/M_{\odot}) = 10.3_{-0.4}^{+0.4}$ and $\log (M_{\rm h}^{\rm max}/M_{\odot}) = 11.1_{-0.4}^{+0.3}$, respectively. More recently, based on the halo occupation distribution (HOD) models, \citet{Ouchi17} also estimated a halo mass of a $z\sim 6.6$ LAE from 873 $z\sim 6.6$ LAE candidates detected using the early data of the Hyper Suprime-Cam Subaru Strategic Program survey to the brighter limit (NB921 $<$ 25.0 or $L({\rm Ly}\alpha) > 7.9 \times 10^{42}$ erg s$^{-1}$ or $\gtrsim L^*$) over 21.2 deg$^2$ area of sky. They estimated the minimum and average halo masses to be $\log (M_{\rm h}^{\rm min}/M_{\odot}) = 9.1_{-1.9}^{+0.7}$ and $\log (\langle M_{\rm h} \rangle /M_{\odot}) = 10.8_{-0.5}^{+0.3}$, respectively. Eventually, $z\sim 6.6$ LAEs could have a halo mass from $<10^{10} M_{\odot}$ to $\sim 10^{11} M_{\odot}$. Thus, formation of $M_{\rm h} < 1$--$3 \times 10^{10} M_{\odot}$ LAEs could be suppressed by the QSO radiation in the QSO proximity region while that of LAEs hosted by higher mass halos is not. The three LAE candidates seen in projection in the QSO proximity region could have $M_{\rm h} > 1$--$3 \times 10^{10} M_{\odot}$. Otherwise, they are located at $r > 3$ physical Mpc foreground from the QSO in nearly line-of-sight direction as the NB921 band have the better sensitivity to the foreground LAEs due to the redshift of the QSO as mentioned in Section 1 and Figure \ref{NB921_LyaPeakDist}.  

On the other hand, our LBG candidates in the QSO field have the rest-frame UV continuum magnitudes (at 1250\AA) of $M_{\rm UV} \leq -20.8$ converted from the limiting magnitude for the LBG selection $z' \leq 26.1$ assuming that LBGs are at $z=6.61$. \citet{Garel15} predicted that halo masses of bright ($-20.8 > M_{1500} > -23.3$) LBGs at $z=6.6$ are $\log (M_{\rm h}/M_{\odot}) = 11.5_{-0.1}^{+0.2}$. Meanwhile, based on the HOD models, \citet{Harikane17} estimated the halo mass of $z\sim6.8$ LBGs with $M_{\rm UV} < -19.5$ to be $\log (M_{\rm h}/M_{\odot}) = 11.00_{-0.08}^{+0.07}$. As the halo masses of the LBGs are sufficiently high (i.e., $M_{\rm h} > 3 \times 10^{10} M_{\odot}$), their formation would not be suppressed by the radiation from the $z=6.61$ QSO even if they are within the QSO proximity region. In addition, as the redshift range of the LBG candidates spans $6 < z < 6.9$, if some of them are far from the QSO in the line-of-signt direction, their formation is of course not affected by the QSO radiation.

All the discussions above can explain the fact that there are much less LAE candidates than LBG candidates within the QSO proximity region. Moreover, this can also be caused by the difference in the probed volume between LAEs and LBGs ($\Delta z \sim 0.1$ for LAEs and $\Delta z \sim 0.9$ for LBGs) as more galaxies are detected in a larger volume. However, the sparsity of LAE candidates in the QSO proximity region compared to the number of LAE candidates outside the proximity region may imply that formation of some of the lower mass LAEs was possibly suppressed by the QSO radiation.  

\subsection{A Bright Extended Ly$\alpha$ Blob Candidate near the Most Overdense Region in the QSO Field\label{LAB}}
We found a $z\sim6.6$ bright extended LAB candidate, VIKING-z66LAB, near the most overdense region consisting of LBG candidates located in the south-west of the QSO field (see Figures \ref{NumberDensityContours-QSO-SDF-each-own-mean-and-sigma} and \ref{NumberDensityContours-QSO-SDF-mean-and-sigma_in_SDF} for its location in the QSO field). VIKING-z66LAB was detected as one of the $z\sim6.6$ LAE candidates in the QSO field and satisfies the LAE selection criteria (\ref{Criteria-1}). Figure \ref{CMDand2CD} shows the locations of VIKING-z66LAB in the $z'-{\rm NB921}$ versus NB921 color-magnitude diagram and the $z'-{\rm NB921}$ versus $i'-z'$ two-color diagram. VIKING-z66LAB is also identified as an object with a very bright NB921 magnitude and its size in NB921 band (Ly$\alpha$ + UV continuum) much extended than stellar sources and any other LAE candidates in the isophotal area versus NB921 magnitude diagram in Figure \ref{IsophotalArea}. Figure \ref{LABFig} also shows VIKING-z66LAB is much more extended than the PSF FWHM size in the NB921 image. 

VIKING-z66LAB also has a very red color of $i'-z'>1.94$ comparable to the $i'-z'>1.8$ color cut of the LBG selection criteria (\ref{Criteria-3}). However, it was not selected as an LBG candidate as it has a $2''$ aperture magnitude of $z'=26.16$ and is slightly fainter than the limiting magnitude $z'=26.1$ for the LBG selection criteria (\ref{Criteria-3}). Hence, VIKING-z66LAB is more likely a Ly$\alpha$ emitting LBG with strong Ly$\alpha$ emission and a faint UV continuum. Actually, it has a high Ly$\alpha$ luminosity of $L({\rm Ly}\alpha) \sim 2.6 \times 10^{43}$ erg s$^{-1}$ and a faint UV continuum magnitude of $M_{\rm UV} \sim -20.63$ mag both estimated from its total NB921 and $z'$ magnitudes and using the equation (\ref{Eqn_LyaUVLum}). Figure \ref{CMD_LBGs_both_fields} shows the location of VIKING-z66LAB in the $i'-z'$ versus $z'$ color-magnitude diagram. Also, the photometric properties of VIKING-z66LAB are shown in Table \ref{LABPhotometry}.     

Figure \ref{LABFig} shows the images of VIKING-z66LAB in $i'$, $z'$ and NB921 bands. It is not detected in $i'$, faintly detected in $z'$ and bright and extended in NB921 (Ly$\alpha$ + UV continuum) with its angular diameter $\gtrsim 3''$ or $\gtrsim 16$ physical kpc. This size in NB921 is comparable to those of the previously found $z\sim6.6$ LABs also detected in NB921 with Subaru Suprime-Cam such as Himiko and CR7 \citep{Ouchi09,Sobral15}. The HST Wide Field Camera 3 (WFC3) near-infrared high resolution images of Himiko and CR7 revealed that each of them consists of three objects merging or interacting within a large Ly$\alpha$ cloud \citep{Ouchi13,Sobral15}. VIKING-z66LAB appears to consist of two sources merging or interacting in the $z'$-band as seen in Figure \ref{LABFig}. Thus, VIKING-z66LAB might turn out to be a merger system if seen in higher resolution images because it is located near the most overdense region in the QSO field, and because merging of galaxies tends to frequently occur in/around such dense environments. Therefore, if VIKING-z66LAB is found to be a multiple merger system at $z\sim6.6$ by follow-up high resolution imaging and spectroscopy, this would support the reality of the most overdense region of the LBG candidates in the south-west of the QSO field, which is a part of the larger scale structure of LBG candidates also containing the $z=6.61$ QSO and its proximity region. 

On the other hand, as mentioned in Section \ref{NC}, there is also the comparably bright (NB921$_{\rm total}$ = 23.69) and extended ($\gtrsim 3''$) $z=6.541$ LAB, SDF J132415.7+273058 \citep{Kodaira03,Taniguchi05}, in the average LAE and LBG density region in the Control Field SDF as seen in Figure \ref{NumberDensityContours-QSO-SDF-each-own-mean-and-sigma}. This LAB can be also identified with its size more extended than stellar sources in the isophotal area versus NB921 magnitude diagram in Figure \ref{IsophotalArea}. \citet{Jiang13} carried out high resolution observations of this LAB in the HST WFC3 near-infrared bands and found that it does not show multiple components or tails but is extended and elongated. They also found that there is a misalignment between the positions of its Ly$\alpha$ and UV continuum emission in this LAB with the Ly$\alpha$ position close to that of the fainter component of the UV continuum emission. Hence, they concluded that this LAB could be in the end of the merging process. If this is the case, this means that galaxy merging also occurs at $z\sim6.6$ in an average galaxy density environment in a general blank field, thereby forming an LAB. Thus, the positional relation between VIKING-z66LAB and the highest overdensity region in the QSO field could be alternatively interpreted as the product of chance. To examine this, follow-up spectroscopy of the LBG candidates in/around the highest overdensity region is required. 

\section{Summary and Conclusion}
We conducted Subaru Suprime-Cam $i'$, $z'$ and NB921 band imaging of a sky area of $\sim 700$ arcmin$^2$ around the $z=6.61$ QSO J0305--3150 hosting an $M_{\rm BH} \sim 1 \times 10^9 M_{\odot}$ SMBH and detect both LAE and LBG candidates in the QSO field. In the same way and to the comparable depths, area and completeness, we also detect LAE and LBG candidates as a control sample in the Control Field (SDF), a general blank sky field where we confirm that there exist neither $z \sim 6.6$ QSOs, clustering of LAEs and LBGs nor over/underdensities of them. This allows us, for the first time, to  probe galaxies with a wide range of masses and ages around a $z>6$ QSO in a large sky area to elucidate potential large scale galaxy overdensities by measuring galaxy densities around the QSO accurately using the control sample as a rigorous baseline for comparison. This makes up for the shortcomings of previous studies that probed only LBGs (biased to massive older galaxies) in small areas (only tens of arcmin$^2$) around $z>6$ QSOs not using consistently constructed control samples as a baseline and possibly causing puzzling results of finding a wide variety of galaxy densities right around $z>6$ QSOs. We compare sky distributions, surface number density contours, number counts, Ly$\alpha$ LFs and ACFs of the LAEs/LBGs in the $z=6.61$ QSO and Control fields. 

The sky distributions and the number density contours indicate that LAE and LBG candidates are spreading on a large scale mostly over a $\sim 30 \times 60$ comoving Mpc$^2$ area in the south half part of the QSO field. Over this area, the number density of LAEs is almost equivalent to the mean to mean$-1\sigma$ desnsity of LAEs in the Control Field. Conversely, over this area, LBGs exhibit a filamentary overdensity structure running from east to west. The LBG structure contains several 3--$7\sigma$ high density excess clumps. On the other hand, LAEs and LBGs are very sparse in the north half of the QSO field, both showing the number densities equivalent to the mean to mean$-1\sigma$ desnsities of LAEs and LBGs in the Conrol Field. 

The QSO and its proximity region (projected circular region around the QSO equivalent to the size of the NB921 filter's FWHM) could be part of the large scale LBG structure but are located at its near-edge region and not exactly at the highest density peaks. In this proximity region of the QSO, LAEs show lower number densities while LBGs exhibit average to $4\sigma$ excess number densities compared to the Control Field. Thus, the QSO may be part of a moderate galaxy overdensity at most. If this environment reflects a halo mass, the QSO may be in a moderately massive halo, not the most massive one.

The number counts of LAEs in the NB921 total magnitude and the Ly$\alpha$ LF of LAEs in the QSO field are consistent with those in the Control Field within statistical errors and cosmic variance at the faintest (NB921 $=$ 25.5--26.0 and $\log L({\rm Ly}\alpha)$/[erg s$^{-1}$] $=$ 42.4--42.6) and bright (NB921 $=$ 23.0--24.5 and $\log L({\rm Ly}\alpha)$/[erg s$^{-1}$] $=$ 43.2--43.6) magnitudes and Ly$\alpha$ luminosities. However, the number counts and the Ly$\alpha$ LF of LAEs are lower in the QSO field than the Control Field at the intermediate (NB921 $=$ 24.5--25.5 and $\log L({\rm Ly}\alpha)$/[erg s$^{-1}$] $=$ 42.6--43.2) magnitudes and Ly$\alpha$ luminosities. This is consistent with the fact that the sky distribution and the number density contours of LAEs in the QSO field exhibit only those equivalent to the mean to mean$-1\sigma$ densities of LAEs in the Control Field.

Meanwhile, the number counts of LBGs in the $z'$ band total magnitude show a clear  excess of the faintest ($z'=25.8$--26.0) LBGs in the QSO field against the Control Field. At the brighter magnitudes ($z'=25.0$--25.8), the number counts are consistent between the QSO and Control fields within statistical errors and cosmic variance. However, though cosistent within the uncertainties, there is a sign that the number counts of LBGs in the QSO field tend to be higher than that in the Control Field at even intermidiate to faint magnitudes ($z'=25.4$--25.8). This suggests that the high density clumps seen in the large scale structure of LBGs in the QSO field would comprise mainly relatively fainter LBGs. We confirm this trend in the sky distribution and the number density contours of LBGs in the QSO field where LBGs are plotted with symbols whose sizes are proportional to brightness ($z'$ band magnitudes) of the LBGs.
 
Moreover, the ACFs indicate that in the QSO field the LAEs are clustering over a wide range of angular scales $\sim 8$--20 comoving Mpc while LBGs small angular scales of $\sim 4$--8 comoving Mpc. The highest LBG density clump located in the west of the LBG large scale structure includes a bright (NB921$_{\rm total} =$ 23.78) and extended (diameter $\gtrsim 3''$ or 16 physical kpc) LAB candidate. As LABs are often found in/around overdense environments such as protoclusters, the highest density clump could be a protocluster at $z\sim 6.6$. This might support the validity of the highest density clump and the large scale LBG overdense structure it is associated with. 

All of those phenomena observed in the QSO field is in stark contrast to the Control Field where the number density distributions of LAEs and LBGs over the field are almost flat within the mean $\pm 1\sigma$ fluctuations, and both LAEs and LBGs exhibit no clustering signals in their ACFs. Hence, the QSO field is quite different from and seems to be more biased than a general blank field in terms of galaxy spatial and density distributions on a large scale and clustering of them.

We also investigate the possible effect of the QSO UV radiation on the formation of LAEs and LBGs. We find that star formation of the LAEs hosted by halos at the lowest mass end ($M_{\rm h}<1$--$3 \times 10^{10} M_{\odot}$) could be suppressed by the QSO UV radiation within $<$ 3 physical Mpc from the QSO (i.e., within the proximity region), but those of LAEs with higher halo masses and LBGs would not be suppressed. This can explain the fact that there are much less LAE candidates than LBG candidates within the QSO proximity region. This may also be caused by the difference in the probed volume between LAEs and LBGs ($\Delta z \sim 0.1$ for LAEs and $\Delta z \sim 0.9$ for LBGs) as more galaxies are detected in a larger volume. However, the sparsity of LAE candidates in the QSO proximity region compared to the number of LAE candidates outside the QSO proximity region may imply that formation of some of the lower mass LAEs was possibly suppressed by the QSO radiation.  

Our result presented in this paper is based on only one QSO. To see whether it is a universal trend of environments of high redshift QSOs or any diversity exists, we need to observe more QSOs. Also, the redshift of the $z=6.61$ QSO J0305--3150 we observed is in the red side of the bandpass of the NB921 filter where the sensitivity to LAEs is lower than nominal (see Figure \ref{NB921_LyaPeakDist}). Hence, we might have missed detecting some fraction of LAEs around the $z=6.61$ QSO, especially those located at the far side of the QSO. This might result in our finding of average or low LAE density in the proximity of the QSO. Therefore, it is important to find $z\sim6.6$ QSOs whose redshifts are located in the blue side of the bandpass of the NB921 filter where the sensitivity to LAEs reaches its peak. There are several ongoing high redshift ($z>6$) QSO searches exploiting wide area multiwavelength survey data. It is possible that those searches will find the QSOs whose redshifts best match the bandpass of the NB921 filter. In fact, at this moment, \citet{Venemans15} and \citet{Banados16} found such a QSO at $z=6.5412$, PSO J036.5078+03.0498, from the Pan-STARRS1 survey. Even though Subaru Suprime-Cam has been decommissioned recently, a similar NB921 filter is also available for its currently working successor Hyper Suprime-Cam that has a seven times wider FoV \citep{Konno17,Ouchi17,Shibuya17a,Shibuya17b}. Hence, if appropriate QSOs are found, it is possible to investigate galaxy densities around $z>6$ QSOs more accurately over even much larger volumes. This will yield a more general picture of a variety of $z>6$ QSO environments.

\acknowledgments

We are grateful to the staff at the Subaru Telescope for their support during our observations. We thank Kazuhiro Shimasaku for his helpful comments. We thank Akie Ichikawa and Tomoe Takeuchi for helping us conduct our observations, Tomoki Morokuma and Masao Hayashi for providing us the $R_c$, $i'$, $z'$ and $J$ band images of the SDF, Masaru Ajiki for providing us the detailed information about selecting the $z=6.6$ LAEs in the SDF, and Tomotsugu Goto and Yousuke Utsumi for providing us the information about their narrowband observations and data analysis. We also thank our referee for carefully reading and examining the manuscript and providing very valuable comments and suggestions that helped us improve the paper significantly. This research has benefitted from the SpeX Prism Spectral Libraries, maintained by Adam Burgasser at http://pono.ucsd.edu/{\textasciitilde}adam/browndwarfs/spexprism. K.O. acknowledges the Kavli Institute Fellowship at the Kavli Institute for Cosmology in the University of Cambridge supported by the Kavli Foundation. B.P.V. and F.W. acknowledge funding through the ERC grant ``Cosmic Dawn''. R.O. received support from CNPq (400738/2014-7) and FAPERJ (E-26/202.876/2015). D.R. acknowledges support from the National Science Foundation under grant number AST-1614213 to Cornell University. The authors recognize and acknowledge the very significant cultural role and reverence that the summit of Mauna Kea has always had within the indigenous Hawaiian community. We are most fortunate to have the opportunity to conduct observations from this mountain.

%

\vspace{5mm}
\facility{Subaru (Suprime-Cam)}


\software{SDFRED2 \citep{Yagi02,Ouchi04}, IRAF (http://iraf.noao.edu/), SExtractor\\ \citep{BA96}}




\appendix

\section{A Check on the Nonexistence of $z \sim 6.6$ QSOs in the Control Field SDF\label{QSOcheck}}
For the SDF to be the Control Field, there should not exist any $z\sim6.6$ QSOs within the field. Thus, we check for the nonexistence of such QSOs in the SDF. We use the SDF version 1.0 public $B$ and $V$ band images \citep[][see footnote 2]{Kashikawa04}, the SDF $R_c$, $i'$ and $z'$ band images deeper than the public images \citep{Poznanski07,Graur11,Toshikawa12} and the SDF $J$ band image \citep[M. Hayashi et al., in preparation;][]{Toshikawa12} taken with WFCAM on UKIRT \citep{Casali07} to see if there are any objects whose colors and magnitudes are consistent with those expected for a $z \sim 6.6$ QSO. The limiting magnitudes ($3\sigma$, $2''$ aperture) of these images are $(B,~V,~R_c,~i',~z',~J)=$ (28.45, 27.74, 28.35, 27.72, 27.09, 23.30--24.80). Note that the limiting magnitude of the SDF $J$ band image is not uniform because it is a mosaic of 9 regions with different depths (23.3--24.8 at $3\sigma$ level), each of which was imaged by one of the WFCAM detectors \citep[M. Hayashi et al., in preparation;][]{Toshikawa12}.


\begin{figure}
\noindent
\epsscale{1.1}
\plotone{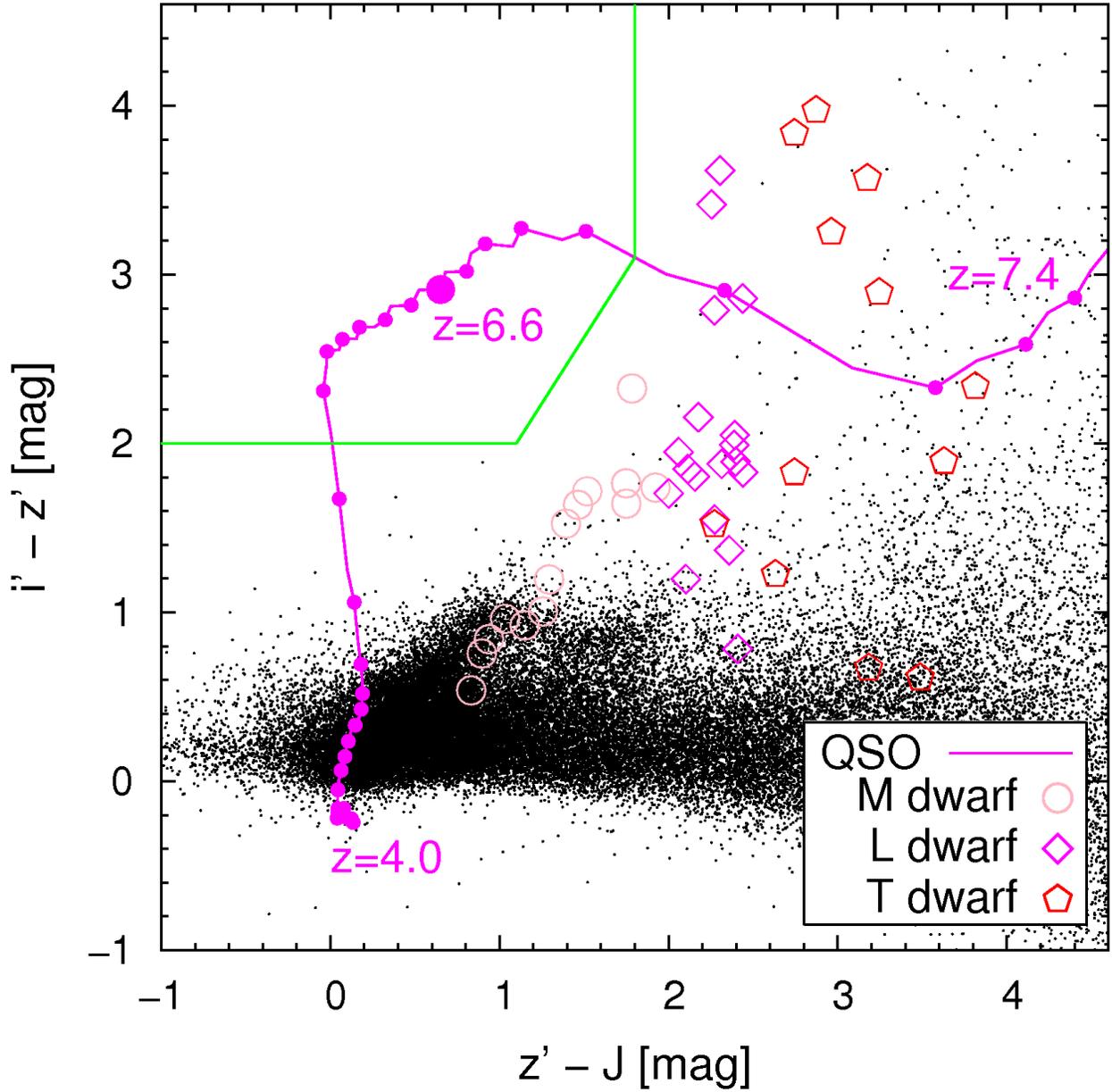}
\caption{The $i'-z'$ versus $z'-J$ color-color diagram of all the objects detected in the SDF $J$ band image to $J < J_{3\sigma}$ (shown by dots). The solid curve shows the redshift evolution track of colors of QSOs at $z=4.0$--7.4 calculated by using the QSO model developed by \citet{Kashikawa15}. On the model QSO track, we denote by filled circles redshits from $z=4.0$ to 7.4 by $\Delta z = 0.1$ step. The larger filled circle indicates the $z=6.6$ QSO colors. We also plot the colors of M/L/T dwarfs (types M3--M9.5, L0--L9.5 and T0--T8) using the spectra taken from \citet{Burgasser04,Burgasser06a,Burgasser06b,Burgasser08,Burgasser10} and \citet{Kirkpatrick10}. The solid line shows the selection window of $z\sim6$--7 QSOs, the criteria (\ref{Criteria-4}) in Appendix A. No object satisfying the QSO selection criteria was found in the SDF.\label{2CD_QSO}}
\end{figure} 

To determine selection criteria of $z\sim6.6$ QSOs, we calculate redshift evolution of $i'-z'$ and $z'-J$ colors of $z=4$--7.4 QSOs by using the model QSO spectra created by \citet{Kashikawa15} and the $i'$, $z'$ and $J$ band filter response curves, and plot it in Figure \ref{2CD_QSO}. In the figure, we also calculate and plot colors of possible contaminants, M/L/T dwarf stars, using their actual spectra provided by \citet{Burgasser04,Burgasser06a,Burgasser06b,Burgasser08,Burgasser10} and \citet{Kirkpatrick10} at the SpeX Prism Spectral Libraries (see footnote 3). As seen in the figure, we can clearly isolate $z\sim6.6$ QSOs (more specifically $z\sim 6$--7 QSOs) from M/L/T dwarfs using the $i'-z'$ and $z'-J$ colors. Based on this, we use the following $z\sim 6$--7 QSO selection criteria (all magnitudes are measured in a $2''$ aperture) to examine if there exist any QSOs at $z\sim6.6$ in the SDF.
\begin{eqnarray}
B > B_{2\sigma},~V > V_{2\sigma},~R_{\rm c} > R_{c2\sigma} \nonumber\\
i' - z' > 2.0 \nonumber\\
z' - J < 1.9 \nonumber\\
i' - z' > 1.6(z' - J) + 0.3 \nonumber\\
J < J_{3\sigma}~{\rm (detection~image)}
\label{Criteria-4}
\end{eqnarray}
Here, $B_{2\sigma}=28.89$, $V_{2\sigma}=28.18$ and $R_{c2\sigma}=28.79$ ($J_{3\sigma}=23.3$--24.8 dependent on the region in the $J$ image) are $2\sigma$ ($3\sigma$) $2''$ aperture limitting magnitudes of the SDF $B$, $V$ and $R_c$ ($J$) band images, respectively. We use the $J$ band as the object detection image because most $z \gtrsim 6$ QSOs previously found by SDSS (Canada-France High-$z$ Quasar Survey (CFHQS)) tend to have brighter magnitudes in $J$ band than in $i$ ($i'$) and $z$ ($z'$) bands \citep{Fan01,Fan03,Fan04,Fan06,Willott07,Willott09,Willott10a}. 

The PSFs of the SDF images are ($B$, $V$, $R_c$, $i'$, $z'$, $J$) $=$ ($0.''98$, $0.''98$, $1.''15$, $0.''93$, $0.''97$, $1.''11$). We convolve the $i'$ and $z'$ band images to $1.''11$ to measure the $i'-z'$ and $z'-J$ colors of objects with common PSF and aperture. Also, the pixel scale and the geometry of the $J$ band (WFCAM) image is matched to those ($0.''202$ pixel$^{-1}$) of the $B$, $V$, $R_c$, $i'$ and $z'$ (Suprime-Cam) images. Running SExtractor version 2.8.6 \citep{BA96}, we first detect objects in the $J$ band image and then measure magnitudes in $B$, $V$, $R_c$, $i'$, $z'$ and $J$ band images to construct the $J$-detected object catalog. We consider an area larger than five contiguous pixels with a flux (mag arcsec$^{-2}$) greater than $2\sigma$ (two times the background rms) to be an object. Finally, we apply the QSO selection criteria (\ref{Criteria-4}) to the $J$-detected object catalog and find no object consistent with $z\sim6$--7 QSOs in the SDF.

We also confirm the nonexistence of $z\sim6.6$ QSOs in SDF in another way. It is known that some fraction of LAEs have active galactic nuclei (AGNs). For example, \citet{Ouchi08} and \citet{Konno16} found that the brightest LAEs at $z=2.2$, 3.1 and 3.7 with Ly$\alpha$ luminosities of $\log L({\rm Ly}\alpha)/({\rm erg~s}^{-1}) \gtrsim 43.4$--43.6 always host AGNs. In SDF, \citet{Taniguchi05} detected 58 $z\sim6.6$ LAE candidates and found no exceptionally bright $z\sim6.6$ LAEs in $z'$ and NB921 bands. More specifically, their magnitudes are $z' \gtrsim 25.7$ and ${\rm NB921} \gtrsim 24.1$ mag compared to $z' = 22.02$ and ${\rm NB921} = 21.93$ of the $z=6.61$ QSO J0305--3150 which we measured using our imaging data of the QSO field. Moreover, \citet{Kashikawa06,Kashikawa11} spectroscopically identified 42 out of the 58 $z\sim6.6$ LAE candidates and found no AGN. Also, there is no $z\sim6.6$ LAE with $\log L({\rm Ly}\alpha)/({\rm erg~s}^{-1}) \gtrsim 43.4$ in SDF. Hence, in the present study, we consider that SDF contains no $z\sim6.6$ QSO at least equivalent to the $z=6.61$ QSO J0305--3150. 

\section{Evaluating the Contamination due to the Lack of B, V and R$_c$ Bands\label{lackBVR}}
Before applying the LAE selection criteria (\ref{Criteria-1}) and (\ref{Criteria-2}) to our NB921-detected object catalogs, we investigated the impact of omitting the criteria $B > B_{3\sigma}$, $V > V_{3\sigma}$ and $R_c > R_{c 3\sigma}$ on the reliable selection of LAEs. For this, we used the public SDF data version 1.0 which are the same SDF images and catalogs used by \citet{Taniguchi05} for their study of $z\sim 6.6$ LAEs (and we also use the same SDF $i'$, $z'$ and NB921 images in this study). 

We first combined the public SDF $B$, $V$, $R_c$, $i'$, $z'$ and NB921 catalogs of NB921-detected objects and applied the same criteria (i.e., the criteria (\ref{Criteria-1}) and (\ref{Criteria-2}) plus $B > B_{3\sigma}$, $V > V_{3\sigma}$ and $R_c > R_{c 3\sigma}$) used by \citet{Taniguchi05} to the combined catalog to see if we can correctly re-select the same 58 $z\sim 6.6$ LAE candidates that \citet{Taniguchi05} previously selected. We confirmed that we could select 57 LAE candidates, of which 56 are the same as the LAE candidates \citet{Taniguchi05} selected. 

The $i'$ and $z'$ band magnitudes of one of the two LAE candidates that we failed to re-select listed in Table 2 in the \citet{Taniguchi05} paper turned out to be wrong (Taniguchi et al.~2015, private communication). This is the object No.~22, SDF J132338.6+272940, and its correct magnitudes are $i'=28.04$ and $z'=28.75$ or $z' > 27.81$ ($1\sigma$ limit). After correcting the magnitudes, this LAE candidate coincided with our remaining one LAE candidate (their coordinates also coincided).

The last \citet{Taniguchi05} LAE candidate we failed to re-select is the same as one of the $z\sim 6.6$ LAEs \citet{Kodaira03} selected in their SDF $z\sim 6.6$ LAE survey (preliminary to the \citet{Taniguchi05} study) by using the previous shallower SDF NB921 image with a better PSF ($0.''9$) than that ($0.''98$) of the public SDF NB921 image and using a smaller diameter aperture ($1.''8$) for photometry than that ($2.''0$) used to produce the SDF public photometric catalogs. The PSF of the public SDF NB921 image is the one convolved to the worse PSF than the original one for the aperture photometry purpose. We failed to re-select this LAE candidate because it blends with its neighboring object in the public SDF NB921 image while \citet{Kodaira03} could select it as it is not blended in the higher resolution NB921 image. According to Taniguchi et al.~(2015 private communication), \citet{Taniguchi05} also used the NB921 image before the convolution so that they could select and include the \citet{Kodaira03} LAE candidate in their final LAE sample. This is the object No.~5, SDF J132418.3+271455, in Table 2 in the \citet{Taniguchi05} paper or the object No.~2 in Table 1 in the \citet{Kodaira03} paper. Hence, excluding this \citet{Kodaira03} LAE, we eventually confirmed that we could correctly re-select the 57 LAE candidates previously selected by \citet{Taniguchi05} using the the criteria (\ref{Criteria-1}) and (\ref{Criteria-2}) plus $B > B_{3\sigma}$, $V > V_{3\sigma}$ and $R_c > R_{c 3\sigma}$. 


\begin{figure}
\epsscale{1.17}
\plotone{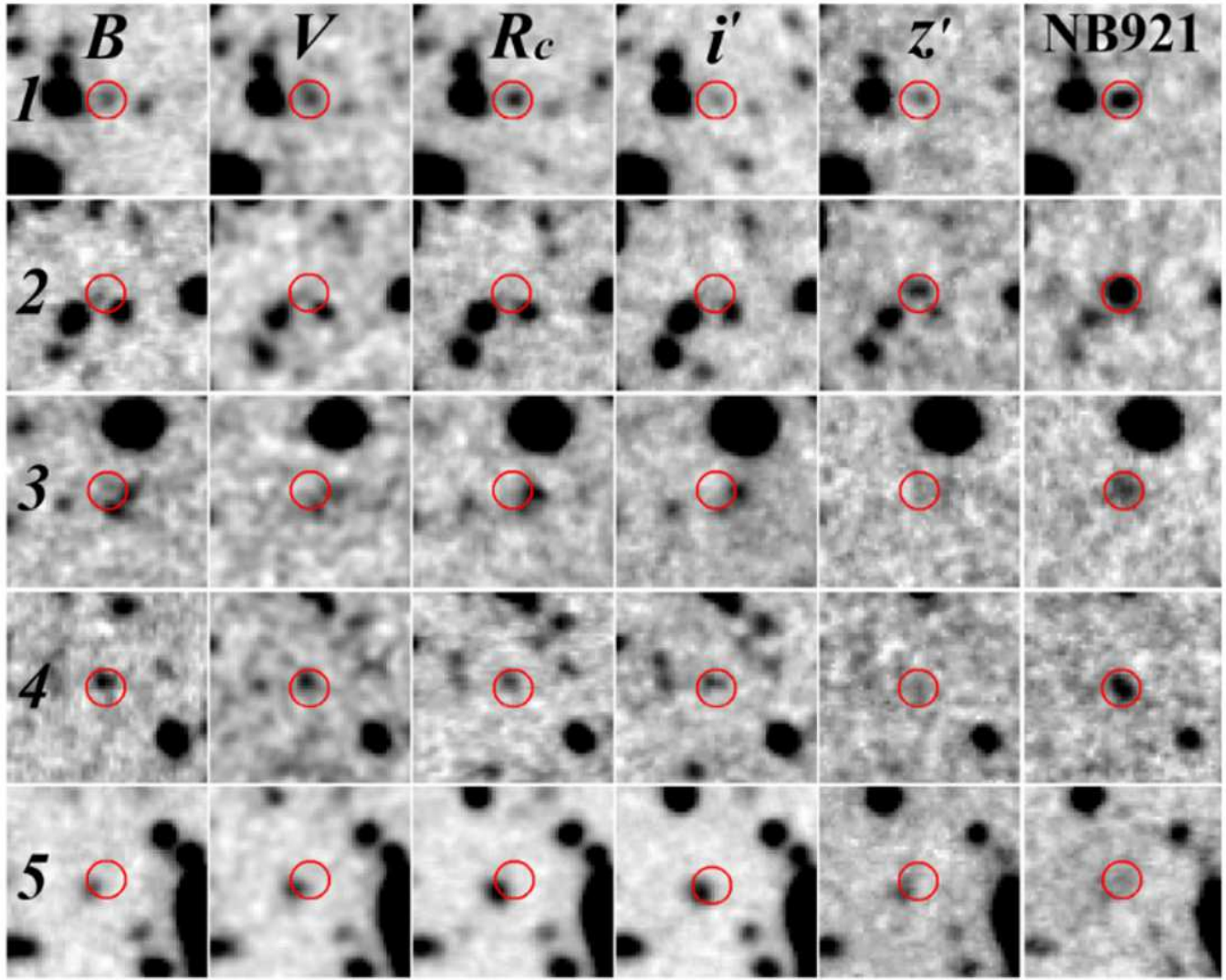}
\caption{Multi-waveband images of the five sources in the Subaru Deep Field (cases 1 to 5) selected by using the citeria (\ref{Criteria-1}) and (\ref{Criteria-2}) (see Section \ref{LAE-Selections}) without imposing the null detections in the wavebands bluewards $z\sim6.6$ Ly$\alpha$ ($B > B_{3\sigma}$, $V > V_{3\sigma}$, and $R_c > R_{c 3\sigma}$). The size of each image is $10'' \times 10''$. North is up and east to the left. The $2''$ aperture circles (the same aperture size used for our photometry to select LAE candidates) are shown at the source detection positions (i.e., The sources are detected in the NB921 image by using SExtractor). \label{woBVR_Objects}}
\end{figure}

Based on this, we applied the criteria (\ref{Criteria-1}) and (\ref{Criteria-2}) without $B > B_{3\sigma}$, $V > V_{3\sigma}$ and $R_c > R_{c 3\sigma}$ to the public SDF NB921-detected object catalog to see what would happen. The purpose of $B > B_{3\sigma}$, $V > V_{3\sigma}$ and $R_c > R_{c 3\sigma}$ (i.e., null detections in the wavebands bluewards of $z\sim6.6$ Ly$\alpha$) is to reduce the contamination from low-$z$ line emitters and stars as the fluxes of $z\sim6.6$ LAEs at these wavelengths should be absorbed by IGM \citep{Madau95}. Hence, omitting these null detection criteria, we expected to select some additional objects, some of which could be contaminants. We actually detected 5 additional objects. We call them cases 1--5 and show their $B$, $V$, $R_c$, $i'$, $z'$ and NB921 band images in Figure \ref{woBVR_Objects}. We carefully visually inspected these images to see if they are any sort of contaminants. 

As for the case 1, an object is seen in all the six band images at the SExtractor detection position (center of each image denoted by a circle of $2''$ aperture used for photometry in Figure \ref{woBVR_Objects}). Hence, it is likely a contaminant unless an object seen in only the $B$, $V$ and $R_c$ images coincidentally exists at the same position as a $z\sim6.6$ LAE seen in only $i'$, $z'$ and NB921 bands.

For each of the cases 2--5, an object is seen at the detection position in either only the $z'$ and NB921 images or only the NB921 image. On the other hand, one or two objects are seen at the locations close to but slightly separate from the detection position centers in $2''$ aperture circles either in the 4--6 band images. They are obviously not the objects detected in the center of the $2''$ apertures by SExtractor, but some parts or most of their fluxes enter the apertures used for the photometry and thereby affect the magnitude measurements to some extent, resulting in $>3\sigma$ detection in either or all the $B$, $V$ and $R_c$ bands. In all the cases 2--5, $i'$ band $2''$ aperture magnitudes are fainter than $2\sigma$ (i.e., $i' > i'_{2\sigma, {\rm SDF}}$), and the neighboring objects are even fainter or not seen in $z'$ and NB921 bands, so their impacts on the photometry on $i'$, $z'$ and NB921 are considered negligible. Hence, in all the cases 2--5, the objects seen at the detection positions are likely $z\sim 6.6$ LAEs. 

We conclude that the contamination rate from additionally selected objects (cases 1--5) by omitting the criteria $B > B_{3\sigma}$, $V > V_{3\sigma}$ and $R_c > R_{c 3\sigma}$ is low (1/5). Similarly to the possible $z\sim6.6$ LAE candidates from the cases 2--5, 57 out of the 58 $z\sim6.6$ LAE candidates in the SDF previously selected by \citet{Taniguchi05} are not significantly detected in $i'$ band (56 have $i' > i'_{2\sigma}$ and 1 has $i' > i'_{3\sigma}$. The remaining one is detected in $i'$ but only very marginally at $3.3\sigma$ level). This implies that whether we impose $B > B_{3\sigma}$, $V > V_{3\sigma}$ and $R_c > R_{c 3\sigma}$ or not, the $i' - z' > 1.3$ color in the criteria (\ref{Criteria-1}) and $i' > i'_{2\sigma}$ in the criteria (\ref{Criteria-2}) preferentially select $z\sim 6.6$ LAE candidates with $i' > i'_{2-3\sigma}$ and by itself effectively reduce contamination. This criterion $i' \gtrsim i'_{2-3\sigma}$ could replace the criteria $B > B_{3\sigma}$, $V > V_{3\sigma}$ and $R_c > R_{c 3\sigma}$ because $z\sim 6.6$ Ly$\alpha$ emission is located in the middle of the $z'$ band wavelengths and the entire $i'$ band is bluewards of $z\sim 6.6$ Ly$\alpha$ (see $i'$, $z'$ and NB921 bands in Figure \ref{FilterTransmission}). Hence, omitting the criteria $B > B_{3\sigma}$, $V > V_{3\sigma}$ and $R_c > R_{c 3\sigma}$ would not significantly increase the number of contaminants. 

Based on this analysis, we decide to adopt the criteria (\ref{Criteria-1}) and (\ref{Criteria-2}), which omits $B > B_{3\sigma}$, $V > V_{3\sigma}$ and $R_c > R_{c 3\sigma}$, to select $z\sim6.6$ LAE candidates in the $z=6.61$ QSO field and the Control Field. This means that as the Control Field LAE sample, we adopt the 63 objects: the 58 SDF $z\sim6.6$ LAE candidates previously selected by \citet{Taniguchi05} and re-selected by us plus the 5 additional SDF $z\sim6.6$ LAE candidates we have selected without the criteria $B > B_{3\sigma}$, $V > V_{3\sigma}$ and $R_c > R_{c 3\sigma}$ (cases 1--5 in Figure \ref{woBVR_Objects}). We include the case 1 object in the Control Field LAE candidate sample even though it could be a contaminant. Also, one of the 58 SDF $z\sim6.6$ LAE candidates has been spectroscopically identified as a low-$z$ [OIII] emitter by \citet{Kashikawa11}. However, we also include it in the Control Field LAE candidate sample. This is because we cannot remove such contaminants from the LAE candidate sample in the QSO field without its $B$, $V$ and $R_c$ band images or spectroscopy. Hence, for the fair comparison of the LAE candidates in the QSO and Control fields, we include such contaminants in the LAE candidate samples in the both fields.    

\end{document}